\documentclass[11pt]{article}
\usepackage{amsmath,epsf,amssymb,latexsym,amsthm,setspace,bbm,array,pifont,enumerate,mathrsfs}
\usepackage{xcolor}
\usepackage{jheppub}
\usepackage{mathtools}
\usepackage{tikz}
\usepackage{subcaption}
\usepackage[vcentermath]{youngtab}
\usepackage{enumitem}
\usepackage{amsmath,amsfonts,amsbsy,amssymb,array,accents,dsfont}
\usepackage{enumerate}
\usepackage{graphicx,verbatim} 
\usepackage{caption} 
\usepackage{subcaption} 
\usepackage{mathrsfs}
\usepackage{upgreek}
\usepackage{bm}
\usepackage{ytableau}
\usepackage{enumitem}
\setlist[itemize,1]{label=\textbullet}

\usepackage{float}
\usepackage{hyperref}
\usepackage{overpic}
\usepackage{xcolor}

\allowdisplaybreaks

\renewcommand{\P}{{\mathbb{P}}}
\newcommand{\R}{{\mathbb{R}}}
\newcommand{\Z}{{\mathbb{Z}}}

\newcommand{\ee}{{\mathrm{e}}}
\newcommand{\ii}{{\mathrm{i}}}
\newcommand{\rep}[1]{{\boldsymbol{{#1}}}}
\newcommand{\repconj}[1]{{\overline{\boldsymbol{{#1}}}}}
\newcommand{\tL}{{\text{\normalfont\tiny{L}}}}
\newcommand{\tR}{{\text{\normalfont\tiny{R}}}}
\newcommand{\Ell}{{\mathbb{E}}}
\newcommand{\Chi}{{\mathcal{X}}}
\newcommand{\wsN}{{\mathrm{N}}}

\title{Rational points in the 6d supergravity landscape and simple current extensions}
\author{Guglielmo Lockhart and Yann Proto}
\affiliation{Bethe Center for Theoretical Physics, Universit\"at Bonn, Bonn, Germany}
\emailAdd{glockhar@uni-bonn.de}
\emailAdd{yproto@uni-bonn.de}
\abstract{
We investigate a recently identified class of six-dimensional supergravities without tensor multiplets whose primitive BPS strings are described by a rational superconformal field theory. Rationality imposes severe constraints on the supersymmetric spectrum of their BPS strings and provides an effective way to study this class of models, despite the absence of a conventional geometric realization within F-theory. We find that, in each model,  the rationality constraints are nontrivially satisfied and uniquely determine the elliptic genus of the strings, providing a new consistency criterion satisfied by this exotic class of  candidate quantum gravity theories. In all cases, the left-moving Kac--Moody--Virasoro algebra on the string worldsheet is extended by higher-spin currents, corresponding to discrete symmetries of the 2d SCFT. This allows us to determine the global form of the six-dimensional gauge group for all the members of this class of models.}

\begin{document}

\maketitle

\section{Introduction}
Rational models play a distinguished role in the study of conformal field theory in two dimensions, providing a setting in which many physical quantities can be computed algebraically and, in certain contexts, allowing for systematic classification. Because of these features, rational CFTs have also had a profound impact in the study of string theory compactifications. Exact backgrounds based on rational worldsheet theories were first constructed by Gepner~\cite{Gepner:1987qi} using tensor products of N=2 minimal models. Gepner models describe special points in the moduli space of Calabi--Yau sigma models where the theory becomes exactly solvable, and they have been instrumental in elucidating various properties of the target space geometry, such as mirror symmetry~\cite{Greene:1990ud}.

A hallmark of rational theories is the presence of additional holomorphic currents, which extend the Virasoro algebra to a larger chiral algebra~\cite{Zamolodchikov:1985wn}. These integer spin currents reorganize the infinitely many Virasoro primaries into a finite set of extended conformal families. The appearance of extra holomorphic currents is typically tied to special points in the parameter space of a conformal field theory. For Calabi--Yau sigma models, these points are associated with distinguished loci in the moduli space of the target space geometry where arithmetic structures emerge~\cite{Gukov:2002nw,Chen:2005gm,Kidambi:2024vwl}. They are further singled out by the attractor mechanism~\cite{Ferrara:1995ih} which, in Calabi--Yau compactifications of type II supergravity, fixes the asymptotic values of complex structure moduli in the near-horizon region of supersymmetric black holes~\cite{Moore:1998pn}; see~\cite{Candelas:2023yrg,Grimm:2024fip,Jockers:2025fgv,Blesse:2025dch} for recent discussion and applications.

Rational conformal field theories have been employed extensively to study the properties of critical strings in string-theoretic compactifications, and it is also natural to ask whether they play a meaningful role in describing non-critical strings, such as the BPS strings of six-dimensional $\mathcal{N}=(1,0)$ supergravity theories. Because of their coupling to the 6d bulk degrees of freedom, these BPS strings are important probes of six-dimensional physics, and indeed they have been instrumental in recent investigations of the 6d supergravity landscape~\cite{Kim:2019vuc,Cheng:2021zjh,Cvetic:2021vsw,Dierigl:2022zll,Hayashi:2023hqa,Basile:2023zng,Kim:2024tdh,Birkar:2025rcg,Becker:2025xgy,Duque:2025kaa,Cheng:2025ikd,Baykara:2025gcc,Lee:2019skh,Tarazi:2021duw,Hamada:2025vga,Hamada:2023zol,Hamada:2024oap,Kim:2024eoa}. Their worldsheet theories are described by a rich set of two-dimensional superconformal field theories with (0,4) supersymmetry. Various properties of these strings can be inferred using anomaly inflow~\cite{Kim:2019vuc} or via constructions in F-theory~\cite{Haghighat:2015ega,Lee:2018urn}. Nevertheless, their dynamics remains largely unexplored. In this regard, locating solvable points in the moduli space of such BPS strings is particularly desirable. A step in this direction was taken in~\cite{Lockhart:2025lea}, which identified several supergravity models whose primitive BPS strings must be described by a rational theory. The goal of this work is to explore these theories in detail, through an analysis of their chiral algebras and supersymmetric spectra, and to examine some target-space implications of rationality.

\subsection{New rational SCFTs for 6d supergravity strings}
In six dimensions the absence of gauge and gravitational anomalies imposes stringent conditions on the allowed gauge groups and matter content of $\mathcal{N} = (1,0)$ supergravity theories. Models with few tensor multiplets are particularly constrained: the set of admissible non-abelian gauge groups is known to be finite~\cite{Kumar:2010ru}, and in recent years much progress has been made toward elucidating its structure~\cite{Kumar:2010ru,Kumar:2010am,Park:2011wv,Morrison:2012np,Morrison:2012js,Lee:2019skh,Tarazi:2021duw,Hamada:2023zol,Hamada:2024oap,Kim:2024eoa,Hamada:2025vga}. However, anomaly cancelation alone does not determine a definitive list of consistent theories. Indeed, there exist infinite families of supergravity models with $T\geq 9$ tensor multiplets~\cite{Schwarz:1995zw,Kumar:2010ru} which, despite being free of perturbative anomalies, are believed not to possess a consistent UV completion. Bottom-up arguments~\cite{Kim:2024eoa} strongly suggest that this is the case.

Six-dimensional $\mathcal{N}=(1,0)$ supergravity is a particularly interesting and nontrivial setting in which the string lamppost principle can be put to the test. This principle, also referred to as string universality~\cite{Kumar:2009us}, is the expectation that every consistent quantum theory of gravity should admit a realization within string theory, and has been shown to hold for supergravity in dimensions $d>6$~\cite{Adams:2010zy,Montero:2020icj,Hamada:2021bbz,Bedroya:2021fbu}. Many of the known anomaly-free six-dimensional supergravity models which are not excluded by the analysis of~\cite{Kim:2024eoa} actually admit a higher-dimensional uplift as F-theory compactifications on elliptically fibered Calabi--Yau threefolds. In this paper we will focus on the specific case of anomaly-free models without tensor multiplets, for which a partial classification has been achieved using graph-theoretic techniques~\cite{Hamada:2023zol,Hamada:2024oap}, yielding an extensive set of candidate theories. Among these, there also exist several models which do not appear to have any known realization within string theory---for instance, they evade standard consistency conditions such as the Kodaira constraint required of F-theory compactifications on elliptic threefolds. Determining whether these theories are consistent at the quantum level is therefore especially instructive, as it might point to novel regions of the landscape that do not (as of yet) have a home within string theory.

An important diagnostic for the consistency of supergravity theories that goes beyond anomaly cancelation is whether their BPS string probes are described by well-defined, unitary worldsheet theories. In particular, in the models we study in this paper, we will focus on a class of BPS strings whose infrared behavior is expected to be captured by a two-dimensional superconformal field theory (SCFT) with (0,4) supersymmetry. The 6d gauge symmetry is realized on the worldsheet as a left-moving current algebra, and the left-moving central charge of the SCFT decomposes as
\begin{align}
c_\tL = c_\tL^{\mathrm{gauge}} + c_\tL^{\mathrm{res}}~,
\end{align}
where the Sugawara contribution $c_\tL^{\mathrm{gauge}}$ is fixed by the 't Hooft anomalies carried by the BPS string. Unitarity of the SCFT imposes a positivity bound $c_\tL^{\mathrm{res}}\geq 0$ on the residual left-moving central charge, which can be used to relegate anomaly-free but inconsistent supergravity models to the swampland~\cite{Kim:2019vuc,Tarazi:2021duw,Lee:2019skh,Hamada:2023zol,Hamada:2024oap,Hamada:2025vga}. In addition to this positivity bound, a more detailed analysis of the BPS string worldsheet theory yields further insights. In particular, for theories without tensor multiplets, the BPS string of minimal charge is a stable object, and the conformal manifold of its SCFT encodes information about the moduli space of six-dimensional supergravity theories with $T=0$~\cite{Lockhart:2025lea}.

The models we consider in this paper are 6d $\mathcal{N} = (1,0)$ supergravity models without tensors for which the BPS string consistency condition is close to being saturated---more precisely, those for which the residual left-moving central charge falls within the range
\begin{align*}
0 < c_\tL^{\mathrm{res}} < 1~.
\end{align*}
In~\cite{Lockhart:2025lea}, five such models were identified within the catalogue of anomaly-free theories of~\cite{Hamada:2023zol}, whose gauge algebras are given respectively by
\begin{align*}
&\mathfrak{e}_6~,&
&\mathfrak{a}_7~,&
&\mathfrak{a}_5+\mathfrak{d}_4~,&
&\mathfrak{a}_3+\mathfrak{a}_3+\mathfrak{e}_7~,&
&\mathfrak{a}_3+\mathfrak{a}_{11}~.
\end{align*}
From a string-theoretic perspective, these theories are far removed from any known perturbative construction: for instance, they fail basic criteria satisfied by Calabi--Yau compactifications of F-theory, such as the Kodaira constraint or the presence of a neutral hypermultiplet which could serve as the universal hypermultiplet. Remarkably, for all five models the residual sector of the left-moving algebra appears to coincide with a unitary Virasoro minimal model. As a consequence, the SCFT is rational, and its left-moving sector possesses a finite number of Kac--Moody--Virasoro (KMV) characters. 

As we will see, a detailed study of the underlying left-moving KMV structure brings significant insight into these models. By making full use of modularity and rationality we will be able to compute the refined elliptic genus for each of the five theories, which encodes information about the BPS spectrum of the probe string.\footnote{The elliptic genus of the model with $\mathfrak{e}_6$ gauge algebra was previously derived in~\cite{Lockhart:2025lea} using vector valued modular forms. Here we reproduce that result by exploiting the $\mathrm{SL}(2,\Z)$ properties of KMV characters.} The computation, although technically involved, only depends on few basic physical assumptions. Notably, for each model, we establish the existence of a \emph{unique} object satisfying the defining properties obeyed by the elliptic genus of the putative probe string SCFT. The very existence of such an object---a highly nontrivial linear combination of hundreds of KMV characters---is a rare feature. Moreover the resulting expressions are sensible and only involves small integer multiplicities for the various characters. We view these results as evidence that the underlying worldsheet SCFTs are well-defined. By extension, we interpret this as evidence that the five associated six-dimensional supergravity theories pass nontrivial consistency checks that go beyond anomaly cancelation, despite their lack of an explicit realization for them within string theory.

\subsection{Higher-spin currents and extensions}
An outcome of our analysis is the appearance, in all five models, of higher-spin currents that enlarge the left-moving KMV algebra. The presence of such simple currents places strong constraints on the structure of the worldsheet theory, projecting out certain primary operators and organizing the remaining ones into extended families. It is perhaps not too surprising that, at highly rational points in the moduli space of the probe string SCFT, such holomorphic fields could appear. In fact, a subset of these simple currents---those acting purely on the gauge sector---admits a natural interpretation: it is in one-to-one correspondence with $\pi_1(G)$, the fundamental group of the spacetime gauge group $G$. In this way, the worldsheet theory encodes information about the global form of the gauge group, which turns out to be non-simply connected in four of the five cases under study. In addition, we observe the presence of ``mixed'' KMV currents, which may also find a natural explanation in terms of one-form symmetries of the six-dimensional theory~\cite{Lockhart:2026gvb}.

More broadly, the appearance of simple currents in the chiral algebra of probe BPS strings is a general phenomenon which occurs in a variety of settings much less exotic than the one studied here---for instance, heterotic compactifications on asymmetric orbifolds. Despite its richness, this structure has often been overlooked in the literature and is worthy of further attention. Its general features, as well as its implications for the swampland program, will be discussed in the separate work~\cite{Lockhart:2026gvb}.

\subsection{Organization}
The rest of this article is organized as follows. We begin in section~\ref{sec:BPSstrings} by reviewing basic aspects of six-dimensional $\mathcal{N} = (1,0)$ supergravities, summarizing the main features of BPS string probes, and introducing the five models that form the focus of this work. In section~\ref{sec:KMValgebra}, we set up our conventions for the KMV algebra underpinning the subsequent discussion, outline the computational methods, and determine the refined elliptic genera of the hyperplane string corresponding to the five models. In section~\ref{sec:simplecurrents} we discuss chiral algebra extensions and Kac--Moody simple currents. In section~\ref{sec:ellipticgenera} we present the extended left-moving algebras for the five theories under study and use them to give their elliptic genera in compact form in terms of extended characters. Finally, section~\ref{sec:discussion} contains our conclusions, including a determination of the global form of the 6d gauge group in our five models, and further directions of research. Technical details on Virasoro minimal models and Kac--Moody algebras are collected in appendices~\ref{app:minimalmodels} and~\ref{app:KMconventions}, respectively. A dataset containing the KMV expansions of the five elliptic genera is provided in~\cite{KMVdata}.

\section{A brief review of BPS strings in 6d supergravities}
\label{sec:BPSstrings}
In this section we recall some general aspects of six-dimensional supergravities with $\mathcal{N}=(1,0)$ supersymmetry, with particular emphasis on theories without tensor multiplets. We review the main features of their BPS string probes and introduce the five models that will be studied in the remainder of this work. 

\subsection{Perturbative anomalies in 6d}
We consider six-dimensional supergravity theories preserving eight supercharges, with a non-abelian gauge group
\begin{align}
G = G_1\times\cdots\times G_N~,
\label{eq:6dgaugegroup}
\end{align}
where each factor $G_\alpha$ is a simple Lie group.\footnote{We ignore subtleties related to the global form of $G$ for the moment; we will return to these in section~\ref{sec:discussion}.} The spectrum comprises four types of supersymmetry multiplets: a single gravity multiplet containing a self-dual two-form, $T$ tensor multiplets each containing an anti-self-dual two-form, $V$ vector multiplets in the adjoint representation of $G$, and $H$ hypermultiplets in a representation $\mathscr{H}$ of $G$. Charges under the two-form fields take values in a full-rank integral self-dual lattice $\Gamma\subset \mathbb{R}^{1,T}$~\cite{Seiberg:2011dr}.

Gravitational and gauge anomalies are encoded in the anomaly eight-form $I_8$, which receives contributions from the symplectic Majorana--Weyl fermions of the above multiplets. Consistency of the theory requires cancelation of these anomalies via the generalized Green--Schwarz mechanism~\cite{Green:1984sg,Green:1984bx,Sagnotti:1992qw}: the anomaly polynomial must factorize as 
\begin{align}
I_8 = \tfrac{1}{2} X_4\cdot X_4~,
\label{eq:GS}
\end{align}
where $\cdot$ denotes the $\mathbb{R}^{1,T}$ inner product. Here, $X_4$ encodes a collection of four-forms constructed from the Riemann and Yang--Mills curvatures $R$ and $F_\alpha$. It takes the form
\begin{align}
X_4=\frac{1}{2}a \operatorname{tr}R^2+\sum_{\alpha=1}^N \frac{2}{\lambda(G_\alpha)}b_\alpha \operatorname{tr}F_\alpha^2~,
\end{align}
for some vectors $a,b_1,\dots,b_{N}\in\mathbb{R}^{1,T}$. The integers $\lambda(G_\alpha)$ depend on the gauge group simple factors $G_\alpha$.\footnote{Explicitly, $\lambda(\mathrm{A}_r)=\lambda(\mathrm{C}_r)=1$, $\lambda(\mathrm{B}_r)=\lambda(\mathrm{D}_r)=\lambda(\mathrm{G}_2)=2$, $\lambda(\mathrm{E}_6)=\lambda(\mathrm{F}_4)=6$, $\lambda(\mathrm{E}_7)=12$, $\lambda(\mathrm{E}_8)=60$.} The factorization~\eqref{eq:GS} imposes strong constraints on the spectrum. Cancelation of the irreducible $\operatorname{tr}R^4$ term yields the familiar condition
\begin{align}
H-V+29\,T = 273~.
\end{align}
Similarly, cancelation of the irreducible $\operatorname{tr}F_\alpha^4$ terms provides $N$ additional relations between the vector and hypermultiplet sectors. The remaining terms in $I_8$ determine the $\mathbb{R}^{1,T}$ inner products
\begin{align}
&a\cdot a~,&
&a\cdot b_\alpha~,&
&b_\alpha\cdot b_\beta~.
\label{eq:anomalylattice}
\end{align}
When a consistent choice of $a$ and $b_\alpha$ exists, the anomaly can be cancelled by including an appropriate Green--Schwarz term in the 6d action. The gravitational and gauge kinetic terms are encoded in a $\R^{1,T}$ vector $j$; their positivity requires
\begin{align}
j\cdot a &< 0 ~,&
j\cdot b_\alpha &> 0 ~.
\end{align}

\subsection{Rigid 6d \texorpdfstring{$\mathcal{N} = (1,0)$}{N=(1,0)} supergravity theories}
Cancelation of the six-dimensional perturbative anomaly~\eqref{eq:GS} significantly constrains the set of possible consistent supergravities. For $T<9$, only a finite number of distinct gauge groups $G$ and matter content charged under the non-abelian factors of $G$ satisfy the anomaly cancelation conditions while admitting physical kinetic terms. A complete enumeration of perturbatively consistent theories is, in principle, possible. Partial classifications have been achieved through various methods; see, for instance,~\cite{Kumar:2010ru,Kumar:2010am,Hamada:2023zol,Hamada:2024oap,Kim:2024eoa,Hamada:2025vga,Baykara:2025gcc}.

A broad class of six-dimensional $\mathcal{N} = (1,0)$ theories may be obtained from F-theory compactified on elliptically fibered Calabi--Yau threefolds~\cite{Morrison:1996na,Morrison:1996pp}. The non-abelian gauge group emerges from the codimension-one singularities of the elliptic fibration, situated along a locus $\cup_\alpha \Sigma_\alpha$ in the base of the Calabi–Yau. The type of singular fiber over a generic point of each component $\Sigma_\alpha$ is characterized by a set of strictly-positive integers $\nu_\alpha$. The resulting six-dimensional theory satisfies the Kodaira condition
\begin{align}
-j\cdot(12a+\sum_{\alpha=1}^N\nu_\alpha b_\alpha) >0~.
\end{align}
Every such theory contains a neutral hypermultiplet, often referred to as the \emph{universal hypermultiplet}, corresponding to the overall volume modulus of the base of the Calabi--Yau manifold.

In theories satisfying the anomaly cancelation conditions---together with certain global anomaly constraints~\cite{Bershadsky:1997sb,Suzuki:2005vu}---the inner products ~\eqref{eq:anomalylattice} of the anomaly polynomial coefficients are all integers~\cite{Kumar:2010ru}. Hence, the vectors $a,b_1,\dots,b_N$ of a  consistent six-dimensional $\mathcal{N} = (1,0)$ supergravity span an integral sublattice $\Lambda\subset\Gamma$ of the charge lattice, called the anomaly lattice. This lattice can be used to recover the topological data for an underlying Calabi--Yau compactification of F-theory, when one exists.

There exist many candidate anomaly-free six-dimensional $\mathcal{N} = (1,0)$ theories that do not appear to admit a standard F-theory realization. This is the case if the Kodaira condition is violated, or if neutral hypermultiplets are absent from the spectrum. The quantum consistency of such models often remains uncertain, but several consistency checks have been proposed in the literature~\cite{Kim:2019vuc,Lee:2019skh,Tarazi:2021duw,Basile:2023zng,Hamada:2025vga}, in relation to the swampland program. A minimal $T=0$ example, discussed in~\cite{Kumar:2010am,Kim:2019vuc}, is a theory with gauge group $G = \mathrm{A}_7$ and a hypermultiplet in the ``box'' representation,
\begin{align}
\mathscr{H} = \rep{336}={\tiny\yng(2,2)}~.
\end{align}
Despite its exotic matter content and the apparent lack of a stringy origin, this model evades all known consistency constraints. In the following, we will show that the worldsheet SCFT of its BPS strings admits a unitary $\mathfrak{a}_7$ current algebra and a well-defined $\mathfrak{a}_7$-refined elliptic genus, lending further support to its consistency. The above theory is an example of a six-dimensional supergravity in which all hypermultiplets are charged under the non-abelian gauge group $G$. In such models, any deformation along a scalar modulus necessarily breaks $G$ to a strictly smaller subgroup. We refer to such theories as \emph{rigid}, following the terminology of~\cite{Kumar:2010am}. 

\subsection{Theories with \texorpdfstring{$T=0$}{T=0} and hyperplane strings}
Six-dimensional supergravity theories admit non-critical strings that act as sources for the (anti-)self-dual two-forms in the gravity and tensor multiplets. Their couplings to these two-form fields are specified by a charge $Q\in\Gamma$ in the charge lattice. Generically, for a given $Q$ the moduli space of low-energy configurations of a BPS string can possess multiple disconnected branches along which the string's dynamics are described by distinct (0,4) SCFTs. In particular if $Q\cdot Q \geq 0$ the string will possess a branch of moduli space along which the low-energy dynamics is described by a two-dimensional (0,4) SCFT whose $\mathfrak{su}(2)$ superconformal R-symmetry is identified with $\mathfrak{su}(2)\subset\mathfrak{so}(4)$ rotations transverse to the string. The left- and right-moving central charges, determined by anomaly inflow~\cite{Kim:2019vuc}, are given by
\begin{align}
c_\tL &= 3 \,Q\cdot Q-9\,a\cdot Q+2~,&
c_\tR &= 3 \,Q\cdot Q-3\,a\cdot Q~.
\label{eq:cc}
\end{align}
We remark that there is an additional contribution $(c_\tL,c_\tR)=(4,6)$ from the center-of-mass degrees of freedom of the BPS string, corresponding to a decoupled free hypermultiplet, which we do not include in~\eqref{eq:cc}.

The 6d gauge symmetry $G$ is realized on the string worldsheet as an affine Kac--Moody (KM) algebra. The chirality of the KM currents is opposite to that of the supersymmetry currents, so the KM algebra is purely left-moving. The levels $k_\alpha$ of the current algebra simple factors are given by
\begin{align}
k_\alpha = Q\cdot b_\alpha~.
\end{align}
The Sugawara contribution to the left-moving central charge, denoted by $c_\tL^{\mathrm{gauge}}$, must satisfy
\begin{align}
c_\tL^{\mathrm{gauge}} < c_\tL~.
\end{align}
This ``brane probe'' unitarity constraint provides an additional criterion for excluding apparently consistent anomaly-free 6d supergravities~\cite{Kim:2019vuc}.

In theories without tensor multiplets, the charge lattice reduces to $\Gamma\simeq\Z$. The gravitational anomaly coefficient is $a=-3$.\footnote{This follows from the anomaly cancelation condition $a\cdot a=9-T$ and the positivity constraint $j\cdot a<0$.} The minimal BPS string of charge $Q=1$ is a primitive object. It possesses a moduli space branch along which it is described by a (0,4) SCFT with central charges
\begin{align}
c_\tL &= 32~,&
c_\tR &= 12~,
\end{align}
and a left-moving current algebra for the group $G$, with levels $k_\alpha=b_\alpha$ determined by the gauge anomaly coefficients.\footnote{The models we focus on in this paper all have $b_\alpha > 1$.  In this case, we expect the string moduli space to consist of a unique branch with the central charges discussed above.} When the 6d gauge algebra is trivial this primitive BPS string, referred to as the \emph{hyperplane string} or \emph{L-string} in~\cite{Lockhart:2025lea}, has a well-understood geometric realization~\cite{Haghighat:2015ega}. Indeed, in this case the $T=0$ supergravity can be realized as F-theory compactified on an elliptically fibered Calabi--Yau with base $\P^2$, and the hyperplane string arises from a D3-brane wrapping the hyperplane class $[L]$ of the $\P^2$ base.

The picture suggested in~\cite{Lockhart:2025lea} is that the (quantum corrected) moduli space of the L-string (0,4) SCFT, $\mathcal{M}^{[L]}$, possesses loci where the current algebra on the string is enhanced, corresponding to 6d supergravity theories with nontrivial gauge symmetry. Some of the resulting models do admit a know realization within F-theory with suitable degenerations of the elliptic fiber (see e.g.~\cite{Klevers:2017aku,Hayashi:2023hqa}), but $\mathcal{M}^{[L]}$ can in principle also possess loci along which the Kodaira condition is violated. The hyperplane string (0,4) SCFT may also be used to study these models and probe for their consistency, regardless of whether they admit any currently known string-theoretic embedding. The partial classification of anomaly-free 6d $\mathcal{N} = (1,0)$ supergravity theories with $T=0$ of ~\cite{Hamada:2023zol,Hamada:2024oap}, which enumerates all models for which the gauge group $G$ has no abelian factor and all simple components $G_\alpha$ are distinct from $\mathrm{A}_1$ and $\mathrm{A}_2$, provides a particularly fruitful testing ground. For this class of theories, the Sugawara contribution to the central charge always satisfies $c_\tL^{\mathrm{gauge}}<32$. Several models come close to saturating the bound. In particular, there are five theories for which $c_\tL^{\mathrm{gauge}}>31$:
\begin{align}
G &= \mathrm{E}_6~, &
\mathscr{H} &= \rep{351}'~,
\nonumber\\
G &= \mathrm{A}_7~, &
\mathscr{H} &= \rep{336}~,
\nonumber\\
G &= \mathrm{A}_5\times \mathrm{D}_4~, &
\mathscr{H} &=  (\rep{56},\rep{1})+(\tfrac{1}{2}\rep{20},\rep{28})~,
\nonumber\\
G &= \mathrm{A}_3\times\mathrm{A}_3\times \mathrm{E}_7~, &
\mathscr{H} &=  (\rep{10},\rep{10},\rep{1})+(\rep{1},\rep{6},\tfrac{1}{2}\rep{56})+(\rep{6},\rep{1},\tfrac{1}{2}\rep{56})~,
\nonumber\\
G &= \mathrm{A}_3\times \mathrm{A}_{11}~, &
\mathscr{H} &=  (\rep{35},\rep{1})+(\rep{6},\rep{66})~.
\label{eq:6dsupergravities}
\end{align}
These theories, identified in~\cite{Lockhart:2025lea}, are rigid---they contain no neutral hypermultiplets---and therefore admit no F-theory realization as they lack a universal hypermultiplet in their spectrum, analogously to the asymmetric orbifold models considered in~\cite{Baykara:2023plc}.

Out of the $19867$ theories enumerated in~\cite{Hamada:2023zol,Hamada:2024oap}, the five cases~\eqref{eq:6dsupergravities} stand out because the residual central charge
\begin{align}
c_\tL^{\text{res}}=c_\tL-c_\tL^{\text{gauge}}
\end{align}
of the hyperplane string SCFT falls in the discrete range $c_\tL^{\text{res}}<1$. If this SCFT is well-defined, this value of $c_\tL^{\text{res}}$ must correspond to that of a Virasoro minimal model, as confirmed in~\cite{Lockhart:2025lea}. A particularly drastic consequence is that the hyperplane string worldsheet theory is rational. We explore some consequences of this fact in the next section.

\section{The \texorpdfstring{$c_\tL=32$}{cL=32} holomorphic sector}
\label{sec:KMValgebra}
We now turn to the worldsheet theory  of BPS strings probing the five 6d supergravities~\eqref{eq:6dsupergravities}, and examine their left-moving sector. We show how the chiral algebra organizes the SCFT Hilbert space into a finite set of primary representations.
We then outline our strategy for computing the elliptic genera of the five theories under consideration and summarize our results.

\subsection{The left-moving KMV algebra}
The six-dimensional gauge group~\eqref{eq:6dgaugegroup} appears in the worldsheet SCFT of the hyperplane string as a left-moving Kac--Moody algebra
\begin{align}
\mathfrak{g} = \mathfrak{g}_1+ \cdots+\mathfrak{g}_N~,
\end{align}
where the level of the simple ADE component $\mathfrak{g}_\alpha$ is $k_\alpha$. By Sugawara decomposition, the $c_\tL=32$ stress tensor $T$ can be split into two commuting pieces
\begin{align}
T = T_{\text{gauge}} + T_{\text{res}}~.
\end{align}
The Sugawara energy-momentum tensor $T_{\text{gauge}}$ has central charge
\begin{align}
c_\tL^{\text{gauge}} = \sum_{\alpha=1}^N \frac{k_\alpha\operatorname{dim}\mathfrak{g}_\alpha}{k_\alpha+h^\vee_\alpha}~,
\end{align}
where $h^\vee_\alpha$ denotes the dual Coxeter number of $\mathfrak{g}_\alpha$. The residual stress tensor $T_{\text{res}}$ commutes with the KM currents, and has central charge $c_\tL^{\text{res}}=c_\tL-c_\tL^{\text{gauge}}$. For the 6d $\mathcal{N} = (1,0)$ theories~\eqref{eq:6dsupergravities} under study, the residual central charge of the BPS string worldsheet theory coincides with the central charge of a unitary Virasoro minimal model:
\begin{align}
c_\tL^{\text{res}} = 1-\frac{6}{p(p-1)} ~,
\end{align}
for some $p\geq 4$, which we denote by $\mathfrak{M}_{(p,p-1)}$. This equality serves as a rudimentary consistency check of the (0,4) SCFT.

The KM currents, together with the stress tensor $T_{\text{res}}$, generate a Kac--Moody--Virasoro (KMV) algebra. This left-moving algebra can be used to organize states in irreducible representations. The KMV primaries are denoted as
\begin{align}
\phi_\lambda^{(r,s)}~.
\label{eq:KMVprimaries}
\end{align}
The superscript $(r,s)$ labels irreducible representations of the $c_{\text{res}}<1$ Virasoro algebra: the integers $r$ and $s$ take values in 
\begin{align}
1\leq  r\leq s\leq p-1~.
\end{align}
The subscript $\lambda$ refers to a dominant weight of the algebra $\mathfrak{g}$ satisfying the $N$ integrability conditions
\begin{align}
\gamma_\alpha\cdot\lambda \leq k_\alpha~,
\end{align}
where $\gamma_\alpha$ is the highest root of the simple algebra $\mathfrak{g}_\alpha$. The weight $\lambda$ admits a decomposition $\lambda = (\lambda_1,\dots,\lambda_N)$, with each $\lambda_\alpha$ defining a unitary highest-weight representation of the $\mathfrak{g}_\alpha$ KM algebra at level $k_\alpha$. We will often denote weights by their Dynkin labels; our conventions for simple Lie algebras are spelled out in appendix~\ref{app:Lie}.

Notably, there is a finite number of distinct left-moving primaries. The total number of irreducible KMV primary representations is given by
\begin{align}
\ell &= \ell(\mathfrak{g}_1)\times \dots\times \ell(\mathfrak{g}_N) \times \ell_{\text{res}} ~,
\end{align}
where $\ell_{\text{res}} = \frac{1}{2}(p-1)(p-2)$, and $\ell(\mathfrak{g}_\alpha)$ is the number of unitary highest weight representations of the $\mathfrak{g}_\alpha$ KM algebra at level $k_\alpha$. For the primitive BPS strings probing the 6d theories~\eqref{eq:6dsupergravities}, the $c_\tL=32$ algebra and the number $\ell$ of KMV primary representations are listed in table~\ref{tab:KMVlist}.

\begin{table}[h]
\centering
\begin{tabular}{ccc}
$\mathfrak{g}$ & $c_\tL^{\text{res}}$ & $\ell$
\\\hline
$\mathfrak{e}_{6,8}$ & $4/5$ & $3720$ \\
$\mathfrak{a}_{7,8}$ & $1/2$ & $19305$ \\
$\mathfrak{a}_{5,6}+\mathfrak{d}_{4,6}$ & $1/2$ & $180180$ \\
$\mathfrak{a}_{3,6}+ \mathfrak{a}_{3,6}+\mathfrak{e}_{7,2}$ & $7/10$ & $254016$ \\
$\mathfrak{a}_{3,10}+ \mathfrak{a}_{11,2}$ & $6/7$ & $334620$
\end{tabular}
\caption{KMV algebras of the hyperplane string probing the 6d theories~\eqref{eq:6dsupergravities}. The first column lists the affine KM algebra, the second column lists the central charge of the residual Virasoro algebra, and the last column indicates the total number of KMV primaries.}
\label{tab:KMVlist}
\end{table}

To lighten notation, we sometimes denote the KMV primaries as $\phi_i$, using a single index running over $i=0,1,\dots,\ell-1$. It is understood that each $i$ uniquely indexes a KMV irreducible representation $(r_i,s_i,\lambda_i)$. The identity operator is $\phi_0 = \phi^{(1,1)}_{[0\cdots 0]}$.

\subsection{Characters}
The KMV descendant states in a primary representation~\eqref{eq:KMVprimaries} are counted by a character
\begin{align}
\chi^{(r,s)}_{\lambda}(\tau,\xi)~,
\end{align}
that depends on the torus complex structure parameter $\tau$, valued in the upper half-plane, and on a variable $\xi=(\xi_1,\dots,\xi_N)$, valued in the complexified Cartan subalgebra of $\mathfrak{g}$. The character $\chi^{(r,s)}_{\lambda}$ is a product
\begin{align}
\chi^{(r,s)}_{\lambda}(\tau,\xi) = \chi^{\mathfrak{g}_1}_{\lambda_1}(\tau,\xi_1)\dots \chi^{\mathfrak{g}_N}_{\lambda_N}(\tau,\xi_N) \chi^{\text{res}}_{(r,s)}(\tau)
\label{eq:KMVcharacter}
\end{align}
of a minimal model character $\chi^{\text{res}}_{(r,s)}$ and of KM characters $\chi^{\mathfrak{g}_\alpha}_{\lambda_\alpha}$ associated with the simple factors $\mathfrak{g}_\alpha\subset \mathfrak{g}$. 

The modular properties of $\chi_\lambda^{(r,s)}$ may be derived from those of the individual factors in~\eqref{eq:KMVcharacter}. Under the $\tau\to\tau + 1$ transformation, the characters pick up a phase
\begin{align}
\chi_{(\lambda_1,\dots,\lambda_N)}^{(r,s)} (\tau+1,\xi) = \ee^{2\pi \ii (h_\tL-c_\tL/24)}\chi_{(\lambda_1,\dots,\lambda_N)}^{(r,s)} (\tau,\xi)~,
\end{align}
where 
\begin{align}
h_\tL = h_{\lambda_1}^{\mathfrak{g}_1}+\dots+ h_{\lambda_N}^{\mathfrak{g}_N} + h^{\text{res}}_{(r,s)}
\end{align}
is the left-moving conformal dimension of the KMV primary $\phi_\lambda^{(r,s)}$. Under the $\tau\to-1/\tau$ transformation, the characters transform as
\begin{align}
\chi_{(\lambda_1,\dots,\lambda_N)}^{(r,s)} (-1/\tau,\xi/\tau) = \ee^{\pi \ii \rho(\xi)/\tau} \sum_{\lambda'_1,\dots,\lambda'_N}\sum_{r',s'}\mathcal{S}^{\mathfrak{g}_1}_{\lambda_1\lambda'_1}\dots \mathcal{S}^{\mathfrak{g}_N}_{\lambda_N\lambda_N'}\mathcal{S}^{\text{res}}_{(r,s)(r',s')} \chi_{(\lambda'_1,\dots,\lambda'_N)}^{(r',s')} (\tau)~,
\end{align}
where $\rho(\xi) = k_1\xi_1\cdot\xi_1+\dots+k_N\xi_N\cdot\xi_N$. Here $\mathcal{S}^{\mathfrak{g}_\alpha}$ denotes the modular matrix of the $\mathfrak{g}_\alpha$ KM algebra at level $k_\alpha$, that can be obtained from the Kac--Peterson formula~\eqref{eq:KacPeterson}, and $\mathcal{S}^{\text{res}}$ is the minimal model modular matrix~\eqref{eq:minimalSmatrix}.

As before, we use the condensed notation $\chi_i$ to denote the character associated to the KMV primary $\phi_i$. The modular properties of the characters are encoded in the two $\ell\times\ell$ matrices $\mathcal{T}$ and $\mathcal{S}$ defined by
\begin{align}
\chi_i(\tau+1,\xi) &= \sum_{j=0}^{\ell-1} \mathcal{T}_{ij}\chi_j(\tau,\xi)~,&
\chi_i(-1/\tau,\xi/\tau) &= \ee^{\pi \ii \rho(\xi)/\tau} \sum_{j=0}^{\ell-1} \mathcal{S}_{ij}\chi_j(\tau,\xi)~.
\end{align}
It will also be convenient for us to introduce the notation $t=\ee^{2\pi\ii(h_\tL-c_\tL/24)}$ for the eigenvalues of $\mathcal{T}$, so that 
\begin{align}
\mathcal{T} = \operatorname{diag}(t_0,t_1,\dots,t_{\ell-1})~.
\end{align}

\subsection{Partition function and Ramond ground states}
The spectrum of the (0,4) worldsheet SCFT is constructed out of the $c_\tL=32$ left-moving sector described above, and a anti-holomorphic sector realizing the small $\wsN$=4 superconformal algebra (SCA)  with central charge $c_\tR=12$. The right-moving SCA contains a $\mathfrak{su}(2)$ KM algebra at level $2$, whose $\mathfrak{u}(1)\subset\mathfrak{su}(2)$ subalgebra generates spectral flow between the Neveu--Schwarz and Ramond sectors. Unitary representations of the $\wsN$=4 SCA with $c_\tR=12$ are characterized by their conformal weight $h_\tR$, and their $\mathfrak{u}(1)$ charge $j_\tR=0,1,2$. In the Ramond sector, massless representations have $h_\tR=\frac{1}{2}$ and $j_\tR=0,1,2$, while massive representations have $h_\tR>\frac{1}{2}$ and $j_\tR=1,2$. The $\wsN$=4 characters in the Ramond sector are denoted by
\begin{align}
\overline{\chi}_{h_\tR,j_\tR}(\bar{\tau},\bar{z}) = \operatorname{tr}_{\text{R}}\left( \bar{q}^{\bar{L}_0-\frac{1}{2}} \bar{y}^{\bar{J}_0}\right)
~.
\end{align}
where $\bar{J}_0$ denotes the zero mode of the $\mathfrak{u}(1)\subset\mathfrak{su}(2)$ current, normalized to have integer eigenvalues $j_\tR$---in particular, the fermion number is $(-1)^{F_\tR}=\ee^{\pi\ii \bar{J}_0}$. Following standard notation, $\bar{q}$ denotes the modular parameter $\bar{q}=\ee^{-2\pi\ii\bar{\tau}}$, and $\bar{y}=\ee^{-2\pi\ii \bar{z}}$ keeps track of the $\mathfrak{u}(1)$ charge of the states.
The Ramond ground states, with $h_\tR=\frac{1}{2}$, have Witten index given by
\begin{align}
\overline{\chi}_{\frac{1}{2},j_\tR}\left(\overline{\tau},1\right) =(-1)^{j_\tR}(3-j_\tR)~,
\label{eq:Wittenindex}
\end{align}
while $\overline{\chi}_{h_\tR,j_\tR}\left(\overline{\tau},1\right)=0$ for $h_\tR>\frac{1}{2}$.

The SCFT spectrum is encoded by the Ramond partition function
\begin{align}
Z(\tau,\xi,\overline{\tau},\overline{z}) = \sum_{i=0}^{\ell-1}  \sum_{h_\tR,j_\tR} M_{i,h_\tR,j_\tR}\;\chi_i(\tau,\xi) \overline{\chi}_{h_\tR,j_\tR}(\overline{\tau},\overline{z})~.
\end{align}
The multiplicities $M_{i,h_\tR,j_\tR}$ are non-negative integers. Although a determination of the spectrum is beyond the scope of our analysis in this paper, we know from unitarity that the vacuum is unique, hence $M_{0,\frac{1}{2},0}=1$. We also assume that $M_{0,\frac{1}{2},1} =M_{0,\frac{1}{2},2} =0 $ so that there is no free right-moving fermion nor additional spin-$1$ current on the anti-holomorphic side.

The elliptic genus $\Ell$ is obtained as the right-moving Witten index
$\Ell(\tau,\xi) = Z\left(\tau,\xi,\overline{\tau},1\right)$. It admits the character expansion
\begin{align}
\Ell= \sum_{i=0}^{\ell-1} C_i \,\chi_i
\label{eq:ellipticgenus}
\end{align}
where the integer coefficients $C_i$ are given by
\begin{align}
C_i = 3\,M_{i,\frac{1}{2},0}-2\,M_{i,\frac{1}{2},1}+M_{i,\frac{1}{2},2}~,
\end{align}
due to~\eqref{eq:Wittenindex}. In particular, the coefficient of the identity character is $C_0=3$. The elliptic genus satisfies the modular properties
\begin{align}
\Ell(\tau+1,\xi) &= \ee^{-\frac{2\pi\ii }{24}c_\tL}\,\Ell(\tau,\xi)~,&
\Ell(-1/\tau,\xi/\tau) &= \ee^{\pi \ii \rho(\xi)/\tau}  \Ell(\tau,\xi)~,
\label{eq:modularconstraints}
\end{align}
where $c_\tL=32$ and $\rho(\xi)=k_1\xi_1\cdot\xi_1+\dots+k_N\xi_N\cdot\xi_N$ as before. In particular, if the conformal weight $h_\tL$ of $\phi_i$  is not an integer, then the first property implies that $C_i=0$.

The elliptic genus admits an expansion in the modular parameter $q=\ee^{2\pi\ii\tau}$. The 6d massless spectrum is partly captured by the $h_\tL=1$ coefficient~\cite{Lockhart:2025lea}:
\begin{align}
\Ell(\tau,\xi) &= 3\,q^{-\frac{4}{3}} + (2\,\mathcal{I}-12)q^{-\frac{1}{3}} +\dots~,
\label{eq:ellipticgenusqexpansion}
\end{align}
where $\mathcal{I}(\xi)$ is the 6d chiral index $\mathcal{I}=2(V+3+T-H)$. The shift by $-12$ is due to fact that we have factored out the decoupled center of mass hypermultiplet, which has the effect of removing the graviton contribution from the index.
In a 6d $\mathcal{N}=(1,0)$ theory without tensor multiplets, the anomaly cancelation conditions imply $\mathcal{I}(0)=-540$. Hence, the first two coefficients of the unrefined elliptic genus $\Ell(\tau,0)$ are known. This, together with the modularity properties~\eqref{eq:modularconstraints}, is enough to uniquely determine the unrefined elliptic genus as
\begin{align}
\Ell(\tau,0) = \frac{E_4(\tau)\left(31 \,E_4(\tau)^3+113\, E_6(\tau)^2\right)}{48\,\eta(\tau)^{32}}~,
\label{eq:unrefinedellipticgenus}
\end{align}
in agreement with topological string computations \cite{Huang:2015sta}. This expression is given in terms of the Dedekind function $\eta(\tau)$ and the first two Eisenstein series $E_4(\tau)$ and $E_6(\tau)$.

\subsection{Imposing modular constraints}
\label{subsec:modularmethod}
Let us summarize the conditions imposed on the elliptic genus $\Ell(\tau,\xi)$ of the (0,4) SCFT.

\begin{enumerate}[label=\textcolor{blue}{\arabic*.}, ref=\arabic*]
\item \emph{Rationality}. The elliptic genus can be written as a linear combination~\eqref{eq:ellipticgenus} of KMV characters $\chi_i(\tau,\xi)$, with integer coefficients $C_i$.
\label{cond:rationality}

\item \emph{Modularity}. The elliptic genus transforms as~\eqref{eq:modularconstraints} under $\mathrm{SL}(2,\Z)$.
\label{cond:modularity}

\item \emph{Relation to the 6d massless spectrum}.
The first two coefficients in the $q$-expansion of the elliptic genus are given by~\eqref{eq:ellipticgenusqexpansion}.
\label{cond:6dspectrum}
\end{enumerate}
These three conditions severely  constrain the expression of $\Ell(\tau,\xi)$. Let us group the elliptic genus coefficients in a vector 
\begin{align}
C = (C_0,C_1,\dots, C_{\ell-1})~,
\end{align}
and introduce a $\ell\times\ell$ diagonal matrix  $\mathcal{P}$, defined by
\begin{align}
\mathcal{P}_{ii} = 
\left\lbrace
\begin{array}{ll}
1 & \text{if }t_i = \ee^{-2\pi\ii c_\tL/24}~, \\
0 & \text{otherwise}~.
\end{array}
\right.
\end{align}
This $\mathcal{P}$ is a projector onto the  $\mathcal{T}$-eigenspace of KMV primaries with integer conformal weight. Let $\ell_0=\operatorname{tr}\mathcal{P}$ denote the number of such primaries. We can obtain an expression compatible with the requirements~\ref{cond:rationality} and~\ref{cond:modularity} by finding a solution $C$ to the linear system
\begin{align}
\mathcal{P} C &= C~,&
\mathcal{S} C &= C~.
\label{eq:linearsystem}
\end{align}
Let us introduce the vector space 
\begin{align}
\mathscr{E} = \operatorname{ker}(\operatorname{id}-\mathcal{P}) \;\cap\; \operatorname{ker}(\operatorname{id}-\mathcal{P})\mathcal{S} \mathcal{P} \quad \subset\mathbb{R}^\ell ~.
\end{align}
By construction, the modular $\mathcal{T}$ matrix acts on $\mathscr{E}$ as multiplication by the phase $\ee^{4\pi \ii/3}$. Using that $\mathcal{S}^4=\mathrm{id}$ and $\mathcal{S}^2=(\mathcal{S}\mathcal{T})^3$, it is not too hard to verify that $\mathcal{S}$ acts trivially on $\mathcal{E}$. Hence, the set of solutions of~\eqref{eq:linearsystem} are in one-to-one correspondence with the discrete subspace $\Z^\ell\cap\mathscr{E}$. The space $\mathscr{E}$ is determined by a linear system of $\ell-\ell_0$ equations in $\ell_0$ variables. Therefore, the set of solutions to conditions~\ref{cond:rationality} and~\ref{cond:modularity} is obtained merely by a linear-algebra computation. 

Let us denote by $\psi^\mu=(\psi^\mu_0,\psi^\mu_1,\dots,\psi^\mu_{\ell-1})$ a basis of $\Z^\ell\cap \mathscr{E}$, with $\mu=1,\dots,\operatorname{dim}\mathscr{E}$. The elliptic genus can be written as a linear combination
\begin{align}
\Ell(\tau,\xi) = \sum_{\mu=1}^{\operatorname{dim}\mathscr{E}} \varrho_\mu\sum_{i=0}^{\ell-1}\psi^\mu_i\chi_i(\tau,\xi)~.
\end{align}
The integers $\varrho_\mu$ are left undetermined by the rationality and modularity constraints. However, for the five $c_\tL=32$ KMV algebras under consideration, listed in table~\ref{tab:KMVlist}, they are completely fixed upon imposing condition~\ref{cond:6dspectrum}.

\subsubsection*{An aside on algorithmic implementation}
Although conceptually the computation of the elliptic genus boils down to a basic linear algebra computation,  in practice it can be challenging to implement numerically, due to the large number of KMV primary representations for the $c_\tL=32$ algebras listed in table~\ref{tab:KMVlist}. However, the method introduced above can be refined whenever the modular $\mathcal{S}$ matrix factorizes as a tensor product of two smaller blocks $\mathcal{S}_1$ and $\mathcal{S}_2$---as is the case for all models considered here.\footnote{The method straightforwardly generalizes to cases where $\mathcal{S}$ factorizes into more than two blocks, further diminishing the computational complexity.} The characters of the product theory then take the factorized form
\begin{align}
\chi(\tau) = \chi_1(\tau)\chi_2(\tau)~,
\end{align}
where $\chi_1$ and $\chi_2$ are respectively character for the $\mathcal{S}_1$ and $\mathcal{S}_2$ theories. Such a character can appear in the elliptic genus only if its $\mathcal{T}_1$ and $\mathcal{T}_2$ eigenvalues, denoted by $t_1$ and $t_2$, satisfy
\begin{align}
t_1 t_2 = \ee^{-2\pi\ii c_\tL/24}~.
\label{eq:charactercompatibility}
\end{align}
Let $\mathcal{P}_1$ be the projector onto the $\mathcal{T}_1$ eigenspaces associated with eigenvalues $t_1$ that satisfy~\eqref{eq:charactercompatibility} for some eigenvalue $t_2$ of $\mathcal{T}_2$. Similarly, let $\mathcal{P}_2$ be the projector onto the $\mathcal{T}_2$ eigenspaces associated to the complementary eigenvalues $t_2$. We define the vector spaces
\begin{align}
\mathscr{E}_1 &= \operatorname{ker}(\operatorname{id}_1-\mathcal{P}_1) \;\cap\; \operatorname{ker}(\operatorname{id}_1-\mathcal{P}_1)\mathcal{S}_1 \mathcal{P}_1  ~,\nonumber\\
\mathscr{E}_2 &= \operatorname{ker}(\operatorname{id}_2-\mathcal{P}_2) \;\cap\; \operatorname{ker}(\operatorname{id}_2-\mathcal{P}_2)\mathcal{S}_2 \mathcal{P}_2  ~.
\end{align}
By construction, $\mathscr{E}_1$ and $\mathscr{E}_2$ are stable under the action of the modular group:
\begin{align}
\mathcal{S}_1\mathscr{E}_1 &= \mathscr{E}_1~,&
\mathcal{S}_2\mathscr{E}_2 &= \mathscr{E}_2~.
\end{align}
Let $\psi_1^{\mu_1}$ be a basis of $\mathscr{E}_1$ and $\psi_2^{\mu_2}$ be a basis of $\mathscr{E}_2$, chosen to diagonalize $\mathcal{T}_1$ and $\mathcal{T}_2$ respectively. The combinations $\psi_1^{\mu_1}\psi_2^{\mu_2}$ transform into each other under $\mathrm{SL}(2,\Z)$. We may treat them as ``characters'' (although in general they have no CFT interpretation), and apply the preceding method to them. This procedure typically reduces the computational cost significantly.

To illustrate, let us take $\mathcal{S}_1$ to be the $372\times 372$ modular matrix of the $\mathfrak{e}_6$ KM algebra at level $8$, and $\mathcal{S}_2$ to be the $10\times 10$ modular matrix of the $c=4/5$ Virasoro minimal model. Out of the $\ell=3720$ KMV primary representations, a total of $\ell_0=104$ have integral conformal weight. A direct implementation of our initial method would require computing the kernel of a  $3616\times 104$ matrix. Alternatively, by exploiting the direct product structure of the modular matrix, we can first project onto the eigenvalues
\begin{align}
t_1\;&\in\;\{ \ee^{\frac{7\pi\ii}{5}},\ee^{\frac{19\pi\ii}{15}},\ee^{\frac{3\pi\ii}{5}},\ee^{\frac{\pi\ii}{15}}\}~,&
t_2\;&\in\;\{ 1, \ee^{\frac{28\pi\ii}{15}}, \ee^{\frac{2\pi\ii}{3}}, \ee^{\frac{6\pi\ii}{5}} \}~,
\end{align}
which selects  $78$ $\mathfrak{e}_6$ KM primaries and $6$ minimal Virasoro primaries. Hence, we only have to compute the kernels of two matrices of dimensions $294\times 78$ and $4\times 6$ respectively. In this way we obtain $\mathscr{E}_1$ and $\mathscr{E}_2$, with $\operatorname{dim}\mathscr{E}_1=\operatorname{dim}\mathscr{E}_2=4$. There is then a single modular covariant quantity that can be constructed from these combinations of characters; it coincides with the elliptic genus of the $G=\mathrm{E}_6$ theory (up to an overall constant factor determined by unitarity). 

\subsection{Results}
\label{subsec:results}
We will now apply the methods outlined in the previous paragraphs to determine the elliptic genera of the five (0,4) SCFTs whose left-moving $c_\tL=32$ algebras are listed in table~\ref{tab:KMVlist}.  As a first step, we compute the modular $\mathcal{T}$ and $\mathcal{S}$ matrices of those theories. The $\mathrm{SL}(2,\Z)$ transformations of minimal model Virasoro characters follow directly from the formulas~\eqref{eq:Kacformula} and~\eqref{eq:minimalSmatrix}. The modular properties of Kac--Moody characters can---at least in principle---be obtained from~\eqref{eq:TmatrixKM} and~\eqref{eq:KacPeterson}. However, the Kac--Peterson formula~\eqref{eq:KacPeterson} involves a sum over the Weyl group of the underlying  Lie algebra, whose size can render any direct numerical implementation impractical. For all but two of the KM factors appearing in table~\ref{tab:KMVlist}, the computation can be carried out with no trouble. We use a different strategy to handle the KM algebras associated with $\mathfrak{a}_{11}$ and $\mathfrak{e}_7$, whose Weyl groups are of order $479001600$ and $2903040$ respectively.
For the $\mathfrak{a}_{11}$ KM algebra at level $2$, we exploit the conformal embedding
\begin{align}
\mathfrak{a}_{11,2}+\mathfrak{a}_{1,12} \subset \mathfrak{a}_{23,1}~,
\end{align}
allowing us to deduce the modular properties of $\mathfrak{a}_{11,2}$ from the known ones of $\mathfrak{a}_{1,12}$ and $\mathfrak{a}_{23,1}$. As for the $\mathfrak{e}_7$ KM algebra at level $2$, we make use of the isomorphism
\begin{align}
\frac{ \mathfrak{e}_{7,1}+ \mathfrak{e}_{7,1} }{ \mathfrak{e}_{7,2} } \simeq \mathfrak{M}_{(5,4)}
\end{align}
between the diagonal coset of $\mathfrak{e}_7$ at level $1$ and the $c=7/10$ minimal Virasoro model $\mathfrak{M}_{(5,4)}$. The modular data of $\mathfrak{e}_{7,2}$ can then be extracted from the corresponding branching relations. Further details are provided in appendices~\ref{app:Sa11lvl2} and \ref{app:Se7lvl2}.

We then determine all modular-covariant combinations $\psi^\mu(\tau,\xi)$ of KMV characters using the method introduced above. These generate a module $\Z^\ell\cap\mathscr{E}$, whose dimension is listed in table~\ref{tab:modularcovariants} for the five cases under study.
\begin{table}
\centering
\begin{tabular}{cc}
KMV algebra & $\operatorname{dim}\mathscr{E}$
\\\hline
$\mathfrak{e}_{6,8} + \mathfrak{M}_{(6,5)}$ & $1$ \\
$\mathfrak{a}_{7,8} + \mathfrak{M}_{(4,3)}$ & $2$ \\
$\mathfrak{a}_{5,6}+\mathfrak{d}_{4,6} + \mathfrak{M}_{(4,3)}$ & $5$ \\
$\mathfrak{a}_{3,6}+ \mathfrak{a}_{3,6}+\mathfrak{e}_{7,2} + \mathfrak{M}_{(5,4)}$ & $9$ \\
$\mathfrak{a}_{3,10}+ \mathfrak{a}_{11,2} + \mathfrak{M}_{(7,6)}$ & $2$ \\
\end{tabular}
\caption{Number of independent modular-covariant linear combinations of KMV characters.}
\label{tab:modularcovariants}
\end{table}
Any linear combination $\sum\varrho_\mu\psi^\mu$ in this module satisfies conditions~\ref{cond:rationality} and~\ref{cond:modularity}. The elliptic genus is obtained by imposing condition~\ref{cond:6dspectrum} on a generic linear combination, which fixes the integer coefficients $\varrho_\mu$ uniquely---up to a subtlety in the theory with $\mathfrak{a}_3+\mathfrak{a}_3+\mathfrak{e}_7$ current algebra, which we address below. 

The resulting multi-variable Jacobi form $\Ell(\tau,\xi)$ admits an exact decomposition as a finite sum of left-moving characters. Its expansion in the modular parameter takes the form
\begin{align}
\Ell(\tau,\xi) = q^{-\frac{4}{3}} \sum_{s=0}^\infty \chi_s(\xi)\,q^s~.
\end{align}
The coefficients $\chi_s$ are finite Weyl characters associated to virtual  representations
\begin{align}
\boldsymbol{R}_s = \sum_{\boldsymbol{r}}C_s({\boldsymbol{r}}) \,\boldsymbol{r}~,
\end{align}
where the sum runs over all finite-dimensional representations $\boldsymbol{r}$ of the corresponding Lie group. The integers $C_s(\boldsymbol{r})$ count the net number of Ramond ground states at each Virasoro level $h_\tL=s$. The virtual dimensions of the $\boldsymbol{R}_s$, defined by 
\begin{align}
|\boldsymbol{R}_s| = \sum_{\boldsymbol{r}}C_s({\boldsymbol{r}})\operatorname{dim}\boldsymbol{r}~,
\end{align}
are determined by modularity: they coincide with the coefficients in the $q$-expansion of the unrefined elliptic genus $\Ell(\tau,0)$, given in~\eqref{eq:unrefinedellipticgenus}. At the first Virasoro levels, one finds
\begin{align}
|\boldsymbol{R}_{0}| &= 3~,&
|\boldsymbol{R}_{1}| &= -1092~,&
|\boldsymbol{R}_{2}| &= 147696~.
\end{align}

In what follows, we outline our results for each of the 6d $\mathcal{N} = (1,0)$ theories listed in~\eqref{eq:6dsupergravities}. For each theory under consideration, we have determined a unique expression for the elliptic genus as a decomposition into KMV characters. We refrain from displaying the full expressions at this stage, as they are typically rather lengthy---for instance, the elliptic genus of the theory with $G=\mathrm{A}_7$ involves an expansion into $233$ distinct KMV characters. Readers interested in the explicit formulas may consult the data files provided in~\cite{KMVdata}.
In subsequent sections, we will introduce extended characters that render the expressions considerably more compact. Before doing so, we briefly describe the salient features of each model.

\subsubsection*{Gauge group $G=\mathrm{E}_6$ }
The left-moving algebra of the first model is constructed from the $\mathfrak{e}_6$ KM algebra at level $8$ together with the minimal model $\mathfrak{M}_{(6,5)}$. The elliptic genus KMV expansion begins as
\begin{align}
\Ell_{\mathrm{E}_6} &= 3\,\chi_{\rep{1} }^{(1,1)}+\chi_{\rep{78}}^{(2,1)}-2\,\chi_{\rep{351}'}^{(3,3)}-2\,\chi_{\repconj{351}'}^{(3,3)}+\dots~,
\end{align}
where we include only the primaries with $h_\tL\leq 1$. We recover the expression discovered in~\cite{Lockhart:2025lea}. The net number of Ramond ground states at the lowest Virasoro levels are given by
\begin{align}
\boldsymbol{R}_{0} &=  
3\cdot\rep{1}~,
\nonumber\\
\boldsymbol{R}_{1} &=  
4\cdot \rep{78}
-2\cdot\rep{351}'
-2\cdot\repconj{351}' ~,
\nonumber\\
\boldsymbol{R}_{2} &=  
7 \cdot  \rep{ 1 } 
+6 \cdot  \rep{ 78 } 
-2 \cdot  \rep{ 351 } 
-2 \cdot  \repconj{ 351 } 
-4 \cdot  \rep{ 351 }' 
-4 \cdot  \repconj{ 351 }' 
+4 \cdot  \rep{ 650 } 
+4 \cdot  \rep{ 2430 } 
+ \rep{ 2925 }
\nonumber\\ 
& \quad 
- \rep{ 7371 } 
- \repconj{ 7371 } 
-2 \cdot  \rep{ 19305 } 
-2 \cdot  \repconj{ 19305 } 
+ \rep{ 34749 } 
+ \rep{ 54054 } 
+ \repconj{ 54054 } 
+ \rep{ 85293 } ~.
\end{align}

\subsubsection*{Gauge group $G=\mathrm{A}_7$ }
The left-moving algebra of the second model is constructed from the $\mathfrak{a}_7$ KM algebra at level $8$ together with the minimal model $\mathfrak{M}_{(4,3)}$. 
The elliptic genus KMV expansion begins as
\begin{align}
\Ell_{\mathrm{A}_7} &=  3\, \chi_{\rep{1} }^{ (1,1)} + \chi_{\rep{63} }^{ (2,1)} -2 \,\chi_{\rep{336} }^{ (2,2)} -2 \,\chi_{\repconj{336} }^{ (2,2)}  + \dots~,
\end{align}
where we include only the primaries with $h_\tL\leq 1$. The net number of Ramond ground states at the lowest Virasoro levels are given by
\begin{align}
\boldsymbol{R}_{0} &=  
3 \cdot  \rep{ 1 }~,
\nonumber\\
\boldsymbol{R}_{1} &=  
4 \cdot  \rep{ 63 } 
 -2 \cdot  \rep{ 336 } 
 -2 \cdot  \repconj{ 336 } ~,
\nonumber\\
\boldsymbol{R}_{2} &=
7 \cdot  \rep{ 1 } 
 +9 \cdot  \rep{ 63 } 
 -4 \cdot  \rep{ 336 } 
 -4 \cdot  \repconj{ 336 } 
 -2 \cdot  \rep{ 378 } 
 -2 \cdot  \repconj{ 378 } 
 -2 \cdot  \rep{ 630 } 
 -2 \cdot  \repconj{ 630 } 
 +4 \cdot  \rep{ 720 } 
 \nonumber\\
 &\quad
 + \rep{ 945 } 
 + \repconj{ 945 } 
 +4 \cdot  \rep{ 1232 } 
 -2 \cdot  \rep{ 7350 } 
 -2 \cdot  \repconj{ 7350 } 
 -2 \cdot  \rep{ 12474 } 
 -2 \cdot  \repconj{ 12474 } 
 + \rep{ 14700 } 
 \nonumber\\
 &\quad
 + \repconj{ 14700 } 
 + \rep{ 24255 } 
 + \rep{ 41580 } 
 + \repconj{ 41580 } 
 + \rep{ 86625 }~.
\end{align}

\subsubsection*{Gauge group $G=\mathrm{A}_5\times\mathrm{D}_4$ }
The left-moving algebra of the third model is constructed from the $\mathfrak{a}_5 +\mathfrak{d}_4$ KM algebra at level $(6,6)$ together with the minimal model $\mathfrak{M}_{(4,3)}$. The elliptic genus KMV expansion begins as
\begin{align}
\Ell_{\mathrm{A}_5\times\mathrm{D}_4 } &=  3\, \chi_{\rep{1},\rep{1} }^{ (1,1) } +\chi_{\rep{35},\rep{1} }^{ (2,1) }+\chi_{\rep{1},\rep{28} }^{ (2,1) }  - 2\,\chi_{\rep{20},\rep{28} }^{ (2,2) }  - 2\,\chi_{\rep{56},\rep{1} }^{ (2,2) }  - 2\,\chi_{\repconj{56},\rep{1} }^{ (2,2) }+\dots~,
\end{align}
where we include only the primaries with $h_\tL\leq 1$. The net number of Ramond ground states at the lowest Virasoro levels are given by
\begin{align}
\boldsymbol{R}_0 &= 
3 ( \rep{ 1 } , \rep{ 1 } ) ~,
\nonumber\\
\boldsymbol{R}_1 &= 
4 ( \rep{ 1 } , \rep{ 28 } )
 +4 ( \rep{ 35 } , \rep{ 1 } )
 -2 ( \rep{ 56 } , \rep{ 1 } )
 -2 ( \repconj{ 56 } , \rep{ 1 } )
 -2 ( \rep{ 20 } , \rep{ 28 } )~,
\nonumber\\
\boldsymbol{R}_2 &=
11 ( \rep{ 1 } , \rep{ 1 } )
 -2 ( \rep{ 20 } , \rep{ 1 } )
 +5 ( \rep{ 1 } , \rep{ 28 } )
 +4 ( \rep{ 1 } , \rep{ 35^v } )
 +4 ( \rep{ 1 } , \rep{ 35^c } )
 +4 ( \rep{ 1 } , \rep{ 35^s } )
 +9 ( \rep{ 35 } , \rep{ 1 } )
 \nonumber\\
 &\quad
 -4 ( \rep{ 56 } , \rep{ 1 } )
 -4 ( \repconj{ 56 } , \rep{ 1 } )
 -2 ( \rep{ 70 } , \rep{ 1 } )
 -2 ( \repconj{ 70 } , \rep{ 1 } )
 +4 ( \rep{ 189 } , \rep{ 1 } )
 + ( \rep{ 280 } , \rep{ 1 } )
 + ( \repconj{ 280 } , \rep{ 1 } )
 \nonumber\\
 &\quad
 +4 ( \rep{ 1 } , \rep{ 300 } )
 + ( \rep{ 1 } , \rep{ 350 } )
 +4 ( \rep{ 405 } , \rep{ 1 } )
 + ( \rep{ 490 } , \rep{ 1 } )
 + ( \repconj{ 490 } , \rep{ 1 } )
 -6 ( \rep{ 20 } , \rep{ 28 } )
 \nonumber\\
 &\quad
 -2 ( \rep{ 700 } , \rep{ 1 } )
 -2 ( \repconj{ 700 } , \rep{ 1 } )
 -2 ( \rep{ 20 } , \rep{ 35^v } )
 -2 ( \rep{ 20 } , \rep{ 35^c } )
 -2 ( \rep{ 20 } , \rep{ 35^s } )
 +5 ( \rep{ 35 } , \rep{ 28 } )
 \nonumber\\
 &\quad
 + ( \rep{ 1050 }'' , \rep{ 1 } )
 + ( \repconj{ 1050 }'' , \rep{ 1 } )
 -2 ( \rep{ 1134 } , \rep{ 1 } )
 -2 ( \repconj{ 1134 } , \rep{ 1 } )
 -2 ( \rep{ 56 } , \rep{ 28 } )
 -2 ( \repconj{ 56 } , \rep{ 28 } )
 \nonumber\\
 &\quad
 -2 ( \rep{ 70 } , \rep{ 28 } )
 -2 ( \repconj{ 70 } , \rep{ 28 } )
 + ( \rep{ 2695 } , \rep{ 1 } )
 + ( \rep{ 175 } , \rep{ 28 } )
 -2 ( \rep{ 20 } , \rep{ 300 } )
 + ( \rep{ 189 } , \rep{ 35^v } )
 \nonumber\\
 &\quad
 + ( \rep{ 189 } , \rep{ 35^c } )
 + ( \rep{ 189 } , \rep{ 35^s } )
 -2 ( \rep{ 20 } , \rep{ 350 } )
 + ( \rep{ 280 } , \rep{ 28 } )
 + ( \repconj{ 280 } , \rep{ 28 } )
 + ( \rep{ 35 } , \rep{ 350 } )
 \nonumber\\
 &\quad
 -2 ( \rep{ 540 } , \rep{ 28 } )
 + ( \rep{ 840 }'' , \rep{ 28 } )
 + ( \repconj{ 840 }'' , \rep{ 28 } )
 + ( \rep{ 189 } , \rep{ 300 } )
 + ( \rep{ 175 } , \rep{ 350 } )~.
\end{align}

\subsubsection*{Gauge group $G=\mathrm{A}_3\times\mathrm{A}_3\times\mathrm{E}_7$ }
The left-moving algebra of the fourth model is constructed from the $\mathfrak{a}_3+\mathfrak{a}_3+\mathfrak{e}_7$ KM algebra at level $(6,6,2)$ together with the minimal model $\mathfrak{M}_{(5,4)}$. Remarkably, this $c_\tL=32$ algebra admits a modular-covariant combination of KMV characters $\psi(\tau,\xi)$ whose $q$-expansion begins as 
\begin{align}
\psi(q,\xi) = \widetilde{\chi}_{2}(\xi)\,q^{\frac{2}{3}}+\dots,
\end{align}
where $\widetilde{\chi}_2$ is the Weyl character of the virtual representation
\begin{alignat}{4}
\widetilde{\boldsymbol{R}}_2 = &&
&(\rep{35},\rep{45},\rep{1}) + (\repconj{35},\repconj{45},\rep{1}) + (\rep{45},\repconj{35},\rep{1})  + (\repconj{45},\rep{35},\rep{1}) 
\nonumber\\
&&
\,-\,&  (\rep{45},\rep{35},\rep{1}) - (\repconj{45},\repconj{35},\rep{1}) - (\rep{35},\repconj{45},\rep{1})  - (\repconj{35},\rep{45},\rep{1}) 
\nonumber\\
&&
\,+\,&  (\rep{45},\rep{10},\rep{56}) + (\repconj{45},\repconj{10},\rep{56}) + (\rep{10},\repconj{45},\rep{56})  + (\repconj{10},\rep{45},\rep{56})  
\nonumber\\
&&
\,-\,&  (\rep{10},\rep{45},\rep{56}) - (\repconj{10},\repconj{45},\rep{56}) - (\rep{45},\repconj{10},\rep{56}) - (\repconj{45},\rep{10},\rep{56}) && ~.
\end{alignat}
Note that $\widetilde{\boldsymbol{R}}_2$ is antisymmetric under the exchange of the two $\mathfrak{a}_3$ factors. In fact, $\psi(\tau,\xi)$ is odd under the exchange of the two $\mathfrak{a}_3$ chemical potentials: $\psi(\tau,\xi_1,\xi_2,\xi_3) = -\psi(\tau,\xi_2,\xi_1,\xi_3)$. In particular, it vanishes identically in the unrefined limit.

The existence of $\psi$ complicates the determination of the elliptic genus. Given a solution $\Ell$ to the three defining conditions, any shifted combination $\Ell' = \Ell + \varrho \psi$ is equally valid: adding $\psi$ affects only BPS states with $h_\tL>1$, leaving the massless spectrum captured by the index unchanged. However, considering that the massless spectrum is invariant under the $\Z_2$ exchange of the two $\mathfrak{a}_3$ factors, it is reasonable to demand that the entire elliptic genus should be even under this exchange:
\begin{align}
\Ell(\tau,\xi_1,\xi_2,\xi_3) = \Ell(\tau,\xi_2,\xi_1,\xi_3)~.
\end{align}
Imposing this requirement, together with condition~\ref{cond:6dspectrum}, yields a unique expression for the elliptic genus. Its KMV expansion begins as
\begin{align}
\Ell_{\mathrm{A}_3\times\mathrm{A}_3\times\mathrm{E}_7 } &=  
3\, \chi_{\rep{1},\rep{1},\rep{1} }^{ (1,1)} + \chi_{\rep{15},\rep{1},\rep{1} }^{(3,2)} + \chi_{\rep{1},\rep{15},\rep{1} }^{(3,2)} + \chi_{\rep{1},\rep{1},\rep{133} }^{(3,3)} -  2\,\chi_{\rep{10},\rep{10},\rep{1} }^{(3,3)} -  2\,\chi_{\repconj{10},\repconj{10},\rep{1} }^{(3,3)} 
\nonumber\\
&\quad
-  2\,\chi_{\rep{6},\rep{1},\rep{56} }^{(2,2)} -  2\,\chi_{\rep{1},\rep{6},\rep{56} }^{(2,2)}+\dots~,
\end{align}
where we include only the primaries with $h_\tL\leq 1$. The net number of Ramond ground states at the lowest Virasoro levels are given by
{\allowdisplaybreaks
\begin{align}
\boldsymbol{R}_0 &=
3 ( \rep{ 1 } , \rep{ 1 } , \rep{ 1 } )~,
\nonumber\\
\boldsymbol{R}_1 &=
4 ( \rep{ 1 } , \rep{ 15 } , \rep{ 1 } )
 +4 ( \rep{ 15 } , \rep{ 1 } , \rep{ 1 } )
 -2 ( \rep{ 10 } , \rep{ 10 } , \rep{ 1 } )
 -2 ( \repconj{ 10 } , \repconj{ 10 } , \rep{ 1 } )
 +4 ( \rep{ 1 } , \rep{ 1 } , \rep{ 133 } )
\nonumber\\
&\quad
 -2 ( \rep{ 6 } , \rep{ 1 } , \rep{ 56 } )
 -2 ( \rep{ 1 } , \rep{ 6 } , \rep{ 56 } )
~,
\nonumber\\
\boldsymbol{R}_2 &=
15 ( \rep{ 1 } , \rep{ 1 } , \rep{ 1 } )
 +9 ( \rep{ 1 } , \rep{ 15 } , \rep{ 1 } )
 +9 ( \rep{ 15 } , \rep{ 1 } , \rep{ 1 } )
 +4 ( \rep{ 1 } , \rep{ 20 }' , \rep{ 1 } )
 +4 ( \rep{ 20 }' , \rep{ 1 } , \rep{ 1 } )
 + ( \rep{ 6 } , \rep{ 6 } , \rep{ 1 } )
 + ( \rep{ 45 } , \rep{ 1 } , \rep{ 1 } )
\nonumber\\
&\quad
 + ( \repconj{ 45 } , \rep{ 1 } , \rep{ 1 } )
 + ( \rep{ 1 } , \rep{ 45 } , \rep{ 1 } )
 + ( \rep{ 1 } , \repconj{ 45 } , \rep{ 1 } )
 -2 ( \rep{ 10 } , \rep{ 6 } , \rep{ 1 } )
 -2 ( \rep{ 6 } , \rep{ 10 } , \rep{ 1 } )
 -2 ( \repconj{ 10 } , \rep{ 6 } , \rep{ 1 } )
 -2 ( \rep{ 6 } , \repconj{ 10 } , \rep{ 1 } )
\nonumber\\
&\quad
 +4 ( \rep{ 1 } , \rep{ 84 } , \rep{ 1 } )
 +4 ( \rep{ 84 } , \rep{ 1 } , \rep{ 1 } )
 -6 ( \rep{ 10 } , \rep{ 10 } , \rep{ 1 } )
 -6 ( \repconj{ 10 } , \repconj{ 10 } , \rep{ 1 } )
 +5 ( \rep{ 1 } , \rep{ 1 } , \rep{ 133 } )
 +5 ( \rep{ 15 } , \rep{ 15 } , \rep{ 1 } )
\nonumber\\
&\quad
 -6 ( \rep{ 6 } , \rep{ 1 } , \rep{ 56 } )
 -6 ( \rep{ 1 } , \rep{ 6 } , \rep{ 56 } )
 -2 ( \rep{ 10 } , \rep{ 1 } , \rep{ 56 } )
 -2 ( \repconj{ 10 } , \rep{ 1 } , \rep{ 56 } )
 -2 ( \rep{ 1 } , \rep{ 10 } , \rep{ 56 } )
 -2 ( \rep{ 1 } , \repconj{ 10 } , \rep{ 56 } )
\nonumber\\
&\quad
 -2 ( \rep{ 10 } , \rep{ 64 } , \rep{ 1 } )
 -2 ( \rep{ 64 } , \rep{ 10 } , \rep{ 1 } )
 -2 ( \repconj{ 10 } , \rep{ 64 } , \rep{ 1 } )
 -2 ( \rep{ 64 } , \repconj{ 10 } , \rep{ 1 } )
 -2 ( \rep{ 10 } , \rep{ 70 } , \rep{ 1 } )
 -2 ( \rep{ 70 } , \rep{ 10 } , \rep{ 1 } )
\nonumber\\
&\quad
 -2 ( \repconj{ 10 } , \repconj{ 70 } , \rep{ 1 } )
 -2 ( \repconj{ 70 } , \repconj{ 10 } , \rep{ 1 } )
 + ( \rep{ 20 }' , \rep{ 45 } , \rep{ 1 } )
 + ( \rep{ 20 }' , \repconj{ 45 } , \rep{ 1 } )
 + ( \rep{ 45 } , \rep{ 20 }' , \rep{ 1 } )
 + ( \repconj{ 45 } , \rep{ 20 }' , \rep{ 1 } )
\nonumber\\
&\quad
 + ( \rep{ 15 } , \rep{ 84 } , \rep{ 1 } )
 + ( \rep{ 84 } , \rep{ 15 } , \rep{ 1 } )
 +5 ( \rep{ 1 } , \rep{ 1 } , \rep{ 1539 } )
 + ( \rep{ 45 } , \rep{ 35 } , \rep{ 1 } )
 + ( \repconj{ 45 } , \repconj{ 35 } , \rep{ 1 } )
 + ( \rep{ 35 } , \rep{ 45 } , \rep{ 1 } )
\nonumber\\
&\quad
 + ( \repconj{ 35 } , \repconj{ 45 } , \rep{ 1 } )
 +5 ( \rep{ 1 } , \rep{ 15 } , \rep{ 133 } )
 +5 ( \rep{ 15 } , \rep{ 1 } , \rep{ 133 } )
 -2 ( \rep{ 64 } , \rep{ 1 } , \rep{ 56 } )
 -2 ( \rep{ 1 } , \rep{ 64 } , \rep{ 56 } )
 + ( \rep{ 6 } , \rep{ 6 } , \rep{ 133 } )
\nonumber\\
&\quad
 -2 ( \rep{ 6 } , \rep{ 15 } , \rep{ 56 } )
 -2 ( \rep{ 15 } , \rep{ 6 } , \rep{ 56 } )
 -2 ( \rep{ 6 } , \rep{ 1 } , \rep{ 912 } )
 -2 ( \rep{ 1 } , \rep{ 6 } , \rep{ 912 } )
 + ( \rep{ 84 } , \rep{ 84 } , \rep{ 1 } )
 +3 ( \rep{ 1 } , \rep{ 1 } , \rep{ 7371 } )
\nonumber\\
&\quad
 + ( \rep{ 15 } , \rep{ 10 } , \rep{ 56 } )
 + ( \rep{ 15 } , \repconj{ 10 } , \rep{ 56 } )
 + ( \rep{ 10 } , \rep{ 15 } , \rep{ 56 } )
 + ( \repconj{ 10 } , \rep{ 15 } , \rep{ 56 } )
 + ( \rep{ 1 } , \rep{ 1 } , \rep{ 8645 } )
 -2 ( \rep{ 10 } , \rep{ 10 } , \rep{ 133 } )
\nonumber\\
&\quad
 -2 ( \repconj{ 10 } , \repconj{ 10 } , \rep{ 133 } )
 + ( \rep{ 15 } , \rep{ 1 } , \rep{ 1463 } )
 + ( \rep{ 1 } , \rep{ 15 } , \rep{ 1463 } )
 + ( \rep{ 10 } , \rep{ 45 } , \rep{ 56 } )
 + ( \repconj{ 10 } , \repconj{ 45 } , \rep{ 56 } )
\nonumber\\
&\quad
 + ( \rep{ 45 } , \rep{ 10 } , \rep{ 56 } )
 + ( \repconj{ 45 } , \repconj{ 10 } , \rep{ 56 } )
 + ( \rep{ 20 }' , \rep{ 1 } , \rep{ 1539 } )
 + ( \rep{ 1 } , \rep{ 20 }' , \rep{ 1539 } )
 -2 ( \rep{ 6 } , \rep{ 1 } , \rep{ 6480 } )
\nonumber\\
&\quad
 -2 ( \rep{ 1 } , \rep{ 6 } , \rep{ 6480 } )
 + ( \rep{ 6 } , \rep{ 6 } , \rep{ 1463 } )
 + ( \rep{ 6 } , \rep{ 6 } , \rep{ 1539 } )~.
\end{align}}

\subsubsection*{Gauge group $G=\mathrm{A}_3\times\mathrm{A}_{11}$ }
The left-moving algebra of the fifth model is constructed from the $\mathfrak{a}_3+\mathfrak{a}_{11}$ at level $(10,2)$ together with the minimal model $\mathfrak{M}_{(7,6)}$. The elliptic genus KMV expansion begins as
\begin{align}
\Ell_{\mathrm{A}_3\times\mathrm{A}_{11} } &=  
3\,\chi_{\rep{1},\rep{1}  }^{ (1,1)}
+\chi_{\rep{15},\rep{1} }^{ (5,4)}
+\chi_{\rep{1}, \rep{143}  }^{ (5,5)}
-2\,\chi_{\rep{35},\rep{1} }^{ (5,5)}
-2\,\chi_{\repconj{35},\rep{1}  }^{ (5,5)}
-2\,\chi_{\rep{6},\rep{66} }^{ (3,3)}
-2\,\chi_{\rep{6},\repconj{66} }^{ (3,3)}+\dots~,
\end{align}
where we include only the primaries with $h_\tL\leq 1$. The net number of Ramond ground states at the lowest Virasoro levels are given by
\begin{align}
\boldsymbol{R}_0 &= 
3 ( \rep{ 1 } , \rep{ 1 } )~,
\nonumber\\
\boldsymbol{R}_1 &= 
4 ( \rep{ 15 } , \rep{ 1 } )
 -2 ( \rep{ 35 } , \rep{ 1 } )
 -2 ( \repconj{ 35 } , \rep{ 1 } )
 +4 ( \rep{ 1 } , \rep{ 143 } )
 -2 ( \rep{ 6 } , \rep{ 66 } )
 -2 ( \rep{ 6 } , \repconj{ 66 } )~,
\nonumber\\
\boldsymbol{R}_2 &= 
11 ( \rep{ 1 } , \rep{ 1 } )
 +10 ( \rep{ 15 } , \rep{ 1 } )
 +4 ( \rep{ 20 }' , \rep{ 1 } )
 -4 ( \rep{ 35 } , \rep{ 1 } )
 -4 ( \repconj{ 35 } , \rep{ 1 } )
 - ( \rep{ 45 } , \rep{ 1 } )
 - ( \repconj{ 45 } , \rep{ 1 } )
 +4 ( \rep{ 84 } , \rep{ 1 } )
 \nonumber\\
 &\quad 
 +9 ( \rep{ 1 } , \rep{ 143 } )
 -2 ( \rep{ 189 } , \rep{ 1 } )
 -2 ( \repconj{ 189 } , \rep{ 1 } )
 -2 ( \rep{ 256 } , \rep{ 1 } )
 -2 ( \repconj{ 256 } , \rep{ 1 } )
 + ( \rep{ 280 } , \rep{ 1 } )
 + ( \repconj{ 280 } , \rep{ 1 } )
 \nonumber\\
 &\quad 
 + ( \rep{ 300 }' , \rep{ 1 } )
 + ( \rep{ 315 } , \rep{ 1 } )
 + ( \repconj{ 315 } , \rep{ 1 } )
 -6 ( \rep{ 6 } , \rep{ 66 } )
 -6 ( \rep{ 6 } , \repconj{ 66 } )
 -2 ( \rep{ 6 } , \rep{ 78 } )
 -2 ( \rep{ 6 } , \repconj{ 78 } )
 \nonumber\\
 &\quad 
 -2 ( \rep{ 10 } , \rep{ 66 } )
 -2 ( \rep{ 10 } , \repconj{ 66 } )
 -2 ( \repconj{ 10 } , \rep{ 66 } )
 -2 ( \repconj{ 10 } , \repconj{ 66 } )
 + ( \rep{ 825 } , \rep{ 1 } )
 +5 ( \rep{ 15 } , \rep{ 143 } )
 + ( \rep{ 1 } , \rep{ 2145 } )
 \nonumber\\
 &\quad 
 + ( \rep{ 1 } , \repconj{ 2145 } )
 + ( \rep{ 20 }' , \rep{ 143 } )
 +4 ( \rep{ 1 } , \rep{ 4212 } )
 -2 ( \rep{ 64 } , \rep{ 66 } )
 -2 ( \rep{ 64 } , \repconj{ 66 } )
 + ( \rep{ 70 } , \rep{ 66 } )
 + ( \rep{ 70 } , \repconj{ 66 } )
 \nonumber\\
 &\quad 
 + ( \repconj{ 70 } , \rep{ 66 } )
 + ( \repconj{ 70 } , \repconj{ 66 } )
 -2 ( \rep{ 35 } , \rep{ 143 } )
 -2 ( \repconj{ 35 } , \rep{ 143 } )
 + ( \rep{ 1 } , \rep{ 5005 } )
 + ( \rep{ 1 } , \repconj{ 5005 } )
 +3 ( \rep{ 1 } , \rep{ 5940 } )
 \nonumber\\
 &\quad 
 + ( \rep{ 15 } , \rep{ 495 } )
 + ( \rep{ 15 } , \repconj{ 495 } )
 + ( \rep{ 140 }'' , \rep{ 66 } )
 + ( \rep{ 140 }'' , \repconj{ 66 } )
 + ( \repconj{ 140 }'' , \rep{ 66 } )
 + ( \repconj{ 140 }'' , \repconj{ 66 } )
 -2 ( \rep{ 6 } , \rep{ 2574 } )
 \nonumber\\
 &\quad 
 -2 ( \rep{ 6 } , \repconj{ 2574 } )
 + ( \rep{ 15 } , \rep{ 1716 } )
 + ( \rep{ 15 } , \repconj{ 1716 } )
 -2 ( \rep{ 6 } , \rep{ 6720 } )
 -2 ( \rep{ 6 } , \repconj{ 6720 } )
 + ( \rep{ 20 }' , \rep{ 2145 } )
 \nonumber\\
 &\quad 
 + ( \rep{ 20 }' , \repconj{ 2145 } )
 + ( \rep{ 15 } , \rep{ 4212 } )
 + ( \rep{ 20 }' , \rep{ 4212 } )
 ~.
\end{align}

\subsubsection*{Remarks on the supersymmetric spectrum}
As we already mentioned, the KMV decomposition yields a finite yet lenghty expansion of the elliptic genus in terms of $c_\tL=32$ characters. This decomposition typically exhibits nontrivial symmetry properties. For instance, the elliptic genus $\Ell_{\mathrm{E}_6}$ was already obtained in~\cite{Lockhart:2025lea}, and two surprising observations were made there:
\begin{itemize}
\item only a subset of the $\mathfrak{M}_{(6,5)}$ minimal model characters appear in $\Ell_{\mathrm{E}_6}$---specifically, those corresponding to the Potts model;
\item the $\Z_3$ outer automorphism of the $\mathfrak{e}_6$ KM algebra, which acts on $\mathfrak{e}_6$ representations as
\begin{equation}
\lambda = [l_1l_2l_3l_4l_5l_6] 
\qquad\to\qquad
\lambda' = [l_6l_3l_5l_4l_2l_0]~,
\end{equation}
where $l_0$ denotes the affine Dynkin label, combines with the $\Z_3$ symmetry of the Potts model to generate a discrete $\Z_3$ permutation of the primaries under which $\Ell_{\mathrm{E}_6}$ is invariant.
\end{itemize}
As we will see, both of these observations admit a unified interpretation: they originate from a \emph{simple current extension} of the $c_\tL=32$ left-moving algebra. Furthermore, we will show that this underlying structure appears to be universal among the (0,4) SCFTs under consideration: the remaining four elliptic genera exhibit analogous patterns, indicating that they too are compatible with an extension of the left-moving KMV algebra. As a consequence of this, in the following sections we will show that the supersymmetric spectrum of the hyperplane string for our five models organizes in terms of characters of the extended left-moving algebra, leading to considerably more compact expressions for the refined elliptic genus. These higher-spin extensions admit a natural spacetime interpretation: they are tied to one-form symmetries of the six-dimensional supergravity theory and, in particular, to the center symmetries of the gauge group~\cite{Lockhart:2026gvb}. We will return to this point in section~\ref{sec:discussion}.

\section{Extended chiral algebras}
\label{sec:simplecurrents}
In this section, we review some standard results on integer spin simple current extensions in two-dimensional CFTs.

\subsection{Simple current generalities}
Consider a unitary CFT with a finite number of primaries $\phi_i$, indexed by $i=0,1,\dots,\ell-1$. We conventionally denote the identity primary by $\phi_0$. A simple current is a primary field $J$ whose fusion with the other primaries is of the form
\begin{align}
J\times \phi_i = \phi_{\sigma(i)}~,
\label{eq:simplecurrent}
\end{align}
where $\sigma$ is a permutation  of the $\ell$ indices. The fusion of $J$ with itself gives a second simple current, due to the associativity of the fusion rules. Iteratively, one defines
\begin{align}
J^{a} = J\times J^{a-1}~.
\end{align}
The order of $J$ is the smallest strictly-positive integer $n$ such that $J^n$ is the identity operator. The fusion rules of the currents generated by $J$ is isomorphic to the cyclic group $\Z_n$. The complete set of simple currents of a CFT form an abelian group, called the center of the CFT. We denote it by $\mathcal{Z}$, with
\begin{align}
\mathcal{Z}\simeq \Z_{n_1} \times \dots \times \Z_{n_d}~,
\end{align}
where the cyclic factors are generated by a set of simple currents $J_1,\dots,J_d$, respectively of order $n_1,\dots,n_d$.

Let $J$ be a simple current of order $n$, generating a cyclic group $\Gamma\subset \mathcal{Z}$.
The monodromy charge $Q_i$ of the primary $\phi_i$ with respect to $J$ is defined as
\begin{align}
Q_i = h_{\sigma(0)} + h_i - h_{\sigma(i)}\text{ mod }1~,
\label{eq:monodromycharge}
\end{align}
where $h_i$ is the conformal weight of $\phi_i$, and $\sigma$ is the permutation defined by~\eqref{eq:simplecurrent}---note in particular that $J=\phi_{\sigma(0)}$. Primaries with $Q_i=0$ are said to be local with respect to $J$. The elements of the modular $\mathcal{S}$ matrix satisfy the identity
\begin{align}
\mathcal{S}_{\sigma(i)j} = \ee^{2\pi\ii Q_j}\mathcal{S}_{ij}~.
\label{eq:Smatrixmonodromy}
\end{align}
The simple current $J$ organizes primaries into orbits, related by the action~\eqref{eq:simplecurrent}. Let $\Gamma_i$ be the stabilizer of $\phi_i$, that is, the set of simple currents in $\Gamma$ that have trivial fusion with $\phi_i$. We denote its order by $|\Gamma_i|$. The primary $\phi_i$ is called a fixed point of $J$ when $|\Gamma_i|>1$.

\subsection{Integral spin extensions}
Let $J$ be a simple current generating a cyclic group $\Gamma$ of order $|\Gamma|$. In order to label the orbits of $J$, we pick a subset of indices $I\subset\{0,1,\dots,\ell-1\}$ such that each $\phi_{i\in I}$ belongs to a distinct orbit, and  such that the fusion product of $J$ with the $\phi_{i\in I}$ generates all primaries.

If $J$ has integral conformal dimension, all primaries within a given orbit share the same monodromy charge $Q_{\sigma(i)}=Q_i$. Moreover, in this case, the condition $Q_i=0$ becomes equivalent to $h_{\sigma(i)} = h_i$ mod $1$. Hence, primaries that are local with respect to $J$ have conformal weight equal modulo $1$ to that of other primaries in their orbit. Consequently, it makes sense to group them in an extended character
\begin{align}
\Chi_i(\tau) = \frac{1}{|\Gamma_i|}\sum_{a=1}^{|\Gamma|}\chi_{\sigma^a(i)}(\tau)~,
\end{align}
that satisfies
\begin{align}
\Chi_i(\tau+1) = \ee^{2\pi\ii(h_i-c/24)}\Chi_i(\tau)~.
\end{align}
Making use of the property~\eqref{eq:Smatrixmonodromy}, it is straightforward to see that 
\begin{align}
\Chi_i(-1/\tau) =\frac{|\Gamma|}{|\Gamma_i|}\sum_{\substack{j\in I \\ Q_j=0}} \mathcal{S}_{ij} \Chi_j(\tau)~.
\end{align}
Hence the extended characters $\Chi_i(\tau)$ associated to orbits with monodromy charge $Q_i=0$ form a representation of the modular group $\mathrm{SL}(2,\Z)$.

The extended characters admit a natural CFT interpretation in terms of an extension of the chiral algebra. The initial chiral algebra is enlarged by the integral spin current $J$. Primaries of the extended algebra must be local with respect to $J$, so only the fields $\phi_i$ with $Q_i=0$ are retained. Moreover, the $\phi_i$ in the same orbit~\eqref{eq:simplecurrent} sit in a single primary representation of the extended algebra, which is encoded in the character $\Chi_i(\tau)$. The currents by which the CFT was extended can be read off the extended identity character $\Chi_0(\tau)$.

As a brief aside, let us recall that, in general, the matrix 
\begin{align}
\widetilde{\mathcal{S}}^0_{ij} &= \frac{|\Gamma|}{|\Gamma_i|}\mathcal{S}_{ij} ~,
\end{align}
appearing in the modular transformation of extended characters is not the modular matrix of the extended theory: in particular it fails to be symmetric whenever the simple current $J$ has fixed points. Indeed, each fixed point gives rise to splitted fields, whose $\mathrm{SL}(2,\Z)$ transformations are not fully determined by the modular $\mathcal{S}$ matrix of the unextended theory, but instead require a fixed point resolution. The full modular matrix of the extended theory involves resolution matrices $\widetilde{\mathcal{S}}^a$ which act nontrivially on the fixed points of $J^a$. We will comment on this further below in section~\ref{sec:fp}.

\subsection{The KM center}
Simple currents of Kac--Moody algebras have been classified~\cite{Fuchs:1990wb}. The simple currents of a KM algebra based on a Lie algebra $\mathfrak{g}$ are in one-to-one correspondence with elements of the center of $G$, the simply connected Lie group associated to $\mathfrak{g}$, up to a single exception.\footnote{The unique counterexample is the $\mathfrak{e}_8$ KM algebra at level $2$, whose $\Z_2$ center is generated by the spin-$3/2$ KM primary associated to the representation $\rep{3875}$.} We now provide some details on the KM simple currents relevant for our analysis. Our Lie algebra conventions can be found in appendix~\ref{app:Lie}.

\subsubsection*{Type $\mathfrak{a}_r$}
The $\Z_{r+1}$ center of the $\mathfrak{a}_r$ KM algebra at level $k$ is generated by the current
\begin{align}
J = \phi_{[k00\cdots00]}~,
\end{align}
whose conformal dimension is $h=\tfrac{1}{2}kr/(r+1)$. Under fusion, it acts on the KM primaries by
\begin{align}
J\times \phi_{[l_1l_2\cdots l_r]} = \phi_{[l_0l_1\cdots l_{r-1}]}~,
\end{align}
where the affine Dynkin label $l_0$ is defined by  the relation $l_0+l_1+\dots +l_r = k$. The monodromy charge of the primary $\phi_{[l_1l_2\cdots l_r]}$ with respect to $J$ is equal to 
\begin{align}
Q_{[l_1\cdots l_r]} = \tfrac{1}{r+1}\sum_{a=1}^r a l_a\; \text{mod }1~.
\end{align}

\subsubsection*{Type $\mathfrak{d}_4$}
The $\Z_2\times\Z_2$ center of the $\mathfrak{d}_4$ KM algebra at level $k$ is generated by the currents
\begin{align}
J&= \phi_{ [k000] }~, &
J'&= \phi_{ [000k] }~,
\end{align}
each of which has conformal dimension $h = k/2$. Under fusion, they act on the KM primaries by
\begin{align}
J\times \phi_{[l_1l_2l_3l_4]} &= \phi_{[l_0l_2l_4l_3]}~,& 
J'\times \phi_{[l_1l_2l_3l_4]} &= \phi_{[l_3l_2l_1l_0]}~,
\end{align}
where the affine Dynkin label $l_0$ is defined by $l_0+l_1+2l_2+l_3+l_4 = k$. The monodromy charges of the primary $\phi_{[l_1l_2l_3l_4]}$ with respect to $J$ and $J'$ are equal to 
\begin{align}
Q_{[l_1l_2l_3l_4]} &= \tfrac{1}{2}(l_3+l_4)\;\text{mod }1 ~,&
Q'_{[l_1l_2l_3l_4]} &=  \tfrac{1}{2}(l_1+l_3)\;\text{mod }1 ~.
\end{align}

\subsubsection*{Type $\mathfrak{e}_6$}
The $\Z_3$ center of the $\mathfrak{e}_6$ KM algebra at level $k$ is generated by the current
\begin{align}
J &= \phi_{[00000k]}
\end{align}
whose conformal dimension is $h=2k/3$. Under fusion, it acts on the KM primaries by
\begin{align}
J\times \phi_{[l_1l_2l_3l_4l_5l_6]} = \phi_{[l_6l_3l_5l_4l_2l_0]}~,
\end{align}
where the affine Dynkin label $l_0$ is defined by $l_0+l_1+2l_2+2l_3+3l_4+2l_5+l_6= k$. The monodromy charge of the primary $\phi_{[l_1l_2l_3l_4l_5l_6]}$ with respect to $J$ is equal to 
\begin{align}
Q_{[l_6l_3l_5l_4l_2l_0]} = \tfrac{1}{3}(2l_1+l_2+2l_5+l_6)\;\text{mod }1 ~.
\end{align}

\subsubsection*{Type $\mathfrak{e}_7$}
The $\Z_2$ center of the $\mathfrak{e}_7$ KM algebra at level $k$ is generated by the current
\begin{align}
J &= \phi_{[000000k]}
\end{align}
whose conformal dimension is $h=3k/4$. Under fusion, it acts on the KM primaries by
\begin{align}
J\times \phi_{[l_1l_2l_3l_4l_5l_6l_7]} = \phi_{[l_6l_2l_5l_4l_3l_1l_0]}~,
\end{align}
where the affine Dynkin label $l_0$ is defined by $l_0+2l_1+2l_2+3l_3+4l_4+3l_5+2l_6+l_7  = k$. The monodromy charge of the primary $\phi_{[l_1l_2l_3l_4l_5l_6]}$ with respect to $J$ is equal to 
\begin{align}
Q_{[l_1l_2l_3l_4l_5l_l6l_7]} =\tfrac{1}{2}(l_2+l_5+l_7)\;\text{mod }1~.
\end{align}

\section{Elliptic genus and extensions}
\label{sec:ellipticgenera}
In this section, we list, for each of the five SCFTs under study, the subgroup of the KMV center that extends the left-moving chiral algebra. For each model, we introduce the corresponding extended characters and present the expansion of the elliptic genus in terms of these characters. The result is an exact expression for this refined supersymmetric index, which allows one to extract the full BPS spectrum, including its representation under the spacetime gauge algebra, at arbitrary Virasoro level.

\subsection{\texorpdfstring{$G=\mathrm{E}_6$}{G=E6}}
The left-moving KMV algebra of the $\mathrm{E}_6$ hyperplane string SCFT has central charges
\begin{align}
c_\tL^{\mathrm{gauge}} &= \frac{156}{5}~,&
c_\tL^{\mathrm{res}} &= \frac{4}{5}~.
\end{align}
The  $\mathfrak{e}_6$ KM subalgebra is at level $8$. The elliptic genus $\Ell_{\mathrm{E}_6}$ admits an expansion 
\begin{align}
\Ell_{\mathrm{E}_6} = \sum_{\lambda,r,s} C_\lambda^{(r,s)}\chi_\lambda^{(r,s)}
\end{align}
into characters of KMV primary representations. This finite sum runs over the minimal model labels $(r,s)$ and the $\mathfrak{e}_6$ highest weights $\lambda=[l_1\cdots l_6]$ that belong to the unitary range
\begin{align}
1 \leq s \leq r &\leq 4~,&
\lambda\cdot\gamma_{\mathfrak{e}_6} &\leq 8~,
\end{align}
where $\gamma_{\mathfrak{e}_6} = [010000]$ is the $\mathfrak{e}_6$ highest root. Only $104$ of the coefficients $C_\lambda^{(r,s)}$ are nontrivial.

We observe that some of the irreducible representations of the $c_\tL^{\text{res}}=4/5$ Virasoro algebra never appear, while others are paired up: the integer coefficients $C_\lambda^{(r,s)}$ satisfy 
\begin{align}
C_\lambda^{(1,1)} &= C_\lambda^{(4,1)}~,&
C_\lambda^{(2,1)} &= C_\lambda^{(3,1)}~,&
C_\lambda^{(2,2)} = C_\lambda^{(3,2)} = C_\lambda^{(4,2)} = C_\lambda^{(4,4)} &= 0~.
\end{align}
 This strongly suggests that the left-moving chiral algebra is extended by the operator
\begin{align}
J_0 = \phi_{[000000]}^{(4,1)}~,
\end{align}
which has conformal dimension $h_\tL = 3$. This $J_0$ is a simple current of order $2$.  It commutes with the $\mathfrak{e}_6$ currents of the $c_\tL=32$ theory, and only acts nontrivially on the minimal model sector. The simple current extension by $J_0$ of the Virasoro minimal model $\mathfrak{M}_{(6,5)}$ leads to the Potts CFT, which has the same central charge as $\mathfrak{M}_{(6,5)}$ but a distinct set of primaries. In particular, the Virasoro algebra generated by the stress tensor $T_{\text{res}}$ is enlarged by the spin-$3$ current $J_0$ to the $W_3$ algebra~\cite{Zamolodchikov:1985wn}. 

\subsubsection*{Extended characters for $J_0$}
In the elliptic genus, the extended characters of the Potts CFT
\begin{align}
\mathcal{X}^{\text{P}}_I &= \chi^{\text{res}}_{(1,1)}+\chi^{\text{res}}_{(4,1)}~,&
\mathcal{X}^{\text{P}}_{\sigma_1} = \mathcal{X}^{\text{P}}_{\sigma_2} &= \chi^{\text{res}}_{(3,3)}~,
\nonumber\\
\mathcal{X}^{\text{P}}_\varepsilon &= \chi^{\text{res}}_{(2,1)}+\chi^{\text{res}}_{(3,1)}~,&
\mathcal{X}^{\text{P}}_{\psi_1} = \mathcal{X}^{\text{P}}_{\psi_2} &= \chi^{\text{res}}_{(4,3)}~,
\label{eq:PottsCharacters}
\end{align}
can be factorized. Here, the labels $\{I,\sigma_1,\sigma_2,\psi_1,\psi_2,\varepsilon \}$ correspond to the six Potts $W_3$ primaries. Note that, while the pairs $\{\sigma_1,\sigma_2\}$ and $\{\psi_1,\psi_2\}$ share the same Virasoro characters, they can be differentiated by their charge under the $\Z_3$ symmetry of the Potts model:
\begin{align}
Q^{\text{P}}_I = Q^{\text{P}}_\varepsilon &= 0~,&
Q^{\text{P}}_{\sigma_1} = Q^{\text{P}}_{\psi_1} &= \tfrac{2}{3}~,&
Q^{\text{P}}_{\sigma_2} = Q^{\text{P}}_{\psi_2} &= \tfrac{1}{3}~.
\label{eq:Z3Potts}
\end{align}
The elliptic genus $\Ell_{\mathrm{E}_6}$ can be expanded in extended characters by making the replacements
\begin{align}
\chi_\lambda^{(1,1)}+\chi_\lambda^{(4,1)} &\to \chi_\lambda^I~,&
\chi_\lambda^{(3,3)} &\to \chi_\lambda^{\sigma_1} \text{ or } \chi_\lambda^{\sigma_2}~,
\nonumber\\
\chi_\lambda^{(2,1)}+\chi_\lambda^{(3,1)} &\to \chi_\lambda^\varepsilon~,&
\chi_\lambda^{(4,3)} &\to \chi_\lambda^{\psi_1} \text{ or } \chi_\lambda^{\psi_2}~.
\label{eq:PottsKMVassignment}
\end{align}
This procedure is ill-defined, as one needs to decide which Potts primary gets assigned to the characters $\chi_\lambda^{(3,3)}$ and $\chi_\lambda^{(4,3)}$. However, there is a natural guess to resolve this ambiguity. Under the center symmetry of $\mathfrak{e}_6$, the KM primaries associated to the representation $\lambda=[l_1l_2l_3l_4l_5l_6]$ carry a $\Z_3$ charge
\begin{align}
Q^{\mathfrak{e}_6}_\lambda = \tfrac{1}{3}(2l_1+l_3+2l_5+l_6) \text{ mod }1~.
\label{eq:Z3e6}
\end{align}
The elliptic genus $\Ell_{\mathrm{E}_6}$ does not carry a definite charge under either factor of the $\Z_3\times\Z_3$ symmetry generated by~\eqref{eq:Z3Potts} and~\eqref{eq:Z3e6}. However, there exists a unique assignment~\eqref{eq:PottsKMVassignment} such that $\Ell_{\mathrm{E}_6}$ becomes neutral under the diagonal combination $\Z_3\subset \Z_3\times\Z_3$ defined by
\begin{align}
Q=Q^{\text{P}}-Q^{\mathfrak{e}_6}~.
\label{eq:Z3e6Potts}
\end{align}
This is nontrivial and relies on some selection rules on the coefficients $C_\lambda^{(r,s)}$ which we observe in the expression of $\Ell_{\mathrm{E}_6}$.\footnote{Explicitly, the coefficients satisfy $C_\lambda^{(1,1)},C_\lambda^{(2,1)}=0$ if $Q_\lambda=\tfrac{1}{3},\tfrac{2}{3}$, and $C_\lambda^{(3,3)},C_\lambda^{(4,3)}=0$ if $Q_\lambda=0$.} 

Our educated guess for the expansion of $\Ell_{\mathrm{E}_6}$  can be verified. Consider the Potts extended characters~\eqref{eq:PottsCharacters}, which transform with the modular matrix $\mathcal{S}_{\text{P}}$ given in~\eqref{eq:SmatrixPotts}. Note that this $\mathcal{S}_{\text{P}}$ cannot simply be obtained from the modular matrix $\mathcal{S}_{\text{res}}$ of the minimal model: it contains additional information to disentangle the characters $\Chi_{\psi_1},\Chi_{\psi_2}$ and $\Chi_{\sigma_1},\Chi_{\sigma_2}$---what is known as a fixed point resolution for the simple current extension by $J_{\text{P}}$. In light of this added input, the methods of section~\ref{subsec:modularmethod} can be applied to construct modular invariants out of $\mathfrak{e}_6$ KM characters and Potts extended characters. We obtain two distinct modular invariants. The first is the expression that we obtained by enforcing the invariance of the elliptic genus under the $\Z_3$ symmetry~\eqref{eq:Z3e6Potts}. The second is related to it by relabeling $\psi_1\leftrightarrow\psi_2$, $\sigma_1\leftrightarrow\sigma_2$, and is neutral under the $\Z_3$ symmetry that assigns a charge $Q=Q^{\text{P}}+Q^{\mathfrak{e}_6}$ to the KMV primaries. Hence, the ambiguity~\eqref{eq:PottsKMVassignment} is fully remedied by modularity.

\subsubsection*{The mixed $\Z_3$ symmetry}
The above analysis has an important lesson: when a discrete symmetry of the $c_\tL=32$ theory acts trivially on the elliptic genus, there could be an extension of the chiral algebra. We saw this explicitly with $J_{\text{P}}$: the primaries with nontrivial monodromy charge under this simple current of order $2$
do not appear in the elliptic genus, which is therefore invariant under a global $\Z_2$ symmetry. This strongly suggests that the chiral algebra is augmented to $W_3$ by the spin-$3$ current $J_{\text{P}}$. This observation motivated us to write the elliptic genus in terms of characters of this extended algebra, which in turn led to the identification of an additional invariance of $\Ell_{\mathrm{E}_6}$ under the $\Z_3$ action defined by~\eqref{eq:Z3e6Potts}.

The symmetry~\eqref{eq:Z3e6Potts} combines the $\Z_3$ center symmetry of $\mathfrak{e}_6$ and the global $\Z_3$ symmetry of the Potts model. It is best described using the simple currents of the $c_\tL=32$ chiral algebra made out of the $\mathfrak{e}_6$ KM algebra and the Potts $W_3$ algebra. The two simple currents
\begin{align}
J_{\mathfrak{e}_6} &= \phi_{[000008]}^{I}~,&
J_{\text{P}} &= \phi_{[000000]}^{\psi_1}~,
\end{align}
of conformal dimension $h_\tL=16/3$ and $h_\tL=2/3$ respectively, generate the $\Z_3\times\Z_3$ center of the theory. Their product $J=J_{\mathfrak{e}_6}\times J_{\text{P}}$, i.e.
\begin{align}
J = \phi_{[000008]}^{\psi_1}~,
\end{align}
is an integral spin simple current of order $3$, which can be used to construct a simple current extension. The monodromy charge of a primary with respect to $J$ coincide with its mixed $\Z_3$ charge~\eqref{eq:Z3e6Potts}. The fusion of the spin-$6$ current $J$ with the other primaries is
\begin{align}
J\times \phi_\lambda^{I} &= \phi_{\lambda'}^{\psi_1}~,&
J\times \phi_\lambda^{\psi_1} &= \phi_{\lambda'}^{\psi_2}~,&
J\times \phi_\lambda^{\psi_2} &= \phi_{\lambda'}^{I}~,\nonumber\\
J\times \phi_\lambda^{\varepsilon} &= \phi_{\lambda'}^{\sigma_1}~,&
J\times \phi_\lambda^{\sigma_1} &= \phi_{\lambda'}^{\sigma_2}~,&
J\times \phi_\lambda^{\sigma_2} &= \phi_{\lambda'}^{\varepsilon}~,
\label{eq:fusione6Potts}
\end{align}
so that it defines extended characters 
\begin{align}
\mathcal{X}_{\lambda }^{ I} = 
\chi^{\mathfrak{e}_{6,8}}_{\lambda}\Chi^{\text{P}}_{I} + \chi^{\mathfrak{e}_{6,8}}_{\lambda'}\Chi^{\text{P}}_{\psi_1} + \chi^{\mathfrak{e}_{6,8}}_{\lambda''}\Chi^{\text{P}}_{\psi_2} ~,
\nonumber\\
\mathcal{X}_{\lambda }^{ \varepsilon} = 
\chi^{\mathfrak{e}_{6,8}}_{\lambda}\Chi^{\text{P}}_{\varepsilon} + \chi^{\mathfrak{e}_{6,8}}_{\lambda'}\Chi^{\text{P}}_{\sigma_1} + \chi^{\mathfrak{e}_{6,8}}_{\lambda''}\Chi^{\text{P}}_{\sigma_2} ~.
\end{align}
Here, the $\mathfrak{e}_6$ representations $\lambda,\lambda',\lambda''$ are related to each other by
\begin{align}
\lambda &= [l_1l_2l_3l_4l_5l_6]~,&
\lambda' &= [l_6l_3l_5l_4l_2l_0]~,&
\lambda'' &= [l_0l_5l_2l_4l_3l_1]~,
\label{eq:e6outersymmetry}
\end{align}
with $l_0 = 8-\lambda\cdot \gamma_{\mathfrak{e}_6}$. The integer $l_0$ is attached to the external node of the $\mathfrak{e}_{6}$ extended Dynkin diagram, and the permutation~\eqref{eq:e6outersymmetry} of integrable $\mathfrak{e}_{6,8}$ highest weight representations  corresponds to the $\Z_3$ symmetry of the diagram.

\subsubsection*{Elliptic genus}
The KMV identity character lies in the extended character
\begin{align}
\mathcal{X}_{[000000] }^{ I} &= \chi^{\mathfrak{e}_{6,8}}_{[000000]}\Chi^{\text{P}}_{{I}} + \chi^{\mathfrak{e}_{6,8}}_{[000008]}\Chi^{\text{P}}_{\psi_1} + \chi^{\mathfrak{e}_{6,8}}_{[800000]}\Chi^{\text{P}}_{\psi_2} ~,
\end{align}
that we interpret as the identity character for the chiral algebra enlarged by the spin-$3$ current $J_{\text{P}}$ and the spin-$6$ current $J$. The elliptic genus can be expanded in extended characters. Only $26$ primaries of the extended chiral algebra appear in the expansion, which is given by\footnote{Note that this expansion of $\Ell_{\mathrm{E}_6}$ in extended characters appears implicitly in equation~(4.32) of~\cite{Lockhart:2025lea}.}
\begin{alignat}{16}
\Ell_{\mathrm{E}_6} = &&
&& 4\,\mathcal{X}_{ [200102] }^{ \varepsilon } &&
&\,+\,& 3\,\mathcal{X}_{ [000000] }^{ I } &&
&\,+\,& 3\,\mathcal{X}_{ [000103] }^{ I } &&
&\,+\,& 3\,\mathcal{X}_{ [101011] }^{ I } &&
&\,+\,& 3\,\mathcal{X}_{ [300100] }^{ I } &&
&\,+\,& 3\,\mathcal{X}_{ [010200] }^{ \varepsilon } &&
&\,+\,& 3\,\mathcal{X}_{ [011010] }^{ \varepsilon } \nonumber\\
&&
&\,+\,& 3\,\mathcal{X}_{ [021002] }^{ \varepsilon } &&
&\,+\,& 3\,\mathcal{X}_{ [220010] }^{ \varepsilon } &&
&\,+\,& 2\,\mathcal{X}_{ [000030] }^{ I } &&
&\,+\,& 2\,\mathcal{X}_{ [003000] }^{ I } &&
&\,+\,& 2\,\mathcal{X}_{ [010022] }^{ I } &&
&\,+\,& 2\,\mathcal{X}_{ [212000] }^{ I } &&
&\,+\,& 2\,\mathcal{X}_{ [001021] }^{ \varepsilon } \nonumber\\
&&
&\,+\,& 2\,\mathcal{X}_{ [102010] }^{ \varepsilon } &&
&\,+\,& \mathcal{X}_{ [001005] }^{ I } &&
&\,+\,& \mathcal{X}_{ [100201] }^{ I } &&
&\,+\,& \mathcal{X}_{ [200002] }^{ I } &&
&\,+\,& \mathcal{X}_{ [500001] }^{ I } &&
&\,+\,& \mathcal{X}_{ [010000] }^{ \varepsilon } &&
&\,+\,& \mathcal{X}_{ [110001] }^{ \varepsilon } \nonumber\\
&&
&\,+\,& \mathcal{X}_{ [201004] }^{ \varepsilon } &&
&\,+\,& \mathcal{X}_{ [400012] }^{ \varepsilon } &&
&\,-\,& 2\,\mathcal{X}_{ [000006] }^{ \varepsilon } &&
&\,-\,& 2\,\mathcal{X}_{ [600000] }^{ \varepsilon } &&
&\,-\,& 4\,\mathcal{X}_{ [040000] }^{ I } && ~.
\end{alignat}

\subsection{\texorpdfstring{$G=\mathrm{A}_7$}{G=A7}}
The left-moving KMV algebra of the $\mathrm{A}_7$ hyperplane string SCFT has central charges
\begin{align}
c_\tL^{\mathrm{gauge}} &= \frac{63}{2}~,&
c_\tL^{\mathrm{res}} &= \frac{1}{2}~.
\end{align}
The  $\mathfrak{a}_{7}$ KM subalgebra is at level $8$. The residual factor is associated to the Ising CFT. The elliptic genus $\Ell_{\mathrm{A}_7}$ admits an expansion 
in KMV characters, indexed by integers $(r,s)$ and $\mathfrak{a}_7$ highest weights $\lambda=[l_1\cdots l_7]$ in the unitary range
\begin{align}
1 \leq s \leq r &\leq 2~,&
\lambda\cdot\gamma_{\mathfrak{a}_7} &\leq 8~,
\end{align}
where $\gamma_{\mathfrak{a}_7} = [1000001]$ is the $\mathfrak{a}_7$ highest root. We also denote the KMV primaries as
\begin{align}
\phi_\lambda^{I} &= \phi_\lambda^{(1,1)}~,&
\phi_\lambda^{\varepsilon} &= \phi_\lambda^{(2,1)}~,&
\phi_\lambda^{\sigma} &= \phi_\lambda^{(2,2)}~,
\label{eq:Isingprimaries}
\end{align}
in reference to the Virasoro primaries of the Ising model.

The chiral algebra has a $\Z_8\times\Z_2$ center generated by the simple currents
\begin{align}
J_{\mathfrak{a}_7} &= \phi_{[8000000]}^{I}~,&
J_{\text{I}} &= \phi_{[0000000]}^{\varepsilon}~,
\end{align}
of conformal dimension $h_\tL=7/2$ and $h_\tL=1/2$ respectively. The KMV primaries have monodromy charges
\begin{align}
Q^{\mathfrak{a}_7}_\lambda &= \tfrac{1}{8}\sum_{a=1}^7 a l_a \text{ mod }1 ~,
\end{align}
with respect to the $\Z_8$ factor, and
\begin{align}
Q^{\text{I}}_{I} = 
Q^{\text{I}}_{\varepsilon} &= 0~,&
Q^{\text{I}}_{\sigma} &= \tfrac{1}{2}~,
\end{align}
with respect to the $\Z_2$ factor.

All of the KMV primaries that appear in the elliptic genus are neutral under the $\Z_8$ symmetry defined by $Q=Q^{\mathfrak{a}_7}+Q^{\text{I}}$. This is the monodromy charge associated to
\begin{align}
J = J_{\mathfrak{a}_7}\times J_{\text{I}}~,
\end{align}
which is a spin-$4$ simple current of order $8$. Its fusion with the left-moving primaries is
\begin{align}
J \times \phi_\lambda^{I} &= \phi_{\lambda'}^{\varepsilon}~,&
J \times \phi_\lambda^{\varepsilon} &= \phi_{\lambda'}^{I}~,&
J \times \phi_\lambda^{\sigma} &= \phi_{\lambda'}^{\sigma}~,
\end{align}
where the $\mathfrak{a}_7$ dominant weights
\begin{align}
\lambda &= [l_1l_2\cdots l_7]~,&
\lambda' &= [l_0l_1\cdots l_6]~,
\end{align}
with $l_0=8-\lambda\cdot\gamma_{\mathfrak{a}_7}$, are related by the $\Z_8$ cyclic symmetry of the $\mathfrak{a}_7$ extended Dynkin diagram.

The $\Z_4$ subgroup generated by the spin-$6$ current $J^2=J_{\mathfrak{a}_7}^2=\phi_{[0800000]}^{I}$ only acts on the $\mathfrak{a}_7$ labels of the primaries. Hence, we can think of the $\Z_8$ extension by $J$ in two steps. First, the KM algebra is augmented by $J_{\mathfrak{a}_7}^2$. The extended $\mathfrak{a}_7$ theory has characters $\Chi_\lambda^{\mathfrak{a}_{7,8}}$, with the identity character given by
\begin{align}
\mathcal{X}^{\mathfrak{a}_{7,8}}_{[0000000]} = 
\chi^{\mathfrak{a}_{7,8}}_{[0000000]}+ \chi^{\mathfrak{a}_{7,8}}_{[0800000]}+ \chi^{\mathfrak{a}_{7,8}}_{[0008000]}+ \chi^{\mathfrak{a}_{7,8}}_{[0000080]}~.
\end{align}
The current $J$ becomes a simple current of order $2$ in the $J^2$-extended theory. After its inclusion in the chiral algebra, the identity character becomes
\begin{align}
\mathcal{X}_{ [0000000] }^{ I} = \mathcal{X}^{\mathfrak{a}_{7,8}}_{[0000000]}\chi^{\text{I}}_I + \mathcal{X}^{\mathfrak{a}_{7,8}}_{[8000000]}\chi^{\text{I}}_\varepsilon~.
\end{align}
The elliptic genus is compatible with a simple current extension by $J$: it can be expanded in $\Z_8$-extended characters as
\begin{alignat}{14}
\Ell_{\mathrm{A}_7} = &&
&& 8\,\mathcal{X}_{ [1111111] }^{ \sigma} &&
&\,+\,& 4\,\mathcal{X}_{ [2101210] }^{ I} &&
&\,+\,& 3\,\mathcal{X}_{ [0000000] }^{ I} &&
&\,+\,& 3\,\mathcal{X}_{ [0300200] }^{ I} &&
&\,+\,& 3\,\mathcal{X}_{ [1110111] }^{ I} &&
&\,+\,& 3\,\mathcal{X}_{ [2201010] }^{ I} \nonumber\\
&&
&\,+\,& 2\,\mathcal{X}_{ [0220010] }^{ I} &&
&\,+\,& 2\,\mathcal{X}_{ [0100220] }^{ I} &&
&\,+\,& 2\,\mathcal{X}_{ [3010200] }^{ I} &&
&\,+\,& 2\,\mathcal{X}_{ [0020103] }^{ I} &&
&\,+\,& 2\,\mathcal{X}_{ [1002110] }^{ \sigma} &&
&\,+\,& 2\,\mathcal{X}_{ [0112001] }^{ \sigma} \nonumber\\
&&
&\,+\,& 2\,\mathcal{X}_{ [0200210] }^{ \sigma} &&
&\,+\,& 2\,\mathcal{X}_{ [0120020] }^{ \sigma} &&
&\,+\,& 2\,\mathcal{X}_{ [2101020] }^{ \sigma} &&
&\,+\,& 2\,\mathcal{X}_{ [0201012] }^{ \sigma} &&
&\,+\,& \mathcal{X}_{ [6100000] }^{ I} &&
&\,+\,& \mathcal{X}_{ [1210000] }^{ I} \nonumber\\
&&
&\,+\,& \mathcal{X}_{ [0000121] }^{ I} &&
&\,+\,& \mathcal{X}_{ [5000110] }^{ I} &&
&\,+\,& \mathcal{X}_{ [01100005] }^{ I} &&
&\,+\,& \mathcal{X}_{ [4110001] }^{ I} &&
&\,+\,& \mathcal{X}_{ [2000301] }^{ I} &&
&\,+\,& \mathcal{X}_{ [1030002] }^{ I} \nonumber\\
&&
&\,+\,& \mathcal{X}_{ [0200020] }^{ I} &&
&\,+\,& \mathcal{X}_{ [1010400] }^{ I} &&
&\,+\,& \mathcal{X}_{ [0040101] }^{ I} &&
&\,+\,& \mathcal{X}_{ [0120210] }^{ I} &&
&\,-\,& 2\,\mathcal{X}_{ [0200000] }^{ \sigma } &&
&\,-\,& 2\,\mathcal{X}_{ [0000020] }^{ \sigma } \nonumber\\
&&
&\,-\,& 4\,\mathcal{X}_{ [0004000] }^{ I} 
&&~.
\end{alignat}

\subsection{\texorpdfstring{$G=\mathrm{A}_5\times\mathrm{D}_4$}{G=A5xD4}}
The left-moving KMV algebra of the $\mathrm{A}_5\times\mathrm{D}_4$ hyperplane string SCFT has central charges
\begin{align}
c_\tL^{\mathrm{gauge}} &= \frac{63}{2}~,&
c_\tL^{\mathrm{res}} &= \frac{1}{2}~.
\end{align}
Both factors of the $\mathfrak{a}_{5}+\mathfrak{d}_4$ KM subalgebra are at level $6$. Just as in the $\mathrm{A}_7$ model, the residual factor is associated to the Ising CFT. As in~\eqref{eq:Isingprimaries}, we use labels $\{I,\varepsilon,\sigma\}$ to denote the three Ising primaries.

The center of the KMV algebra is isomorphic to $\Z_6\times\Z_2\times\Z_2\times\Z_2$. It is generated by the four currents
\begin{align}
J_{\mathfrak{a}_5}&= \phi_{ [60000,0000] }^I~, &
J_{\mathfrak{d}_4}&= \phi_{ [00000,6000] }^I~, &
J_{\mathfrak{d}_4}'&= \phi_{ [00000,0006] }^I~, &
J_{\text{I}}&= \phi_{ [00000,0000] }^\varepsilon~.
\end{align}
After examination, we see that the elliptic genus $\Ell_{ \mathrm{A}_5\times\mathrm{D}_4 }$ has zero monodromy charge under the $\Z_6\times\Z_2\times\Z_2$ subgroup generated by
\begin{align}
J_1 &= J_{\mathfrak{a}_5}\times J_{\text{I}} ~, &
J_2 &= J_{\mathfrak{d}_4}~, &
J_3 &= J_{\mathfrak{d}_4}'~,
\end{align}
which are all spin-3 currents. Note that since $J_1^2$ commutes with the $\mathfrak{d}_4$ currents and with the stress tensor $T_{\text{res}}$,  it defines a simple current extension of the $\mathfrak{a}_5$ KM algebra. Similarly, the currents $J_2$ and $J_3$ extend the $\mathfrak{d}_4$ KM algebra.

The identity character of the $c_\tL=32$ left-moving chiral algebra takes the form
\begin{align}
\Chi^I_{[00000,0000]} = \Chi^{\mathfrak{a}_{5,6}}_{[00000]} \Chi^{\mathfrak{d}_{4,6}}_{[0000]} \chi^{\text{I}}_{I} + \Chi^{\mathfrak{a}_{5,6}}_{[60000]} \Chi^{\mathfrak{d}_{4,6}}_{[0000]} \chi^{\text{I}}_{\varepsilon}~,
\end{align}
where 
\begin{align}
\Chi^{\mathfrak{a}_{5,6}}_{[00000]} &= \chi^{\mathfrak{a}_{5,6}}_{[00000]} + \chi^{\mathfrak{a}_{5,6}}_{[06000]} + \chi^{\mathfrak{a}_{5,6}}_{[00060]}~,
\nonumber\\
\Chi^{\mathfrak{a}_{5,6}}_{[60000]} &= \chi^{\mathfrak{a}_{5,6}}_{[60000]} + \chi^{\mathfrak{a}_{5,6}}_{[00600]} + \chi^{\mathfrak{a}_{5,6}}_{[00006]}~,
\nonumber\\
\Chi^{\mathfrak{d}_{4,6}}_{[0000]} &= \chi^{\mathfrak{d}_{4,6}}_{[0000]} + \chi^{\mathfrak{d}_{4,6}}_{[6000]} + \chi^{\mathfrak{d}_{4,6}}_{[0060]} + \chi^{\mathfrak{d}_{4,6}}_{[0006]}~.
\end{align}
The elliptic genus of the (0,4) SCFT is given by
\begin{alignat}{12}
\Ell_{ \mathrm{A}_5\times\mathrm{D}_4 } = &&
&& 12\,\Chi_{ [11111 , 1111] }^{ \sigma } &&
&\,+\,& 4\,\Chi_{ [11100 , 1111] }^{ I } &&
&\,+\,& 4\,\Chi_{ [00111 , 1111] }^{ I } &&
&\,+\,& 4\,\Chi_{ [00200 , 0300] }^{ I } &&
&\,+\,& 3\,\Chi_{ [00000 , 0000] }^{ I } \nonumber\\
&&
&\,+\,& 3\,\Chi_{ [20202 , 0100] }^{ I } &&
&\,+\,& 3\,\Chi_{ [10112 , 1011] }^{ I } &&
&\,+\,& 3\,\Chi_{ [20202 , 2020] }^{ I } &&
&\,+\,& 2\,\Chi_{ [11200 , 1011] }^{ \sigma } &&
&\,+\,& 2\,\Chi_{ [01110 , 0022] }^{ \sigma } \nonumber\\
&&
&\,+\,& 2\,\Chi_{ [30000 , 0300] }^{ \sigma } &&
&\,+\,& 2\,\Chi_{ [00211 , 0300] }^{ \sigma } &&
&\,+\,& 2\,\Chi_{ [01021 , 0200] }^{ \sigma } &&
&\,+\,& 2\,\Chi_{ [00022 , 0102] }^{ I } &&
&\,+\,& 2\,\Chi_{ [10201 , 0102] }^{ I } \nonumber\\
&&
&\,+\,& 2\,\Chi_{ [00022 , 0120] }^{ I } &&
&\,+\,& 2\,\Chi_{ [10201 , 0120] }^{ I } &&
&\,+\,& 2\,\Chi_{ [22000 , 2100] }^{ I } &&
&\,+\,& 2\,\Chi_{ [10201 , 2100] }^{ I } &&
&\,+\,& 2\,\Chi_{ [01102 , 0102] }^{ \sigma } \nonumber\\
&&
&\,+\,& 2\,\Chi_{ [01102 , 0120] }^{ \sigma } &&
&\,+\,& 2\,\Chi_{ [20110 , 2100] }^{ \sigma } &&
&\,+\,& 2\,\Chi_{ [02001 , 1111] }^{ \sigma } &&
&\,+\,& 2\,\Chi_{ [10020 , 1111] }^{ \sigma } &&
&\,+\,& \Chi_{ [00014 , 0000] }^{ I } \nonumber\\
&&
&\,+\,& \Chi_{ [00014 , 1011] }^{ I } &&
&\,+\,& \Chi_{ [30100 , 0022] }^{ I } &&
&\,+\,& \Chi_{ [00103 , 2020] }^{ I } &&
&\,+\,& \Chi_{ [30100 , 0100] }^{ I } &&
&\,+\,& \Chi_{ [00103 , 0100] }^{ I } \nonumber\\
&&
&\,+\,& \Chi_{ [00006 , 0100] }^{ I } &&
&\,+\,& \Chi_{ [02004 , 0100] }^{ I } &&
&\,+\,& \Chi_{ [11011 , 2020] }^{ I } &&
&\,+\,& \Chi_{ [00030 , 2020] }^{ I } &&
&\,+\,& \Chi_{ [01010 , 0200] }^{ I } \nonumber\\
&&
&\,+\,& \Chi_{ [02101 , 0200] }^{ I } &&
&\,+\,& \Chi_{ [10104 , 0002] }^{ I } &&
&\,+\,& \Chi_{ [21012 , 0002] }^{ I } &&
&\,+\,& \Chi_{ [10104 , 0020] }^{ I } &&
&\,+\,& \Chi_{ [21012 , 0020] }^{ I } \nonumber\\
&&
&\,+\,& \Chi_{ [10104 , 2000] }^{ I } &&
&\,+\,& \Chi_{ [21012 , 2000] }^{ I } &&
&\,+\,& \Chi_{ [30003 , 0000] }^{ I } &&
&\,+\,& \Chi_{ [00200 , 1011] }^{ I } &&
&\,-\,& 2\,\Chi_{ [00003 , 0000] }^{ \sigma } \nonumber\\
&&
&\,-\,& 2\,\Chi_{ [00100 , 0100] }^{ \sigma } &&
&\,-\,& 4\,\Chi_{ [30003 , 0300] }^{ I } &&~.
\end{alignat}

\subsection{\texorpdfstring{$G=\mathrm{A}_3\times\mathrm{A}_3\times\mathrm{E}_7$}{G=A3xA3xE7}}
\label{subsec:modelA3xA3xE7}
The left-moving KMV algebra of the $\mathrm{A}_3\times\mathrm{A}_3\times\mathrm{E}_7$ hyperplane string SCFT has central charges
\begin{align}
c_\tL^{\mathrm{gauge}} &= \frac{313}{10}~,&
c_\tL^{\mathrm{res}} &= \frac{7}{10}~.
\end{align}
Both $\mathfrak{a}_3$ factors of KM subalgebra are at level $6$, while the $\mathfrak{e}_7$ factor is at level $2$. The residual factor is associated to the tricritical Ising CFT. We label the Virasoro primaries as
\begin{align}
\phi_{I} &= \phi_{(1,1)}~,&
\phi_{\sigma} &= \phi_{(2,2)}~,&
\phi_{\varepsilon} &= \phi_{(3,3)}~,\nonumber\\
\phi_{\sigma'} &= \phi_{(2,1)}~,&
\phi_{\varepsilon'} &= \phi_{(3,2)}~,&
\phi_{\varepsilon''} &= \phi_{(3,1)}~.
\end{align}
There is a $\Z_4\times\Z_4\times\Z_2\times\Z_2$ center generated by
\begin{align}
J_{\mathfrak{a}_3}&= \phi^{I}_{[600,000,0000000]}~,&
J_{\mathfrak{a}_3}'&=  \phi^{I}_{[000,600,0000000]}~,
\nonumber\\
J_{\mathfrak{e}_7}&= \phi^{I}_{[000,000,0000002]}~,&
J_{\text{tI}} &= \phi^{\varepsilon''}_{[000,000,0000000]}~.
\end{align}
The currents $J_{\mathfrak{a}_3}$ and $J_{\mathfrak{a}_3}'$ have conformal dimension $h_\tL=9/4$, while $J_{\mathfrak{e}_7}$ and $J_{\text{tI}}$ have $h_\tL=3/2$.

The elliptic genus is neutral under a $\Z_4\times\Z_2\times\Z_2$ subgroup generated by the spin-6 current $J_1$ and the spin-$3$ currents $J_2$ and $J_3$ defined as
\begin{align}
J_1 &= J_{\mathfrak{a}_3}\times J_{\mathfrak{a}_3}'\times J_{\mathfrak{e}_7}~,&
J_2 &= (J_{\mathfrak{a}_3})^2~,&
J_3 &= J_{\mathfrak{e}_7}\times J_{\text{tI}}~.
\end{align}
Note that $(J_{\mathfrak{a}_3}')^2=(J_1)^2\times J_2$. 

The identity character of the $c_\tL=32$ left-moving chiral algebra takes the form
\begin{alignat}{4}
\Chi^{I}_{[000,000,0000000]} = &&
&& ( \chi^{\mathfrak{a}_{3,6}}_{[000]} + \chi^{\mathfrak{a}_{3,6}}_{[060]} )( \chi^{\mathfrak{a}_{3,6}}_{[000]} + \chi^{\mathfrak{a}_{3,6}}_{[060]} )( \chi^{\mathfrak{e}_{7,2}}_{[0000000]}\chi^{\text{tI}}_{I} + \chi^{\mathfrak{e}_{7,2}}_{[0000002]}\chi^{\text{tI}}_{\varepsilon''} )
\nonumber\\
&&
&\,+\,& ( \chi^{\mathfrak{a}_{3,6}}_{[600]} + \chi^{\mathfrak{a}_{3,6}}_{[006]} )( \chi^{\mathfrak{a}_{3,6}}_{[600]} + \chi^{\mathfrak{a}_{3,6}}_{[006]} )( \chi^{\mathfrak{e}_{7,2}}_{[0000000]}\chi^{\text{tI}}_{\varepsilon''} + \chi^{\mathfrak{e}_{7,2}}_{[0000002]}\chi^{\text{tI}}_{I} )&&~.
\end{alignat}
The elliptic genus of the (0,4) SCFT is given by
{\allowdisplaybreaks
\begin{alignat}{10}
\Ell_{ \mathrm{A}_3\times\mathrm{A}_3\times\mathrm{E}_7 } = &&
&& 4\,\Chi_{ [212 , 212 , 0000000] }^{ \varepsilon } &&
&\,+\,& 4\,\Chi_{ [121 , 212 , 0100000] }^{ \sigma } &&
&\,+\,& 3\,\Chi_{ [000 , 000 , 0000000] }^{ I } &&
&\,+\,& 3\,\Chi_{ [113 , 113 , 0000000] }^{ I }  \nonumber\\ 
&&
&\,+\,& 3\,\Chi_{ [202 , 202 , 1000000] }^{ \varepsilon } &&
&\,+\,& 3\,\Chi_{ [111 , 111 , 1000000] }^{ \varepsilon' } &&
&\,+\,& 3\,\Chi_{ [000 , 006 , 0100000] }^{ \sigma' } &&
&\,+\,& 3\,\Chi_{ [111 , 113 , 0100000] }^{ \sigma' }  \nonumber\\ 
&&
&\,+\,& 3\,\Chi_{ [202 , 022 , 0000001] }^{ \sigma } &&
&\,+\,& 3\,\Chi_{ [111 , 113 , 0000001] }^{ \sigma } &&
&\,+\,& 2\,\Chi_{ [105 , 202 , 0000000] }^{ I } &&
&\,+\,& 2\,\Chi_{ [004 , 303 , 0000000] }^{ I }  \nonumber\\ 
&&
&\,+\,& 2\,\Chi_{ [400 , 303 , 0000000] }^{ I } &&
&\,+\,& 2\,\Chi_{ [121 , 012 , 1000000] }^{ \varepsilon } &&
&\,+\,& 2\,\Chi_{ [121 , 210 , 1000000] }^{ \varepsilon } &&
&\,+\,& 2\,\Chi_{ [012 , 121 , 1000000] }^{ \varepsilon }  \nonumber\\ 
&&
&\,+\,& 2\,\Chi_{ [210 , 121 , 1000000] }^{ \varepsilon } &&
&\,+\,& 2\,\Chi_{ [303 , 121 , 1000000] }^{ \varepsilon } &&
&\,+\,& 2\,\Chi_{ [121 , 303 , 0000000] }^{ I } &&
&\,+\,& 2\,\Chi_{ [303 , 004 , 0000000] }^{ I }  \nonumber\\ 
&&
&\,+\,& 2\,\Chi_{ [303 , 400 , 0000000] }^{ I } &&
&\,+\,& 2\,\Chi_{ [101 , 121 , 0000010] }^{ I } &&
&\,+\,& 2\,\Chi_{ [010 , 202 , 0100000] }^{ \sigma' } &&
&\,+\,& 2\,\Chi_{ [121 , 030 , 0100000] }^{ \sigma' }  \nonumber\\ 
&&
&\,+\,& 2\,\Chi_{ [303 , 002 , 0100000] }^{ \sigma' } &&
&\,+\,& 2\,\Chi_{ [303 , 200 , 0100000] }^{ \sigma' } &&
&\,+\,& 2\,\Chi_{ [121 , 103 , 0000001] }^{ \sigma } &&
&\,+\,& 2\,\Chi_{ [121 , 301 , 0000001] }^{ \sigma }  \nonumber\\ 
&&
&\,+\,& 2\,\Chi_{ [012 , 212 , 0000001] }^{ \sigma } &&
&\,+\,& 2\,\Chi_{ [210 , 212 , 0000001] }^{ \sigma } &&
&\,+\,& 2\,\Chi_{ [030 , 121 , 0000001] }^{ \sigma } &&
&\,+\,& 2\,\Chi_{ [002 , 303 , 0100000] }^{ \sigma' }  \nonumber\\ 
&&
&\,+\,& 2\,\Chi_{ [200 , 303 , 0100000] }^{ \sigma' } &&
&\,+\,& 2\,\Chi_{ [204 , 030 , 0000000] }^{ \varepsilon } &&
&\,+\,& 2\,\Chi_{ [020 , 030 , 0100000] }^{ \sigma } &&
&\,+\,& 2\,\Chi_{ [101 , 212 , 0000001] }^{ \sigma' }  \nonumber\\ 
&&
&\,+\,& 2\,\Chi_{ [303 , 121 , 0000000] }^{ I } &&
&\,+\,& 2\,\Chi_{ [121 , 303 , 1000000] }^{ \varepsilon } &&
&\,+\,& 2\,\Chi_{ [030 , 121 , 0100000] }^{ \sigma' } &&
&\,+\,& 2\,\Chi_{ [121 , 030 , 0000001] }^{ \sigma }  \nonumber\\ 
&&
&\,+\,& 2\,\Chi_{ [202 , 105 , 0000000] }^{ I } &&
&\,+\,& 2\,\Chi_{ [121 , 101 , 0000010] }^{ I } &&
&\,+\,& 2\,\Chi_{ [103 , 103 , 1000000] }^{ I } &&
&\,+\,& 2\,\Chi_{ [301 , 301 , 1000000] }^{ I }  \nonumber\\ 
&&
&\,+\,& 2\,\Chi_{ [202 , 010 , 0100000] }^{ \sigma' } &&
&\,+\,& 2\,\Chi_{ [030 , 204 , 0000000] }^{ \varepsilon } &&
&\,+\,& 2\,\Chi_{ [030 , 020 , 0100000] }^{ \sigma } &&
&\,+\,& 2\,\Chi_{ [212 , 101 , 0000001] }^{ \sigma' }  \nonumber\\ 
&&
&\,+\,& 2\,\Chi_{ [103 , 210 , 0000001] }^{ \sigma' } &&
&\,+\,& 2\,\Chi_{ [301 , 012 , 0000001] }^{ \sigma' } &&
&\,+\,& \Chi_{ [010 , 010 , 0000000] }^{ \varepsilon'' } &&
&\,+\,& \Chi_{ [010 , 010 , 1000000] }^{ \varepsilon' }  \nonumber\\ 
&&
&\,+\,& \Chi_{ [004 , 210 , 1000000] }^{ \varepsilon } &&
&\,+\,& \Chi_{ [400 , 012 , 1000000] }^{ \varepsilon } &&
&\,+\,& \Chi_{ [012 , 004 , 0000000] }^{ I } &&
&\,+\,& \Chi_{ [210 , 400 , 0000000] }^{ I }  \nonumber\\ 
&&
&\,+\,& \Chi_{ [014 , 002 , 1000000] }^{ I } &&
&\,+\,& \Chi_{ [410 , 200 , 1000000] }^{ I } &&
&\,+\,& \Chi_{ [020 , 000 , 0000010] }^{ I } &&
&\,+\,& \Chi_{ [020 , 202 , 0000010] }^{ I }  \nonumber\\ 
&&
&\,+\,& \Chi_{ [012 , 200 , 0100000] }^{ \sigma' } &&
&\,+\,& \Chi_{ [210 , 002 , 0100000] }^{ \sigma' } &&
&\,+\,& \Chi_{ [105 , 010 , 0000001] }^{ \sigma } &&
&\,+\,& \Chi_{ [105 , 010 , 0100000] }^{ \sigma' }  \nonumber\\ 
&&
&\,+\,& \Chi_{ [200 , 210 , 0000001] }^{ \sigma } &&
&\,+\,& \Chi_{ [002 , 012 , 0000001] }^{ \sigma } &&
&\,+\,& \Chi_{ [101 , 000 , 0000000] }^{ \varepsilon' } &&
&\,+\,& \Chi_{ [101 , 202 , 0000000] }^{ \varepsilon' }  \nonumber\\ 
&&
&\,+\,& \Chi_{ [020 , 012 , 0000000] }^{ \varepsilon' } &&
&\,+\,& \Chi_{ [020 , 210 , 0000000] }^{ \varepsilon' } &&
&\,+\,& \Chi_{ [410 , 000 , 0100000] }^{ \sigma } &&
&\,+\,& \Chi_{ [101 , 022 , 0100000] }^{ \sigma }  \nonumber\\ 
&&
&\,+\,& \Chi_{ [020 , 103 , 0100000] }^{ \sigma } &&
&\,+\,& \Chi_{ [020 , 301 , 0100000] }^{ \sigma } &&
&\,+\,& \Chi_{ [020 , 022 , 0000001] }^{ \sigma' } &&
&\,+\,& \Chi_{ [101 , 200 , 0000001] }^{ \sigma' }  \nonumber\\ 
&&
&\,+\,& \Chi_{ [101 , 002 , 0000001] }^{ \sigma' } &&
&\,+\,& \Chi_{ [204 , 000 , 0000001] }^{ \sigma' } &&
&\,+\,& \Chi_{ [000 , 000 , 1000000] }^{ \varepsilon } &&
&\,+\,& \Chi_{ [202 , 202 , 0000000] }^{ I }  \nonumber\\ 
&&
&\,+\,& \Chi_{ [000 , 006 , 0000001] }^{ \sigma } &&
&\,+\,& \Chi_{ [202 , 022 , 0100000] }^{ \sigma' } &&
&\,+\,& \Chi_{ [000 , 020 , 0000010] }^{ I } &&
&\,+\,& \Chi_{ [002 , 014 , 1000000] }^{ I }  \nonumber\\ 
&&
&\,+\,& \Chi_{ [200 , 014 , 1000000] }^{ I } &&
&\,+\,& \Chi_{ [202 , 101 , 0000000] }^{ \varepsilon' } &&
&\,+\,& \Chi_{ [012 , 020 , 0000000] }^{ \varepsilon' } &&
&\,+\,& \Chi_{ [210 , 020 , 0000000] }^{ \varepsilon' }  \nonumber\\ 
&&
&\,+\,& \Chi_{ [220 , 101 , 0100000] }^{ \sigma } &&
&\,+\,& \Chi_{ [301 , 020 , 0100000] }^{ \sigma } &&
&\,+\,& \Chi_{ [103 , 020 , 0100000] }^{ \sigma } &&
&\,+\,& \Chi_{ [000 , 204 , 0000001] }^{ \sigma' }  \nonumber\\ 
&&
&\,+\,& \Chi_{ [200 , 101 , 0000001] }^{ \sigma' } &&
&\,+\,& \Chi_{ [002 , 101 , 0000001] }^{ \sigma' } &&
&\,+\,& \Chi_{ [202 , 020 , 0000010] }^{ I } &&
&\,+\,& \Chi_{ [000 , 101 , 0000000] }^{ \varepsilon' }  \nonumber\\ 
&&
&\,+\,& \Chi_{ [000 , 014 , 0100000] }^{ \sigma } &&
&\,+\,& \Chi_{ [220 , 020 , 0000001] }^{ \sigma' } &&
&\,+\,& \Chi_{ [004 , 012 , 0000000] }^{ I } &&
&\,+\,& \Chi_{ [400 , 210 , 0000000] }^{ I }  \nonumber\\ 
&&
&\,+\,& \Chi_{ [012 , 400 , 1000000] }^{ \varepsilon } &&
&\,+\,& \Chi_{ [210 , 004 , 1000000] }^{ \varepsilon } &&
&\,+\,& \Chi_{ [002 , 210 , 0100000] }^{ \sigma' } &&
&\,+\,& \Chi_{ [200 , 012 , 0100000] }^{ \sigma' }  \nonumber\\ 
&&
&\,+\,& \Chi_{ [210 , 200 , 0000001] }^{ \sigma } &&
&\,+\,& \Chi_{ [012 , 002 , 0000001] }^{ \sigma } &&
&\,-\,& 2\,\Chi_{ [105 , 000 , 1000000] }^{ \varepsilon } &&
&\,-\,& 2\,\Chi_{ [010 , 000 , 0000001] }^{ \sigma }  \nonumber\\ 
&&
&\,-\,& 2\,\Chi_{ [000 , 105 , 1000000] }^{ \varepsilon } &&
&\,-\,& 2\,\Chi_{ [000 , 010 , 0000001] }^{ \sigma } &&
&\,-\,& 2\,\Chi_{ [200 , 200 , 0000000] }^{ \varepsilon } &&
&\,-\,& 2\,\Chi_{ [002 , 002 , 0000000] }^{ \varepsilon }  \nonumber\\ 
&&
&\,-\,& 2\,\Chi_{ [004 , 200 , 0100000] }^{ \sigma } &&
&\,-\,& 2\,\Chi_{ [400 , 002 , 0100000] }^{ \sigma } &&
&\,-\,& 4\,\Chi_{ [030 , 030 , 1000000] }^{ I } &&
&\,-\,& 4\,\Chi_{ [030 , 303 , 0000001] }^{ \sigma' } &&~.
\end{alignat}}
It is symmetric under the exchange of the two $\mathfrak{a}_3$ factors of the KM algebra. As discussed in section~\ref{subsec:results}, there is also a unique modular-covariant combination of KMV characters that is antisymmetric under this exchange. This quantity, denoted by $\psi_{ \mathrm{A}_3\times\mathrm{A}_3\times\mathrm{E}_7 }$, also admits a decomposition in extended characters:
\begin{alignat}{10}
\psi_{ \mathrm{A}_3\times\mathrm{A}_3\times\mathrm{E}_7 } = &&
&& \Chi_{ [004 , 012 , 0000000] }^{ I } &&
&\,+\,& \Chi_{ [400 , 210 , 0000000] }^{ I } &&
&\,+\,& \Chi_{ [012 , 400 , 1000000] }^{ \varepsilon } &&
&\,+\,& \Chi_{ [210 , 004 , 1000000] }^{ \varepsilon }  
\nonumber\\ 
&&
&\,+\,& \Chi_{ [002 , 210 , 0100000] }^{ \sigma' } &&
&\,+\,& \Chi_{ [200 , 012 , 0100000] }^{ \sigma' } &&
&\,+\,& \Chi_{ [210 , 200 , 0000001] }^{ \sigma } &&
&\,+\,& \Chi_{ [012 , 002 , 0000001] }^{ \sigma }  
\nonumber\\ 
&&
&\,+\,& \Chi_{ [004 , 012 , 1000000] }^{ \varepsilon } &&
&\,+\,& \Chi_{ [400 , 210 , 1000000] }^{ \varepsilon } &&
&\,+\,& \Chi_{ [012 , 400 , 0000000] }^{ I } &&
&\,+\,& \Chi_{ [210 , 004 , 0000000] }^{ I }  
\nonumber\\ 
&&
&\,+\,& \Chi_{ [210 , 200 , 0100000] }^{ \sigma' } &&
&\,+\,& \Chi_{ [012 , 002 , 0100000] }^{ \sigma' } &&
&\,+\,& \Chi_{ [200 , 012 , 0000001] }^{ \sigma } &&
&\,+\,& \Chi_{ [002 , 210 , 0000001] }^{ \sigma }  
\nonumber\\ 
&&
&\,-\,& \Chi_{ [004 , 210 , 1000000] }^{ \varepsilon } &&
&\,-\,& \Chi_{ [400 , 012 , 1000000] }^{ \varepsilon } &&
&\,-\,& \Chi_{ [012 , 004 , 0000000] }^{ I } &&
&\,-\,& \Chi_{ [210 , 400 , 0000000] }^{ I }  
\nonumber\\ 
&&
&\,-\,& \Chi_{ [210 , 002 , 0100000] }^{ \sigma' } &&
&\,-\,& \Chi_{ [012 , 200 , 0100000] }^{ \sigma' } &&
&\,-\,& \Chi_{ [200 , 210 , 0000001] }^{ \sigma } &&
&\,-\,& \Chi_{ [002 , 012 , 0000001] }^{ \sigma }  
\nonumber\\ 
&&
&\,-\,& \Chi_{ [004 , 210 , 0000000] }^{ I } &&
&\,-\,& \Chi_{ [400 , 012 , 0000000] }^{ I } &&
&\,-\,& \Chi_{ [012 , 004 , 1000000] }^{ \varepsilon } &&
&\,-\,& \Chi_{ [210 , 400 , 1000000] }^{ \varepsilon }  
\nonumber\\ 
&&
&\,-\,& \Chi_{ [210 , 002 , 0000001] }^{ \sigma } &&
&\,-\,& \Chi_{ [012 , 200 , 0000001] }^{ \sigma } &&
&\,-\,& \Chi_{ [200 , 210 , 0100000] }^{ \sigma' } &&
&\,-\,& \Chi_{ [002 , 012 , 0100000] }^{ \sigma' } &&~.
\end{alignat}

\subsection{\texorpdfstring{$G=\mathrm{A}_3\times\mathrm{A}_{11}$}{G=A3xA11}}
The left-moving KMV algebra of the $\mathrm{A}_3\times\mathrm{A}_{11}$ hyperplane string SCFT has central charges
\begin{align}
c_\tL^{\mathrm{gauge}} &= \frac{218}{7}~,&
c_\tL^{\mathrm{res}} &= \frac{6}{7}~.
\end{align}
The $\mathfrak{a}_3$ KM subalgebra is at level $10$, and the $\mathfrak{a}_{11}$ factor is at level $2$. The residual factor is associated to the fourth unitary Virasoro minimal model. The latter contains an integral spin simple current
\begin{align}
J_0 = \phi_{(5,1)}~,
\end{align}
of order two. The extension of the minimal model by this spin-$5$ current leads to the tricritical Potts model, which has nine primary fields
\begin{align}
\mathcal{X}^{\text{tP}}_{I} &= \chi_{(1,1)} + \chi_{(5,1)}~,&
\mathcal{X}^{\text{tP}}_{\sigma_1} = \mathcal{X}^{\text{tP}}_{\sigma_2} &= \chi_{(3,3)}~,
\nonumber\\
\mathcal{X}^{\text{tP}}_{\varepsilon} &= \chi_{(5,2)} + \chi_{(5,5)}~,&
\mathcal{X}^{\text{tP}}_{\psi_1} = \mathcal{X}^{\text{tP}}_{\psi_2} &= \chi_{(3,1)}~,
\nonumber\\
\mathcal{X}^{\text{tP}}_{\varepsilon'} &= \chi_{(5,3)} + \chi_{(5,4)}~,&
\mathcal{X}^{\text{tP}}_{Z_1} = \mathcal{X}^{\text{tP}}_{Z_2} &= \chi_{(3,2)}~.
\label{eq:tPcharacters}
\end{align}
Only the above combinations of characters appear in the elliptic genus, so we can identify the residual factor of the left-moving algebra with the tricritical Potts model. In other words, the SCFT is extended by the spin-$5$ current $J_0$, and its $c_\tL^{\mathrm{res}} = 6/7$ Virasoro subalgebra is enlarged to the $W_{2,5}$ algebra. 

The tricritical Potts model has a $\Z_3$ center generated by the current $J_{\text{tP}} = \phi_{\psi_1}$, with $(J_{\text{tP}})^2= \phi_{\psi_2}$. The decomposition of the elliptic genus in extended characters~\eqref{eq:tPcharacters} is ambiguous, due to the conformal dimension degeneracy of the primaries. This ambiguity is fixed by demanding that the elliptic genus is neutral under the $\Z_3$ symmetry that assigns charges $Q=4Q^{\mathfrak{a}_{11}}-Q^{\text{tP}}$ to the $c_\tL=32$ primaries. This is the unique possibility allowed by modular invariance.\footnote{The other choice of $\Z_3$ charge $Q=4Q^{\mathfrak{a}_{11}}+Q^{\text{tP}}$ is also valid, but amounts to a relabelling of the tricritical Potts primaries.}

The $c_\tL=32$ algebra has center $\Z_4\times\Z_{12}\times\Z_3$, generated by the currents 
\begin{align}
J_{\mathfrak{a}_3} &= \phi_{[1\hspace{-1pt}0\hspace{0.6pt}0\hspace{0.6pt}0,00000000000]}^{I}~,&
J_{\mathfrak{a}_{11}} &= \phi_{[000,20000000000]}^{I}~,&
J_{\text{tP}} &= \phi_{[000,00000000000]}^{\psi_1}~,
\end{align}
of conformal dimension $h_\tL= 15/4 $, $h_\tL= 11/12$ and $h_\tL= 4/3$ respectively.

The elliptic genus is neutral under the $\Z_4\times\Z_6$ subgroup generated by the spin-$6$ current $J_1$ and the spin-$3$ current $J_2$ defined as
\begin{align}
J_1 &= J_{\mathfrak{a}_3}\times (J_{\mathfrak{a}_{11}})^3~,&
J_2 &= (J_{\mathfrak{a}_{11}})^2\times J_{\text{tP}}  ~.
\end{align}
Note that $(J_1)^2(J_2)^3 = (J_{\mathfrak{a}_3})^2$ only acts on the $\mathfrak{a}_3$ index, while $(J_2)^3= (J_{\mathfrak{a}_{11}})^6$ only acts on the $\mathfrak{a}_{11}$ index.

The identity character of the $c_\tL=32$ left-moving chiral algebra takes the form
\begin{alignat}{4}
\Chi_{[000,00000000000]}^I = &&
&& \left( \chi^{\mathfrak{a}_{3,10}}_{[000]}+ \chi^{\mathfrak{a}_{3,10}}_{[0\hspace{+0.6pt}1\hspace{-1pt}0\hspace{+0.6pt}0]} \right)\left( \chi^{\mathfrak{a}_{11,2}}_{[00000000000]}+\chi^{\mathfrak{a}_{11,2}}_{[00000200000]} \right) \Chi^{\text{tP}}_I \nonumber\\
&& &\,+\,& \left( \chi^{\mathfrak{a}_{3,10}}_{[000]}+ \chi^{\mathfrak{a}_{3,10}}_{[0\hspace{+0.6pt}1\hspace{-1pt}0\hspace{+0.6pt}0]} \right)\left( \chi^{\mathfrak{a}_{11,2}}_{[02000000000]}+\chi^{\mathfrak{a}_{11,2}}_{[00000002000]} \right) \Chi^{\text{tP}}_{\psi_1} \nonumber\\
&& &\,+\,& \left( \chi^{\mathfrak{a}_{3,10}}_{[000]}+ \chi^{\mathfrak{a}_{3,10}}_{[0\hspace{+0.6pt}1\hspace{-1pt}0\hspace{+0.6pt}0]} \right)\left( \chi^{\mathfrak{a}_{11,2}}_{[00020000000]}+\chi^{\mathfrak{a}_{11,2}}_{[00000000020]} \right) \Chi^{\text{tP}}_{\psi_2} \nonumber\\
&& &\,+\,&  \left( \chi^{\mathfrak{a}_{3,10}}_{[1\hspace{-1pt}0\hspace{+0.6pt}0\hspace{+0.6pt}0]}+ \chi^{\mathfrak{a}_{3,10}}_{[0\hspace{+0.6pt}0\hspace{+0.6pt}1\hspace{-1pt}0]} \right)\left( \chi^{\mathfrak{a}_{11,2}}_{[00200000000]}+\chi^{\mathfrak{a}_{11,2}}_{[00000000200]} \right) \Chi^{\text{tP}}_I \nonumber\\
&& &\,+\,&  \left( \chi^{\mathfrak{a}_{3,10}}_{[1\hspace{-1pt}0\hspace{+0.6pt}0\hspace{+0.6pt}0]}+ \chi^{\mathfrak{a}_{3,10}}_{[0\hspace{+0.6pt}0\hspace{+0.6pt}1\hspace{-1pt}0]} \right)\left( \chi^{\mathfrak{a}_{11,2}}_{[00002000000]}+\chi^{\mathfrak{a}_{11,2}}_{[00000000002]} \right) \Chi^{\text{tP}}_{\psi_1} \nonumber\\
&& &\,+\,&  \left( \chi^{\mathfrak{a}_{3,10}}_{[1\hspace{-1pt}0\hspace{+0.6pt}0\hspace{+0.6pt}0]}+ \chi^{\mathfrak{a}_{3,10}}_{[0\hspace{+0.6pt}0\hspace{+0.6pt}1\hspace{-1pt}0]} \right)\left( \chi^{\mathfrak{a}_{11,2}}_{[20000000000]}+\chi^{\mathfrak{a}_{11,2}}_{[00000020000]} \right) \Chi^{\text{tP}}_{\psi_2} &&~.
\end{alignat}
The elliptic genus of the (0,4) SCFT is given by
{\allowdisplaybreaks
\begin{alignat}{10}
\Ell_{\mathrm{A}_3\times\mathrm{A}_{11} } = &&
&& 4\,\Chi_{ [232 , 00000100000] }^{ \varepsilon'  } &&
&\,+\,& 4\,\Chi_{ [414 , 00000100000] }^{ \varepsilon } &&
&\,+\,& 3\,\Chi_{ [000 , 00000000000] }^{ I } &&
&\,+\,& 3\,\Chi_{ [303 , 01000000010] }^{ \varepsilon } \nonumber\\
&&
&\,+\,& 3\,\Chi_{ [222 , 10000000001] }^{ \varepsilon'  } &&
&\,+\,& 2\,\Chi_{ [307 , 00000000000] }^{ I } &&
&\,+\,& 2\,\Chi_{ [212 , 00000010001] }^{ \varepsilon } &&
&\,+\,& 2\,\Chi_{ [206 , 10000000001] }^{ I } \nonumber\\
&&
&\,+\,& 2\,\Chi_{ [602 , 10000000001] }^{ I } &&
&\,+\,& 2\,\Chi_{ [040 , 00100000100] }^{ \varepsilon'  } &&
&\,+\,& 2\,\Chi_{ [123 , 00000010001] }^{ \varepsilon'  } &&
&\,+\,& 2\,\Chi_{ [321 , 00000010001] }^{ \varepsilon'  } \nonumber\\
&&
&\,+\,& 2\,\Chi_{ [202 , 00100000100] }^{ \varepsilon } &&
&\,+\,& 2\,\Chi_{ [121 , 00100000100] }^{ I } &&
&\,+\,& 2\,\Chi_{ [323 , 10000000001] }^{ I } &&
&\,+\,& 2\,\Chi_{ [220 , 00000100000] }^{ \varepsilon'  } \nonumber\\
&&
&\,+\,& 2\,\Chi_{ [022 , 00000100000] }^{ \varepsilon'  } &&
&\,+\,& 2\,\Chi_{ [141 , 01000000010] }^{ \varepsilon'  } &&
&\,+\,& 2\,\Chi_{ [123 , 00000100000] }^{ \varepsilon } &&
&\,+\,& 2\,\Chi_{ [321 , 00000100000] }^{ \varepsilon } \nonumber\\
&&
&\,+\,& 2\,\Chi_{ [006 , 00000100000] }^{ I } &&
&\,+\,& 2\,\Chi_{ [600 , 00000100000] }^{ I } &&
&\,+\,& 2\,\Chi_{ [505 , 00000000000] }^{ \varepsilon } &&
&\,+\,& \Chi_{ [016 , 10000000001] }^{ \varepsilon } \nonumber\\
&&
&\,+\,& \Chi_{ [610 , 10000000001] }^{ \varepsilon } &&
&\,+\,& \Chi_{ [404 , 10000000001] }^{ \varepsilon } &&
&\,+\,& \Chi_{ [101 , 00000000000] }^{ \varepsilon'  } &&
&\,+\,& \Chi_{ [230 , 00000000000] }^{ \varepsilon'  } \nonumber\\
&&
&\,+\,& \Chi_{ [032 , 00000000000] }^{ \varepsilon'  } &&
&\,+\,& \Chi_{ [014 , 00000010001] }^{ I } &&
&\,+\,& \Chi_{ [410 , 00000010001] }^{ I } &&
&\,+\,& \Chi_{ [020 , 01000000010] }^{ I } \nonumber\\
&&
&\,+\,& \Chi_{ [016 , 00000000000] }^{ I } &&
&\,+\,& \Chi_{ [610 , 00000000000] }^{ I } &&
&\,+\,& \Chi_{ [404 , 00000000000] }^{ I } &&
&\,+\,& \Chi_{ [000 , 10000000001] }^{ \varepsilon } \nonumber\\
&&
&\,+\,& \Chi_{ [303 , 00000000000] }^{ \varepsilon'  } &&
&\,+\,& \Chi_{ [101 , 01000000010] }^{ \varepsilon } &&
&\,+\,& \Chi_{ [230 , 01000000010] }^{ \varepsilon } &&
&\,+\,& \Chi_{ [032 , 01000000010] }^{ \varepsilon } \nonumber\\
&&
&\,+\,& \Chi_{ [222 , 01000000010] }^{ I } &&
&\,+\,& \Chi_{ [501 , 10000000001] }^{ \varepsilon'  } &&
&\,+\,& \Chi_{ [105 , 10000000001] }^{ \varepsilon'  } &&
&\,+\,& \Chi_{ [020 , 10000000001] }^{ \varepsilon'  } \nonumber\\
&&
&\,-\,& 2\,\Chi_{ [010 , 00000001010] }^{ \varepsilon'  } &&
&\,-\,& 2\,\Chi_{ [400 , 00000000000] }^{ \varepsilon } &&
&\,-\,& 2\,\Chi_{ [004 , 00000000000] }^{ \varepsilon } &&
&\,-\,& 4\,\Chi_{ [050 , 00000100000] }^{ I }
&&~.
\end{alignat}}

\subsection{Comments on fixed point resolutions}
\label{sec:fp}
In all models, we find strong evidence that the left-moving sector of the (0,4) theory is extended by higher-spin KMV currents. For our purposes, the extended characters introduced above simply served as a convenient way of repackaging the KMV characters associated with simple current orbits into a more compact form. However, they carry a deeper significance: they should be interpreted as characters of primary representations of the extended chiral algebra, encoding descendants obtained by acting not only with the KMV modes but also with the modes of the simple currents.

To admit such an interpretation, some care is required when the original KMV algebra contains fixed points, that is, primary representations left invariant under the action of a simple current. In the extended theory, such fixed points typically split into several distinct extended primary representations. By construction, these primaries belong to the same KMV representation and can only be distinguished by additional discrete symmetries of the extended algebra. Disentangling these representations notably requires determining the modular matrix of the extended theory, a procedure known as \emph{fixed point resolution}.

Let us illustrate this point with some examples. The theory with $\mathfrak{e}_6$ algebra introduced above is extended by a spin-$6$ simple current $J=J_{\mathfrak{e}_6}\times J_{\text{P}}$, which combines the center of $\mathrm{E}_6$ with the $\Z_3$ symmetry of the Potts model. This simple current acts as in~\eqref{eq:fusione6Potts}, so it has no fixed points among the KMV primaries. As a result, the modular properties of the extended theory can be obtained straightforwardly from those of the original KMV algebra. One finds that the extended theory contains $248$ primary representations, and that the resulting modular matrix is consistent, in particular yielding well-defined fusion rules.

Conversely, consider the model with $\mathfrak{a}_7$ current algebra. In this case, the left-moving algebra is extended by a spin-$4$ simple current $J=J_{\mathfrak{a}_7}J_{\text{I}}$ of order $8$, which ties the center symmetry of $\mathrm{SU}(8)$ to the $\Z_2$ symmetry of the Ising model. There are $2429$ KMV primaries with vanishing monodromy charge $Q=Q^{\mathfrak{a}_7}+Q^{\text{I}}$ with respect to this mixed current. Under the action of $J$, these primaries organize into $309$ distinct orbits. In contrast to the $\mathfrak{e}_6$ case, the simple current $J$ admits fixed points: some of the orbits have length smaller than $8$. More explicitly, there is one fixed point of length $1$:
\begin{align}
\Chi_{[1111111]}^\sigma = \chi_{[1111111]}^\sigma~,
\end{align}
two fixed points of length $2$:
\begin{align}
\Chi_{[0202020]}^I &= \chi_{[0202020]}^I + \chi_{[2020202]}^\varepsilon~,&
\Chi_{[2020202]}^I &= \chi_{[2020202]}^I + \chi_{[0202020]}^\varepsilon~,
\end{align}
and six fixed points of length $4$:
\begin{align}
\Chi_{[0004000]}^I &= \chi_{[0004000]}^I + \chi_{[0040004]}^\varepsilon + \chi_{[0400040]}^I + \chi_{[4000400]}^\varepsilon~,
\nonumber\\
\Chi_{[0040004]}^I &= \chi_{[0040004]}^I + \chi_{[0400040]}^\varepsilon + \chi_{[4000400]}^I + \chi_{[0004000]}^\varepsilon~,
\nonumber\\
\Chi_{[0121012]}^I &= \chi_{[0121012]}^I + \chi_{[1210121]}^\varepsilon + \chi_{[2101210]}^I + \chi_{[1012101]}^\varepsilon~,
\nonumber\\
\Chi_{[1210121]}^I &= \chi_{[1210121]}^I + \chi_{[2101210]}^\varepsilon + \chi_{[1012101]}^I + \chi_{[0121012]}^\varepsilon~,
\nonumber\\
\Chi_{[0022002]}^\sigma &= \chi_{[0022002]}^\sigma + \chi_{[0220022]}^\sigma + \chi_{[2200220]}^\sigma + \chi_{[2002200]}^\sigma~,
\nonumber\\
\Chi_{[0103010]}^\sigma &= \chi_{[0103010]}^\sigma + \chi_{[1030103]}^\sigma + \chi_{[0301030]}^\sigma + \chi_{[3010301]}^\sigma~.
\end{align}
To fully determine the structure of the extended chiral algebra, one should work out the resolution of these $\Z_8$ fixed points.

In general, the resolution of simple-current fixed points is a notoriously difficult problem. A systematic treatment exists only for a limited class of theories, notably (generalized) diagonal coset models~\cite{Fuchs:1995tq} and Kac--Moody algebras~\cite{Fuchs:1995zr,Fuchs:1995hc,Fuchs:1996dd,Schellekens:1999yg}. The procedure involves determining, for each simple current $J$, a fixed point resolution matrix $\mathcal{S}^J$ satisfying a set of nontrivial consistency conditions, from which the modular data of the extended theory can be reconstructed~\cite{Fuchs:1996dd}. The structure of these matrices is often intricate; for instance, in the case of Kac--Moody algebras they are expressed in terms of orbit Lie algebras obtained by folding the extended Dynkin diagram~\cite{Fuchs:1995zr}. Beyond these special cases, however, no general construction is known. In particular, existing methods do not apply to the mixed KMV simple current extensions encountered in this work.

\subsubsection*{An example with mixed KMV extension}
Let us consider a simple example illustrating the preceding discussion. The model with gauge symmetry $\mathfrak{a}_3+\mathfrak{a}_3+\mathfrak{e}_7$ contains in its left-moving chiral algebra a $c_\tL=14$ subfactor, built from the $\mathfrak{e}_7$ current algebra and the tricritical Ising model. Both components possess a $\Z_2$ center symmetry, associated respectively with the center of $\mathrm{E}_7$ and with the ``fermion parity'' symmetry of the minimal model.\footnote{Recall that the Virasoro algebra of the $c=7/10$ minimal model admits a superconformal extension, in which the spin-$\frac{3}{2}$ primary representation $\varepsilon''$ is identified with the supercurrent $G(z)$.} The corresponding generators, denoted by $J_{\mathfrak{e}_7}$ and $J_{\text{tI}}$, both have conformal dimension $h_\tL=\frac{3}{2}$. Their product
\begin{align}
J = J_{\mathfrak{e}_7} J_{\text{tI}}
\label{eq:mixedcurrente7tI}
\end{align}
is a spin-$3$ simple current. As observed in subsection~\ref{subsec:modelA3xA3xE7}, this higher-spin holomorphic operator appears in the spectrum of the probe string SCFT. We outline here the main features of the extended algebra; further details and notation can be found in appendix~\ref{app:mixedKMV}.

The $c_\tL=14$ KMV algebra contains $36$ primary representations, given by the tensor products of the six integrable representations of $\mathfrak{e}_7$ at level $2$ with the six Virasoro primaries of the tricritical Ising model. Among these, $20$ primaries are local with respect to the higher-spin current $J$. Under fusion with $J$, they organize into eight orbits of length $2$, and four fixed points. The latter correspond to the KMV primary representations
\begin{align}
&\rep{56}\otimes\sigma~,&
&\rep{56}\otimes\sigma'~,&
&\rep{912}\otimes\sigma~,&
&\rep{912}\otimes\sigma'~.
\end{align}
Upon resolving these $\Z_2$ fixed points, each splits into two distinct irreducible representations of the extended chiral algebra.

As a result, after extension by the spin-$3$ current~\eqref{eq:mixedcurrente7tI}, the number of primary representations of the $c_\tL=14$ algebra is reduced to $16$. For completeness, we have explicitly determined the modular properties of the fixed point resolutions; the details are presented in appendix~\ref{app:mixedKMV}. From this data, one can readily extract the fusion rules of the extended theory and, in particular, identify its simple currents. As expected, the extended primary
\begin{align}
\rep{1}\otimes\varepsilon''+\rep{1463}\otimes I~,
\end{align}
defines a simple current, inherited from the KMV simple currents $J_{\mathfrak{e}_7}$ and $J_{\text{tI}}$ of the unextended theory. More unexpectedly, the two spin-$\frac{7}{4}$ extended primaries contained in the KMV representation $\rep{912}\otimes\sigma'$ also generate simple currents of the extended algebra. Consequently, the extended chiral algebra has a $\Z_2\times\Z_2$ center, which is not fully captured by the structure of the unextended KMV algebra.

As this example illustrates, the resolution of simple current fixed points can reveal additional structure in the extended theory. In particular, performing the resolution may uncover new simple currents that are not visible in the unextended theory. For the five models at hand, our focus was restricted on KMV currents. It is possible that further simple currents, arising from the resolution of KMV fixed points, are missed in our study. A systematic characterization of the fixed point resolution matrices for the mixed KMV extensions studied in this work would be certainly desirable. Apart from its intrinsic interest in the study of rational conformal field theories, such an analysis would provide further insight into the discrete symmetries of the SCFTs considered in this work. 

\section{Discussion}
\label{sec:discussion}
In this work we have studied six-dimensional $T=0$ supergravity theories with eight supercharges through the (0,4) SCFT of BPS strings. We focused on five models for which the worldsheet theory is rational, characterized by a left-moving $c_\tL=32$ KMV algebra with a finite set of unitary representations. Exploiting the constraints imposed by rationality, we have determined the complete BPS spectrum of the SCFT in all five cases. In each model, we found compelling evidence for a simple current extension of the chiral algebra by higher-spin KMV currents.  Our analysis lends support to the quantum consistency of the five theories examined here. The refined elliptic genus of the (0,4) SCFT, which encodes its BPS spectrum, is subject to extremely restrictive modular constraints. It is therefore a nontrivial consistency check that, in each case, one can assemble a modular object with the correct transformation properties from KMV characters, which furthermore satisfies all expected physical requirements, including a match to the spacetime massless spectrum.

Although our analysis focused on a rather small set of examples---identified in the classification of~\cite{Hamada:2023zol,Hamada:2024oap} as those with largest Sugawara contribution to the left-moving central charge---both our methods and conclusions are expected to extend to a much broader class of models. Beyond modularity, the only essential input in our approach is rationality, and it is natural to expect many additional $T=0$ supergravity theories whose associated BPS strings are described by a rational SCFT. We believe that rationality of the worldsheet theory is implied by the rigidity of the spacetime model, that is, by the absence of neutral hypermultiplets and, consequently, of marginal scalar deformations. Extending our analysis to this broader class of rigid theories would be very interesting. The principal obstacle is the identification of the residual chiral algebra, that is, the rational CFT that completes the current algebra to the full $c_\tL=32$ left-moving theory. In the examples studied here, the residual central charge falls in the discrete $c<1$ range, effectively fixing a unique choice. For residual theories with $c>1$, however, additional input is required to pinpoint the appropriate rational CFT. A deeper understanding of the BPS string dynamics or of its moduli space may provide precisely such guidance.

A striking outcome of our analysis is the systematic appearance of simple currents in the probe BPS string SCFT. While we have emphasized their worldsheet constraints in the present work, their spacetime interpretation will be explored in detail elsewhere~\cite{Lockhart:2026gvb}. For the theories studied here, the upshot is that the presence of KM simple currents in the BPS string spectrum signals the non-simply-connectedness of the continuous gauge group $G$ in the six-dimensional spacetime theory. Concretely, the KM center of the CFT corresponds to one-form symmetries associated with the center of the simply-connected cover of $G$, and the worldsheet algebra is extended by a simple current precisely when the corresponding one-form symmetry is gauged---that is, for each element of $\pi_1(G)$. This correspondence extends well beyond the present setting: for example, analogous structures appear in higher-dimensional supergravity theories, such as asymmetric orbifold compactifications of the heterotic string. More broadly, these observations shed new light on the swampland no–global-symmetries conjecture; further details will be presented in~\cite{Lockhart:2026gvb}.

Applying these observations to the models studied in this work, we can use the simple current extensions analysis of section~\ref{sec:ellipticgenera} to unambiguously determine the global form of the six-dimensional gauge group $G$. In all but one case, it is non–simply connected, and $\pi_1(G)$ is realized through a simple current extension of the KM algebra. Accordingly, the list of theories given in~\eqref{eq:6dsupergravities} can be refined to
\begin{align}
G &= \mathrm{E}_6~, &
\mathscr{H} &= \rep{351}'~,
\nonumber\\
G &= \mathrm{SU}(8)/\Z_4~, &
\mathscr{H} &= \rep{336}~,
\nonumber\\
G &= \mathrm{SU}(6)/\Z_3 \times \mathrm{SO}(8)/\Z_2~, &
\mathscr{H} &=  (\rep{56},\rep{1})+(\tfrac{1}{2}\rep{20},\rep{28})~,
\nonumber\\
G &= \left(\mathrm{SU}(4)/\Z_2 \times \mathrm{SU}(4)/\Z_2 \times \mathrm{E}_7\right)/\Z_2~, &
\mathscr{H} &=  (\rep{10},\rep{10},\rep{1})+(\rep{1},\rep{6},\tfrac{1}{2}\rep{56})+(\rep{6},\rep{1},\tfrac{1}{2}\rep{56})~,
\nonumber\\
G &= \left(\mathrm{SU}(4)/\Z_2 \times \mathrm{SU}(12)/\Z_2\right)/\Z_2~, &
\mathscr{H} &=  (\rep{35},\rep{1})+(\rep{6},\rep{66})~,
\end{align}
where ``$G=\dots$'' is now understood as an equality of gauge groups, rather than merely of gauge algebras.

Perhaps the most immediate extension of our analysis is to determine the full spectrum of the worldsheet (0,4) SCFT. This requires identifying the appropriate right-moving characters and combining them with the left-moving KMV characters into a partition function with the correct $\mathrm{SL}(2,\Z)$ properties. While this problem is not straightforward, supersymmetry may provide useful guidance. The right-moving sector, with central charge $c_\tR=12$, realizes the small $ \wsN$=4 superconformal algebra, which contains an $\mathfrak{su}(2)$ current algebra at level $2$. This structure could be exploited to constrain the modular properties of the right-moving characters. Moreover, due to rationality of the SCFT, a bootstrap approach may be feasible. One possible avenue would be to construct the right-moving sector from simpler building blocks, such as $\wsN$=2 minimal models, in the spirit of the Gepner construction~\cite{Gepner:1987vz}.

Although originally developed for (2,2) theories, several features of the Gepner construction suggest that it may be relevant to the present (0,4) setting. Simple current extensions play a central role in Gepner models: they implement the projection onto states with integral R-charge and enforce the GSO projection that ensures the existence of a spectral flow operator. Various (0,2) generalizations of the construction have been studied~\cite{Schellekens:1989wx,Blumenhagen:1995tt,Gato-Rivera:2010flr,Israel:2013wwa,Israel:2015efa,Blumenhagen:2016rof}. Further motivation comes from considerations of hyper-Kähler geometry. A large class of (4,4) SCFTs arises as supersymmetric non-linear sigma models on hyper-Kähler manifolds. The prototypical example is the $c_\tL=c_\tR=6$ sigma model on a K3 surface, which admits a Gepner model description at special points in its moduli space~\cite{Eguchi:1988vra,Nahm:1999ps,Nahm:2001kh}. At these solvable points, the theory is rational and decomposes into a finite set of $\wsN$=2 characters. A formulation directly in terms of $\wsN$=4 characters is considerably more subtle, and underlies the discovery of Mathieu moonshine~\cite{Eguchi:2010ej}. One could similarly realize the $c_\tR=12$ $\wsN$=4 superconformal algebra by considering a sigma model on a hyper-Kähler fourfold at a suitable rational point. This possibility is particularly encouraging, as the (0,4) hyperplane string SCFT is expected, in certain large-volume regions of its moduli space, to be well-approximated by a sigma model on a hyper-Kähler fourfold equipped with a stable bundle~\cite{Lockhart:2025lea}.

One may more generally draw on techniques from heterotic sigma models. These provide a natural framework for (0,4) theories as large-radius sigma models on hypercomplex manifolds~\cite{Howe:1984fak,Gates:1984nk,Hull:1986kz,Strominger:1986uh,Papadopoulos:2024uvi}. A canonical exemple is the heterotic sigma model on a K3 surface. The internal SCFT, with  $c_\tL=20$ and $c_\tR=6$, contains sixteen left-moving fermions associated to a holomorphic vector bundle over K3, with structure group inside $\mathrm{E}_8\times\mathrm{E}_8$. While standard embedding of the spin connection enhances the worldsheet supersymmetry to (4,4), theories with only (0,4) supersymmetry require stable holomorphic bundles and their associated Hermitian--Yang--Mills connections. These constructions have been extensively studied, yielding a large class of models relevant for string phenomenology and often exhibiting rich connections with algebraic geometry and non-Kähler spaces~\cite{Anderson:2007nc,Anderson:2008uw,Anderson:2013sia,Anderson:2014xha,Anderson:2019agu,GrootNibbelink:2015dvi,GrootNibbelink:2016lre,Honecker:2006qz,Louis:2011hp,Melnikov:2010pq,Melnikov:2014ywa,Israel:2023tjw,Becker:2009df,delaOssa:2014cia,Candelas:2018lib}. We emphasize that these methods not only admit a broader range of applications in the study of non-critical BPS strings, but appear to be precisely the tools required to construct sigma model realizations of class of SCFTs investigated in this work.

Another possible direction is to search for an explicit field theory realization of the hyperplane string SCFT. Although such a realization should exist, constructing it appears challenging, largely due to the difficulty of realizing higher-level current algebras. Still, a number of examples are known in the heterotic context~\cite{Lewellen:1989qe,Font:1990uw,Dienes:1996yh}. Among these, the fibered Wess--Zumino--Witten models of~\cite{Distler:2007av} may provide a well-suited framework for the class of theories studied here. One may also exploit the fact that several of the KM factors appearing in our models admit conformal embeddings into larger level $1$ current algebras, namely
\begin{align}
\mathfrak{a}_{3,6} &\subset\mathfrak{a}_{9,1} ~,&
\mathfrak{a}_{5,6} &\subset\mathfrak{c}_{10,1} ~,&
\mathfrak{a}_{7,8} &\subset\mathfrak{b}_{31,1} ~.
\end{align}
In addition, it would be desirable to obtain a more natural description of the residual factor. Indeed, minimal models often admit multiple, seemingly distinct realizations---for example as diagonal cosets---and a particular presentation may better reflect the underlying physics. Finally, it may be fruitful to investigate the discrete symmetries of the $c_\tL=32$ chiral algebra further. In this regard, it may be informative that one of our models, with current algebra $\mathfrak{e}_6$, bears similarities to one of the monstralizer pairs discussed in~\cite{Bae:2020pvv}. While it is tempting to look for
analogous moonshine phenomena in the present context, such questions lie beyond the scope of this work.

A more ambitious direction would be to study the moduli space of deformations of the superconformal field theory describing the hyperplane string. The expectation is that this BPS string is stable, and that probe strings associated with distinct six-dimensional supergravities can be connected through marginal deformations~\cite{Lockhart:2025lea}, and in particular one can interpolate between different six-dimensional gauge groups related by the Higgsing mechanism. For instance, a sequence of Higgs deformations of six-dimensional $(1,0)$ supergravity theories was identified in~\cite{Lockhart:2025lea} and is depicted below. On the worldsheet, the associated current algebra is realized uniformly at level $8$, leading to the indicated Sugawara central charges.
\begin{center}
$\begin{array}{rccccccccccccccc}
\mathfrak{g}= &\quad\mathfrak{e}_6\quad & \supset & \quad\mathfrak{f}_4\quad & \supset & \quad\mathfrak{b}_4\quad & \supset & \quad\mathfrak{d}_4\quad & \supset & \quad\mathfrak{b}_3\quad & \supset & \quad\mathfrak{g}_2\quad & \supset & \quad\mathfrak{a}_2\quad & \supset & \quad\mathfrak{a}_1\quad
\nonumber\\
c_\tL^{\text{gauge}}= & \tfrac{156}{5} & > & \tfrac{416}{17} & > & \tfrac{96}{5} & > & 16 & > & \tfrac{168}{13} & > & \tfrac{28}{3} & > & \tfrac{64}{11} & > & \tfrac{12}{5} 
\end{array}$
\end{center}
Only a subset of these anomaly-free supergravities admits an F-theoretic realization, and it would be particularly interesting to understand Higgsing chains that interpolate between them and seemingly non-geometric theories---such as the $\mathrm{E}_6$ theory encountered in this work. More broadly, we hope that this line of investigation can shed light on the global structure of the moduli space of six-dimensional supergravity theories without tensor multiplets.

\section*{Acknowledgements}
We are grateful to T. Gannon, D. Israël, I. Melnikov, and K. Xu for useful discussions, and especially to L. Novelli for collaboration on related projects. The work of GL and YP has received funding from the European Research Council (ERC) under the Horizon Europe (grant agreement No. 101078365) research and innovation program. 

\appendix

\section{Minimal models}
\label{app:minimalmodels}

In this appendix, we provide some details on Virasoro minimal models.
\subsection{Generalities}
Unitary minimal models are classified by their central charge
\begin{align}
c = 1-\frac{6}{p(p-1)}~,
\label{eq:minimalcentralcharge}
\end{align}
where $p\geq 4$ is an integer. The Virasoro primaries $\phi_{(r,s)}$ are indexed by two integers valued in
\begin{align}
1 \leq r &\leq p-2~,&
1 \leq s &\leq p-1~,
\end{align}
they are subject to the field identification $\phi_{(r,s)}=\phi_{(p-1-r,p-s)}$. This redundancy can be eliminated by requiring $(p-1)s < pr$, which reduces the set of indices $(r,s)$ to 
\begin{align}
1 \leq s \leq r \leq p-2~.
\end{align}
The conformal dimension of the primary $\phi_{(r,s)}$ is given by the Kac formula:
\begin{align}
h_{(r,s)} = \frac{\left(pr-(p-1)s\right)^2-1}{4p(p-1)}~.
\label{eq:Kacformula}
\end{align}
The primary $J_0 = \phi_{(p-2,1)}$ 
is a simple current of conformal weight $h_{(p-2,1)} = \tfrac{1}{4}(p-2)(p-3)$. It generates a $\Z_2$ center: its fusion with other Virasoro primaries takes the form
\begin{align}
J_0\times \phi_{(r,s)} = \phi_{(r,p-s)}~.
\end{align}
The monodromy charge of $\phi_{(r,s)}$ is given by
\begin{align}
Q_{(r,s)} = \tfrac{1}{2}\left(p r+(p-1)s+1 \right) \text{ mod }1~.
\end{align}

\subsubsection*{Modular properties}
The Virasoro characters $\chi_{(r,s)}(\tau)$ of the $\mathfrak{M}_{(p,p-1)}$ minimal model transform under $\mathrm{SL}(2,\Z)$ with the modular matrices
\begin{align}
\chi_{(r,s)}(\tau+1) &=
\sum_{r'=1}^{p-2}
\sum_{s'=1}^{r'}
\mathcal{T}_{(r,s)(r',s')} \chi_{(r',s')}(\tau)~,\nonumber\\
\chi_{(r,s)}(-1/\tau) &=
\sum_{r'=1}^{p-2}
\sum_{s'=1}^{r'}
\mathcal{S}_{(r,s)(r',s')} \chi_{(r',s')}(\tau)~.
\end{align}
The modular $\mathcal{T}$ matrix is diagonal:
\begin{align}
\mathcal{T}_{(r,s)(r',s')} = \delta_{rr'}\delta_{ss'} \,\ee^{2\pi \ii(h_{(r,s)}-c/24)}~, 
\end{align}
where $c$ and $h_{(r,s)}$ are given by~\eqref{eq:minimalcentralcharge} and~\eqref{eq:Kacformula}. The modular $\mathcal{S}$ matrix takes the form
\begin{align}
\mathcal{S}_{(r,s)(r',s')} = 2\sqrt{\frac{2}{p(p-1)}}(-1)^{1+ rs'+r's }\sin\left(\pi rr' p/(p-1)\right)\sin\left(\pi ss'(p-1)/p\right)~.
\label{eq:minimalSmatrix}
\end{align}

\subsection{Potts and tricritical Potts models}

\subsubsection*{Potts model}
The Potts model $\mathfrak{M}_{\text{P}}$ has central charge $c_{\text{P}}=4/5$. It is constructed as an extension of the minimal model $\mathfrak{M}_{(6,5)}$ by the by spin-$3$ current $\phi_{(4,1)}$. There are six primaries of the extended algebra, with characters
\begin{align}
\mathcal{X}^{\text{P}}_I &= \chi_{(1,1)}+\chi_{(4,1)}~,&
\mathcal{X}^{\text{P}}_{\sigma_1} = \mathcal{X}^{\text{P}}_{\sigma_2} &= \chi_{(3,3)}~,
\nonumber\\
\mathcal{X}^{\text{P}}_\varepsilon &= \chi_{(2,1)}+\chi_{(3,1)}~,&
\mathcal{X}^{\text{P}}_{\psi_1} = \mathcal{X}^{\text{P}}_{\psi_2} &= \chi_{(4,3)}~.
\end{align}
The modular matrix $\mathcal{S}_{\text{P}}$ takes the form 
\begin{align}
\mathcal{S}_{\text{P}} = 
\frac{1}{
\sqrt{15}}
\begin{pmatrix}
\text{s}_1 & \text{s}_2 & \text{s}_2 & \text{s}_2 & \text{s}_1 & \text{s}_1 \\
\text{s}_2 & -\zeta \text{s}_1 & -\bar{\zeta} \text{s}_1 & - \text{s}_1 & \zeta \text{s}_2 & \bar{\zeta} \text{s}_2\\
\text{s}_2 & -\bar{\zeta} \text{s}_1 & -\zeta \text{s}_1 & - \text{s}_1 & \bar{\zeta} \text{s}_2 & \zeta \text{s}_2\\
\text{s}_2 & -\text{s}_1 & -\text{s}_1 & -\text{s}_1 & \text{s}_2 & \text{s}_2 \\
\text{s}_1 & \zeta\text{s}_2 & \bar{\zeta}\text{s}_2 & \text{s}_2 & \zeta\text{s}_1 & \bar{\zeta}\text{s}_1\\
\text{s}_1 & \bar{\zeta}\text{s}_2 & \zeta\text{s}_2 & \text{s}_2 & \bar{\zeta}\text{s}_1 & \zeta\text{s}_1\\
\end{pmatrix}~,
\label{eq:SmatrixPotts}
\end{align}
where we use the shorthand $\zeta=\ee^{\frac{2\pi\ii}{3}}$ and $\text{s}_l=\sin(\pi l/5)$. The primaries $\{I,\sigma_1,\sigma_2,\varepsilon ,\psi_1,\psi_2\}$ are ordered by conformal dimension.

The fusion algebra can be derived from $\mathcal{S}_{\text{P}}$ using the Verlinde formula. The Potts model has a $\Z_3$ center generated by the current
\begin{align}
J_{\text{P}} = \phi_{\psi_1}~,
\end{align}
with conformal dimension $h=2/3$. This current satisfies $( J_{\text{P}})^2 = \phi_{\psi_2}$ and
\begin{align}
J_{\text{P}}\times\phi_{\sigma_1} &= \phi_{\sigma_2}~,&
J_{\text{P}}\times\phi_{\sigma_2} &= \phi_{\varepsilon}~,&
J_{\text{P}}\times\phi_{\varepsilon} &= \phi_{\sigma_1}~.
\end{align}
The monodromy charges of the extended primaries with respect to $J_{\text{P}}$ are given by
\begin{align}
Q^{\text{P}}_I=Q^{\text{P}}_\varepsilon &=0~,&
Q^{\text{P}}_{\psi_1} = Q^{\text{P}}_{\sigma_1} &=\tfrac{2}{3}~,&
Q^{\text{P}}_{\psi_2} = Q^{\text{P}}_{\sigma_2} &=\tfrac{1}{3}~.
\end{align}

\subsubsection*{Tricritical Potts model}
The Potts model $\mathfrak{M}_{\text{tP}}$ has central charge $c_{\text{tP}}=6/7$. It is constructed as an extension of the minimal model $\mathfrak{M}_{(7,6)}$ by the by spin-$5$ current $\phi_{(5,1)}$. There are nine primaries of the extended algebra, with characters
\begin{align}
\mathcal{X}^{\text{tP}}_{I} &= \chi_{(1,1)} + \chi_{(5,1)}~,&
\mathcal{X}^{\text{tP}}_{\sigma_1} = \mathcal{X}^{\text{tP}}_{\sigma_2} &= \chi_{(3,3)}~,
\nonumber\\
\mathcal{X}^{\text{tP}}_{\varepsilon} &= \chi_{(5,2)} + \chi_{(5,5)}~,&
\mathcal{X}^{\text{tP}}_{\psi_1} = \mathcal{X}^{\text{tP}}_{\psi_2} &= \chi_{(3,1)}~,
\nonumber\\
\mathcal{X}^{\text{tP}}_{\varepsilon'} &= \chi_{(5,3)} + \chi_{(5,4)}~,&
\mathcal{X}^{\text{tP}}_{Z_1} = \mathcal{X}^{\text{tP}}_{Z_2} &= \chi_{(3,2)}~.
\end{align}
The modular matrix $\mathcal{S}_{\text{tP}}$ takes the form 
\begin{align}
S_{\text{tP}} = 
\frac{2}{\sqrt{21}}
\begin{pmatrix}
\text{s}_1 & \text{s}_3 & \text{s}_3 & \text{s}_2 & \text{s}_1 & \text{s}_1 & \text{s}_2 & \text{s}_2 & \text{s}_3 \\
\text{s}_3 & -\zeta\text{s}_2 & -\bar{\zeta}\text{s}_2 & \text{s}_1 & \zeta\text{s}_3 & \bar{\zeta}\text{s}_3 & \zeta\text{s}_1 & \bar{\zeta}\text{s}_1 & \text{s}_2 \\
\text{s}_3 & -\bar{\zeta}\text{s}_2 & -\zeta\text{s}_2 & \text{s}_1 & \bar{\zeta}\text{s}_3 & \zeta\text{s}_3 & \bar{\zeta}\text{s}_1 & \zeta\text{s}_1 & \text{s}_2 \\
\text{s}_2 & \text{s}_1 & \text{s}_1 & -\text{s}_3 & \text{s}_2 & \text{s}_2 & -\text{s}_3 & -\text{s}_3 & \text{s}_1 \\
\text{s}_1 & \zeta\text{s}_3 & \bar{\zeta}\text{s}_3 & \text{s}_2 & \zeta\text{s}_1 & \bar{\zeta}\text{s}_1 & \zeta\text{s}_2 & \bar{\zeta}\text{s}_2 & \text{s}_3 \\
\text{s}_1 & \bar{\zeta}\text{s}_3 & \zeta\text{s}_3 & \text{s}_2 & \bar{\zeta}\text{s}_1 & \zeta\text{s}_1 & \bar{\zeta}\text{s}_2 & \zeta\text{s}_2 & \text{s}_3 \\
\text{s}_2 & \zeta\text{s}_1 & \bar{\zeta}\text{s}_1 & -\text{s}_3 & \zeta\text{s}_2 & \bar{\zeta}\text{s}_2 & -\zeta\text{s}_3 & -\bar{\zeta}\text{s}_3 & \text{s}_1 \\
\text{s}_2 & \bar{\zeta}\text{s}_1 & \zeta\text{s}_1 & -\text{s}_3 & \bar{\zeta}\text{s}_2 & \zeta\text{s}_2 & -\bar{\zeta}\text{s}_3 & -\zeta\text{s}_3 & \text{s}_1 \\
\text{s}_3 & -\text{s}_2 & -\text{s}_2 & \text{s}_1 & \text{s}_3 & \text{s}_3 & \text{s}_1 & \text{s}_1 & -\text{s}_2 \\
\end{pmatrix}~,
\end{align}
where we use the shorthand $\zeta=\ee^{\frac{2\pi\ii}{3}}$ and $\text{s}_l=\sin(\pi l/7)$. The primaries $\{I,\sigma_1,\sigma_2, \varepsilon,\psi_1,\psi_2,Z_1,Z_2,\varepsilon' \}$ are ordered by conformal dimension.
The tricritical Potts model has a $\Z_3$ center generated by the current
\begin{align}
J_{\text{tP}} = \phi_{\psi_1}~,
\end{align}
with conformal dimension $h=4/3$. This current satisfies $( J_{\text{tP}})^2 = \phi_{\psi_2}$ and
\begin{align}
J_{\text{tP}}\times\phi_{\varepsilon} &= \phi_{Z_1}~,&
J_{\text{tP}}\times\phi_{Z_1} &= \phi_{Z_2}~,&
J_{\text{tP}}\times\phi_{Z_2} &= \phi_{\varepsilon}~,
\nonumber\\
J_{\text{tP}}\times\phi_{\varepsilon'} &= \phi_{\sigma_1}~,&
J_{\text{tP}}\times\phi_{\sigma_1} &= \phi_{\sigma_2}~,&
J_{\text{tP}}\times\phi_{\sigma_2} &= \phi_{\varepsilon'}~.
\end{align}
The monodromy charges of the extended primaries with respect to $J_{\text{tP}}$ are given by
\begin{align}
Q^{\text{tP}}_{I} = Q^{\text{tP}}_{\varepsilon} = Q^{\text{tP}}_{\varepsilon'} &= 0~,&
Q^{\text{tP}}_{\sigma_1} = Q^{\text{tP}}_{\psi_1} = Q^{\text{tP}}_{Z_1} &= \tfrac{1}{3}~,&
Q^{\text{tP}}_{\sigma_2} = Q^{\text{tP}}_{\psi_2} = Q^{\text{tP}}_{Z_2} &= \tfrac{2}{3}~.
\end{align}

\section{Kac--Moody algebras}
\label{app:KMconventions}
\subsection{Lie algebra conventions}
\label{app:Lie}
\begin{figure}[h!]
\centering
\begin{subfigure}{0.48\linewidth}
\centering
\begin{tikzpicture}[scale=.175]
\draw[thick] (-4,0) circle (0.5) node [shift={(0,-0.3)}] {$1$};
\draw[thick] (0,0) circle (0.5) node [shift={(0,-0.3)}] {$2$};
\draw[thick] (8,0) circle (0.5) node [shift={(0.0,-0.3)}] {$r-1$};
\draw[thick,fill] (4,4) circle (0.5) node [shift={(0.3,0.1)}] {$0$};
\draw[thick] (12,0) circle (0.5) node [shift={(0.0,-0.3)}] {$r$};
\draw (-4,0) node [shift={(0,-0.65)}] {{\small \textcolor{pink}{$1$}}};
\draw (0,0) node [shift={(0,-0.65)}] {{\small \textcolor{pink}{$1$}}};
\draw (8,0) node [shift={(0,-0.65)}] {{\small \textcolor{pink}{$1$}}};
\draw (12,0) node [shift={(0,-0.65)}] {{\small \textcolor{pink}{$1$}}};
\draw (4,4) node [shift={(0.55,0.1)}] {{\small \textcolor{pink}{$1$}}};
\draw[thick] (-3.5,0) -- (-0.5,0);
\draw[thick] (0.5,0) -- (2.5,0);
\draw[thick,dotted] (2.5,0) -- (5.5,0);
\draw[thick] (5.5,0) -- (7.5,0);
\draw[thick] (8.5,0) -- (11.5,0);
\draw[thick] (-3.65,0.3) -- (4,4);
\draw[thick] (11.65,0.3) -- (4,4);
\end{tikzpicture}
\caption{The $\mathfrak{a}_r$ diagram.}
\label{fig:EDDa}
\end{subfigure}
\begin{subfigure}{0.48\linewidth}
\centering
\begin{tikzpicture}[scale=.175]
\draw[thick] (0,0) circle (0.5) node [shift={(0,-0.3)}] {$2$};
\draw[thick] (3,3) circle (0.5) node [shift={(0.3,0)}] {$3$};
\draw[thick,fill] (-3,3) circle (0.5) node [shift={(-0.3,0)}] {$0$};
\draw[thick] (3,-3) circle (0.5) node [shift={(0.3,0)}] {$4$};
\draw[thick] (-3,-3) circle (0.5) node [shift
={(-0.3,0)}] {$1$};
\draw (0,0) node [shift={(0,-0.65)}] {{\small \textcolor{pink}{$2$}}};
\draw (3,3) node [shift={(0.55,0)}] {{\small \textcolor{pink}{$1$}}};
\draw (3,-3) node [shift={(0.55,0)}] {{\small \textcolor{pink}{$1$}}};
\draw (-3,3) node [shift={(-0.55,0)}] {{\small \textcolor{pink}{$1$}}};
\draw (-3,-3) node [shift={(-0.55,0)}] {{\small \textcolor{pink}{$1$}}};
\draw[thick] (0.35,0.35) -- (2.65,2.65);
\draw[thick] (-0.35,0.35) -- (-2.65,2.65);
\draw[thick] (0.35,-0.35) -- (2.65,-2.65);
\draw[thick] (-0.35,-0.35) -- (-2.65,-2.65);
\end{tikzpicture}
\caption{The $\mathfrak{d}_4$ diagram.}
\label{fig:EDDd4}
\end{subfigure}

\begin{subfigure}{0.48\linewidth}
\centering
\begin{tikzpicture}[scale=.175]
\draw[thick] (-4,0) circle (0.5) node [shift={(0,-0.3)}] {$1$};
\draw[thick] (0,0) circle (0.5) node [shift={(0,-0.3)}] {$3$};
\draw[thick] (4,0) circle (0.5) node [shift={(0.0,-0.3)}] {$4$};
\draw[thick] (8,0) circle (0.5) node [shift={(0.0,-0.3)}] {$5$};
\draw[thick] (4,4) circle (0.5) node [shift={(+0.3,0.0)}] {$2$};
\draw[thick,fill] (4,8) circle (0.5) node [shift
={(+0.3,0.0)}] {$0$};
\draw[thick] (12,0) circle (0.5) node [shift={(0.0,-0.3)}] {$6$};
\draw (-4,0) node [shift={(0,-0.65)}] {{\small \textcolor{pink}{$1$}}};
\draw (0,0) node [shift={(0,-0.65)}] {{\small \textcolor{pink}{$2$}}};
\draw (4,0) node [shift={(0,-0.65)}] {{\small \textcolor{pink}{$3$}}};
\draw (8,0) node [shift={(0,-0.65)}] {{\small \textcolor{pink}{$2$}}};
\draw (12,0) node [shift={(0,-0.65)}] {{\small \textcolor{pink}{$1$}}};
\draw (4,4) node [shift={(0.55,0)}] {{\small \textcolor{pink}{$2$}}};
\draw (4,8) node [shift={(0.55,0)}] {{\small \textcolor{pink}{$1$}}};
\draw[thick] (-3.5,0) -- (-0.5,0);
\draw[thick] (0.5,0) -- (3.5,0);
\draw[thick] (4.5,0) -- (7.5,0);
\draw[thick] (8.5,0) -- (11.5,0);
\draw[thick] (4,0.5) -- (4,3.5);
\draw[thick] (4,4.5) -- (4,7.5);
\end{tikzpicture}
\caption{The $\mathfrak{e}_6$ diagram.}
\label{fig:EDDe6}
\end{subfigure}
\begin{subfigure}{0.48\linewidth}
\centering
\begin{tikzpicture}[scale=.175]
\draw[thick,fill] (-8,0) circle (0.5) node [shift={(0,-0.3)}] {$0$};
\draw[thick] (-4,0) circle (0.5) node [shift={(0,-0.3)}] {$1$};
\draw[thick] (0,0) circle (0.5) node [shift={(0,-0.3)}] {$3$};
\draw[thick] (4,0) circle (0.5) node [shift={(0.0,-0.3)}] {$4$};
\draw[thick] (8,0) circle (0.5) node [shift={(0.0,-0.3)}] {$5$};
\draw[thick] (4,4) circle (0.5) node [shift={(+0.3,0.0)}] {$2$};
\draw[thick] (12,0) circle (0.5) node [shift={(0.0,-0.3)}] {$6$};
\draw[thick] (16,0) circle (0.5) node [shift={(0.0,-0.3)}] {$7$};
\draw (-8,0) node [shift={(0,-0.65)}] {{\small \textcolor{pink}{$1$}}};
\draw (-4,0) node [shift={(0,-0.65)}] {{\small \textcolor{pink}{$2$}}};
\draw (0,0) node [shift={(0,-0.65)}] {{\small \textcolor{pink}{$3$}}};
\draw (4,0) node [shift={(0,-0.65)}] {{\small \textcolor{pink}{$4$}}};
\draw (8,0) node [shift={(0,-0.65)}] {{\small \textcolor{pink}{$3$}}};
\draw (12,0) node [shift={(0,-0.65)}] {{\small \textcolor{pink}{$2$}}};
\draw (16,0) node [shift={(0,-0.65)}] {{\small \textcolor{pink}{$1$}}};
\draw (4,4) node [shift={(0.55,0)}] {{\small \textcolor{pink}{$2$}}};
\draw[thick] (-7.5,0) -- (-4.5,0);
\draw[thick] (-3.5,0) -- (-0.5,0);
\draw[thick] (0.5,0) -- (3.5,0);
\draw[thick] (4.5,0) -- (7.5,0);
\draw[thick] (8.5,0) -- (11.5,0);
\draw[thick] (12.5,0) -- (15.5,0);
\draw[thick] (4,0.5) -- (4,3.5);
\end{tikzpicture}
\caption{The $\mathfrak{e}_7$ diagram.}
\label{fig:EDDe7}
\end{subfigure}

\caption{Extended Dynkin diagrams of the simple Lie algebras appearing in the main text. The Kac marks are shown in color.}
\label{fig:EDD}
\end{figure}
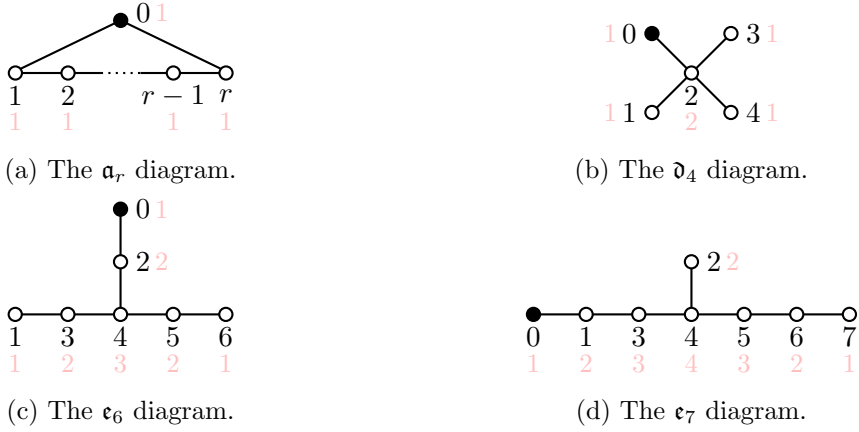

Let $\mathfrak{g}$ be a simple simply-laced Lie algebra of rank $r$. We pick a set of simple roots $\alpha_a$ labelled by $a=1,\dots,r$. The Killing form, denoted by $\cdot$, is normalized so that roots have length squared $2$. The Weyl vector $\rho_{\mathfrak{g}}$ is the half-sum of all the positive roots. The height of a root $\alpha$ is given by $\alpha\cdot\rho_{\mathfrak{g}}$. There is a unique root of maximal height, the highest root $\gamma_{\mathfrak{g}}$. The Kac marks $\kappa_a$ are the coefficients in the expansion
\begin{align}
\gamma_{\mathfrak{g}} = \kappa_1 \alpha_1+\dots+\kappa_r\alpha_r~.
\end{align}
The mark of the lowest root $\alpha_0=-\gamma_{\mathfrak{g}}$ is $\kappa_0=1$. 
The dual Coxeter number $h^\vee_{\mathfrak{g}}$ is obtained as the sum $h^\vee_{\mathfrak{g}}=\kappa_0+\kappa_1+\dots+\kappa_r$ of the Kac marks. The extended Cartan matrix $d_{ab}=\alpha_a\cdot\alpha_b$, with $a,b=0,1,\dots, r$, is encoded in an extended Dynkin diagram. The diagrams relevant in our analysis are depicted in figure~\ref{fig:EDD}.

The fundamental weights, denoted by $\omega_a$, are defined by $\omega_a\cdot\alpha_b = \delta_{ab}$. A weight $\lambda$ is characterized by its Dynkin labels $l_1,\dots,l_r$, which appear in the expansion
\begin{align}
\lambda = l_1\omega_1+\dots+l_r\omega_r~.
\end{align}
We use the shorthand $\lambda=[l_1l_2\cdots l_r]$. A weight is dominant if all of its Dynkin labels are non-negative. Every dominant weight $\lambda$ corresponds to a finite-dimensional irreducible representation $\boldsymbol{r}$ of $\mathfrak{g}$, with $\lambda$ as its highest weight. By abuse of notation, we use $\lambda$ or $\boldsymbol{r}$ interchangeably to refer to this representation. 

For a semi-simple Lie algebra
\begin{align}
\mathfrak{g}=\mathfrak{g}_1+\dots+\mathfrak{g}_N~,
\end{align}
we denote irreducible representations by $\rep{r}=(\rep{r}_1,\dots,\rep{r}_N)$. The corresponding highest weights are grouped in $\lambda=(\lambda_1,\dots,\lambda_N)$. When using the Dynkin label notation, we separate the several simple factors by a comma. For instance, if $\lambda_1=[l_1\cdots l_r]$ and $\lambda_2=[l_{r+1}\cdots l_{r+r'}]$, then we write $\lambda=[l_1\cdots l_{r},l_{r+1}\cdots l_{r+r'}]$.

\subsection{Highest weight representations}
Let $\mathfrak{g}$ be, as before, a simple ADE-type Lie algebra. The associated Kac--Moody algebra at level $k$, denoted by $\mathfrak{g}_k$, has central charge
\begin{align}
c_{\mathfrak{g}_k} = \frac{k\operatorname{dim}\mathfrak{g}}{k+h^\vee_\mathfrak{g}}~.
\end{align}
The KM primary fields $\phi_\lambda$ correspond to dominant weights $\lambda$ satisfying the unitarity condition
\begin{align}
\lambda\cdot\gamma_{\mathfrak{g}}\leq k~.
\end{align}
In terms of the Dynkin labels 
$\lambda = [l_1l_2\cdots l_r]$, this condition reads
\begin{align}
\sum_{a=1}^r \kappa_a l_a\leq k~.
\end{align}
Hence, the number $\ell(\mathfrak{g}_k)$ of primary representations of $\mathfrak{g}_k$ is finite. For the affine algebras relevant to our discussion, this number is
\begin{align}
\ell (\mathfrak{a}_{r,k}) &=\frac{(k+r)!}{k!\,r!} ~,&
\ell(\mathfrak{d}_{4,6}) &= 130~,&
\ell(\mathfrak{e}_{6,8}) &= 372~,&
\ell(\mathfrak{e}_{7,2}) &= 6 ~.
\label{eq:numberKMVprimaries}
\end{align}
The conformal dimension $h_\lambda$ of a primary field $\phi_\lambda$ is given by
\begin{align}
h_\lambda = \frac{\lambda\cdot(\lambda+2\rho_{\mathfrak{g}}) }{2(k+h^\vee_\mathfrak{g})}~,
\end{align}
and the corresponding KM characters $\chi_\lambda$ transform under $\mathrm{SL}(2,\Z)$ as
\begin{align}
\chi_\lambda(\tau+1,\xi) &= \sum_{\lambda'} \mathcal{T}_{\lambda\lambda'} \chi_{\lambda'}(\tau,\xi)~,&
\chi_\lambda(-1/\tau,\xi/\tau) &= \ee^{\pi\ii k\xi\cdot\xi/\tau}\sum_{\lambda'} \mathcal{S}_{\lambda\lambda'} \chi_{\lambda'}(\tau,\xi)~.
\end{align}
Here, $\mathcal{T}$ is the diagonal matrix
\begin{align}
\mathcal{T}_{\lambda\nu} = \ee^{2\pi\ii(h_\lambda-c_{\mathfrak{g}_k}/24)}\delta_{\lambda\nu}~,
\label{eq:TmatrixKM}
\end{align}
while the modular $\mathcal{S}$ matrix is given by the Kac--Peterson formula
\begin{align}
\mathcal{S}_{\lambda\nu} = \frac{\ee^{\pi\ii|\Delta_{\mathfrak{g}}|/4}}{\sqrt{|\Lambda_{\mathfrak{g}}|(k+h^\vee_{\mathfrak{g}})}}\sum_{R\in W_{\mathfrak{g}}} \epsilon_R\; \ee^{-2\pi\ii(\lambda+\rho_{\mathfrak{g}})\cdot R(\nu+\rho_{\mathfrak{g}})/(k+h^\vee_{\mathfrak{g}}) }~.
\label{eq:KacPeterson}
\end{align}
In this expression, $|\Delta_{\mathfrak{g}}|$  denotes the number of roots of $\mathfrak{g}$, $|\Lambda_{\mathfrak{g}}|$ is the determinant of the root lattice, $W_{\mathfrak{g}}$ is the Weyl group of $\mathfrak{g}$, and $\epsilon_R$ is the signature of the Weyl element $R$.

\subsection{Modular data of \texorpdfstring{$\mathfrak{a}_{11,2}$}{a11,2}}
\label{app:Sa11lvl2}
To determine the modular matrix of the $\mathfrak{a}_{11}$ KM algebra at level $2$, we rely on the conformal embedding\footnote{This is a special case of the general conformal embedding $\mathfrak{a}_{l-1,k} + \mathfrak{a}_{k-1,l} \subset \mathfrak{a}_{kl-1,1}$.}
\begin{align}
\mathfrak{a}_{1,12}+\mathfrak{a}_{11,2} \;\subset\; \mathfrak{a}_{23,1}~.
\end{align}
The branching of characters reads
\begin{align}
\chi^{\mathfrak{a}_{23,1}}_{\lambda}(\tau) &= \sum_{\mu,\nu} b_{\lambda\mu\nu}\;\chi^{\mathfrak{a}_{1,12}}_{\mu}(\tau)\chi^{\mathfrak{a}_{11,2}}_{\nu}(\tau)~,
\end{align}
where $\lambda$, $\mu$ and $\nu$ respectively denote integrable highest weight representations of $\mathfrak{a}_{1,12}$, $\mathfrak{a}_{11,2}$, and $\mathfrak{a}_{23,1}$. The constant coefficients $b_{\lambda\mu\nu}$ are easily determined from a known formula~\cite{Altschuler:1989nm}. They enter the relation
\begin{align}
\sum_{\lambda',\mu',\nu'}b_{\lambda'\mu'\nu'}\,\overline{\mathcal{S}}^{\mathfrak{a}_{23,1}}_{\lambda\lambda'}\mathcal{S}^{\mathfrak{a}_{1,12}}_{\mu\mu'}\mathcal{S}^{\mathfrak{a}_{11,2}}_{\nu\nu'}= b_{\lambda\mu\nu}~.
\label{eq:branchingSmatrix}
\end{align}
In this expression, the modular matrices of $\mathfrak{a}_{23,1}$ and $\mathfrak{a}_{1,12}$ are straightforwardly determined: they take the form 
\begin{align}
\mathcal{S}^{\mathfrak{a}_{23,1}}_{ii'} &= \frac{1}{\sqrt{24}}\ee^{2\pi \ii \,i i'/kl}~,&
\mathcal{S}^{\mathfrak{a}_{1,12}}_{jj'} &= \frac{1}{\sqrt{7}}\sin\left(\tfrac{\pi}{14}(j+1)(j'+1)\right)~,
\end{align}
with $i,i'=0,1,\dots,24$ and $j,j'=0,1,\dots,12$. Hence, the relation~\eqref{eq:branchingSmatrix} can be interpreted as a linear system determining the elements of $\mathcal{S}^{\mathfrak{a}_{11,2}}$.

The $\mathfrak{a}_{11}$ KM algebra at level $2$ possesses $78$ unitary highest weight representations. Its center, isomorphic to $\Z_{12}$, is generated by the simple current $\phi_\rep{78}=\phi_{[20000000000]}$. The primary fields can be grouped into seven distinct $\Z_{12}$  orbits, obtained by the action of $\phi_\rep{78}$ on
\begin{align}
\rep{1} &= [00000000000]~,&
\rep{12} &= [10000000000]~,&
\rep{66} &= [01000000000]~,\nonumber\\
\rep{220} &= [00100000000]~,&
\rep{495} &= [00010000000]~,&
\rep{792} &= [00001000000]~,\nonumber\\
\rep{924} &= [00000100000]~.
\end{align}
The entries of the modular matrix $\mathcal{S}^{\mathfrak{a}_{11,2}}$ can be extracted from~\eqref{eq:branchingSmatrix}. For convenience, we write them as
\begin{align}
\mathcal{S}^{\mathfrak{a}_{11,2}}_{\rep{r},\rep{r}'} = \frac{\text{x}_{\rep{r},\rep{r}'}}{\sqrt{168}}~.
\end{align}
Our computation yields
\begin{align}
\text{x}_{\rep{1},\rep{1}} &= \text{x}_{\rep{66},\rep{495}}\zeta^{28} = \zeta^{36}+\zeta^{132}~,
\nonumber\\
\text{x}_{\rep{1},\rep{12}} &= \text{x}_{\rep{12},\rep{792}}\zeta^{133} = \text{x}_{\rep{66},\rep{220}}\zeta^{126} = \text{x}_{\rep{220},\rep{220}}\zeta^{21} = \text{x}_{\rep{495},\rep{792}}\zeta^{28} = \zeta^{30}+\zeta^{138}~,
\nonumber\\
\text{x}_{\rep{1},\rep{66}} &= \text{x}_{\rep{495},\rep{495}}\zeta^{140} =  \zeta^{24}+\zeta^{144}~,
\nonumber\\
\text{x}_{\rep{1},\rep{220}} &= \text{x}_{\rep{12},\rep{12}}\zeta^{161} = \text{x}_{\rep{12},\rep{495}}\zeta^{140} = \text{x}_{\rep{66},\rep{792}}\zeta^{14} = \text{x}_{\rep{220},\rep{792}}\zeta^{147} = \zeta^{18}+\zeta^{150}~,
\nonumber\\
\text{x}_{\rep{1},\rep{495}} &= \text{x}_{\rep{66},\rep{66}}\zeta^{140} = \zeta^{12}+\zeta^{156}~,
\nonumber\\
\text{x}_{\rep{1},\rep{792}} &= \text{x}_{\rep{12},\rep{66}}\zeta^{154} = \text{x}_{\rep{12},\rep{220}}\zeta^{147} = \text{x}_{\rep{220},\rep{495}} = \text{x}_{\rep{792},\rep{792}}\zeta^{161} = \zeta^{6}+\zeta^{162}~,
\nonumber\\
\text{x}_{\rep{1},\rep{924}} &= \text{x}_{\rep{66},\rep{924}} = \text{x}_{\rep{495},\rep{924}} = \text{x}_{\rep{924},\rep{924}} = 2~,
\nonumber\\
\text{x}_{\rep{12},\rep{924}} &= \text{x}_{\rep{220},\rep{924}} = \text{x}_{\rep{792},\rep{924}} =0~,
\end{align}
where $\zeta=\ee^{2\pi\ii/168}$. The remaining elements are derived from property~\eqref{eq:Smatrixmonodromy}.

\subsection{Modular data of \texorpdfstring{$\mathfrak{e}_{7,2}$}{e7,2}}
\label{app:Se7lvl2}
The $\mathfrak{e}_7$ KM algebra at level $1$ has central charge $c_{\mathfrak{e}_{7,1}} = 7$. It admits two primary representations, denoted $\rep{1}$ and $\rep{56}$, with conformal dimension
\begin{align}
h^{\mathfrak{e}_{7,1}}_\rep{1} &= 0~,&
h^{\mathfrak{e}_{7,1}}_\rep{56} &= \tfrac{3}{4}~.
\end{align}
The primary field $\phi_{\rep{56}}$ is a simple current of order two. Exploiting the symmetry~\eqref{eq:Smatrixmonodromy} of the modular matrix $\mathcal{S}_{\mathfrak{e}_{7,1}}$ under the action of this simple current, one readily obtains
\begin{align}
\mathcal{S}_{\mathfrak{e}_{7,1}} = 
\frac{1}{\sqrt{2}}
\begin{pmatrix}
1&1\\
1&-1\\
\end{pmatrix}~.
\end{align}
At level $2$, the $\mathfrak{e}_7$ KM algebra has central charge $c_{\mathfrak{e}_{7,2}} =133/10$. It admits six primary representations,
\begin{align}
\rep{1} &= [0000000]~,&
\rep{56} &= [0000001]~,&
\rep{133} &= [1000000]~,\nonumber\\
\rep{912} &= [0100000]~,&
\rep{1463} &= [0000002]~,&
\rep{1539} &= [0000010]~,
\end{align}
whose conformal dimensions are
\begin{align}
h^{\mathfrak{e}_{7,2}}_\rep{1} &= 0~,&
h^{\mathfrak{e}_{7,2}}_\rep{56} &= \tfrac{57}{80}~,&
h^{\mathfrak{e}_{7,2}}_\rep{133} &= \tfrac{9}{10}~,&
h^{\mathfrak{e}_{7,2}}_\rep{912} &= \tfrac{21}{16}~,&
h^{\mathfrak{e}_{7,2}}_\rep{1463} &= \tfrac{3}{2}~,&
h^{\mathfrak{e}_{7,2}}_\rep{1539} &= \tfrac{7}{5}~.
\end{align}
The diagonal coset
\begin{align}
\frac{\mathfrak{e}_{7,1}+ \mathfrak{e}_{7,1}}{\mathfrak{e}_{7,2}}
\label{eq:e7coset}
\end{align}
has central charge $c=7/10$: it is isomorphic to the tricritical Ising model $\mathfrak{M}_{\text{tI}} = \mathfrak{M}_{(5,4)}$. The latter has six primary fields
\begin{align}
\phi^{\text{tI}}_{I} &= \phi^{\text{tI}}_{(1,1)}~,&
\phi^{\text{tI}}_{\sigma} &= \phi^{\text{tI}}_{(2,2)}~,&
\phi^{\text{tI}}_{\varepsilon} &= \phi^{\text{tI}}_{(3,3)}~,\nonumber\\
\phi^{\text{tI}}_{\sigma'} &= \phi^{\text{tI}}_{(2,1)}~,&
\phi^{\text{tI}}_{\varepsilon'} &= \phi^{\text{tI}}_{(3,2)}~,&
\phi^{\text{tI}}_{\varepsilon''} &= \phi^{\text{tI}}_{(3,1)}~,
\end{align}
with conformal weights
\begin{align}
h^{\text{tI}}_I &= 0~,&
h^{\text{tI}}_\sigma &= \tfrac{3}{80}~,&
h^{\text{tI}}_\varepsilon &= \tfrac{1}{10}~,&
h^{\text{tI}}_{\sigma'} &= \tfrac{7}{16}~,&
h^{\text{tI}}_{\varepsilon'} &= \tfrac{3}{5}~,&
h^{\text{tI}}_{\varepsilon''} &= \tfrac{3}{2}~.
\end{align}
The branching of characters from $\mathfrak{e}_{7,1}+\mathfrak{e}_{7,1}$ to $\mathfrak{e}_{7,2}+\mathfrak{M}_{\text{tI}}$ reads
\begin{align}
\chi^{\mathfrak{e}_{7,1}}_{\rep{1}}\chi^{\mathfrak{e}_{7,1}}_{\rep{1}} &= \chi^{\mathfrak{e}_{7,2}}_{\rep{1}}\chi^{\text{tI}}_I+\chi^{\mathfrak{e}_{7,2}}_{\rep{133}}\chi^{\text{tI}}_\varepsilon+\chi^{\mathfrak{e}_{7,2}}_{\rep{1463}}\chi^{\text{tI}}_{\varepsilon''}+\chi^{\mathfrak{e}_{7,2}}_{\rep{1539}}\chi^{\text{tI}}_{\varepsilon'}~,
\nonumber\\
\chi^{\mathfrak{e}_{7,1}}_{\rep{1}}\chi^{\mathfrak{e}_{7,1}}_{\rep{56}} &= \chi^{\mathfrak{e}_{7,2}}_{\rep{56}}\chi^{\text{tI}}_\sigma+\chi^{\mathfrak{e}_{7,2}}_{\rep{912}}\chi^{\text{tI}}_{\sigma'}~,
\nonumber\\
\chi^{\mathfrak{e}_{7,1}}_{\rep{56}}\chi^{\mathfrak{e}_{7,1}}_{\rep{56}} &= \chi^{\mathfrak{e}_{7,2}}_{\rep{1}}\chi^{\text{tI}}_{\varepsilon''}+\chi^{\mathfrak{e}_{7,2}}_{\rep{133}}\chi^{\text{tI}}_{\varepsilon'}+\chi^{\mathfrak{e}_{7,2}}_{\rep{1463}}\chi^{\text{tI}}_I+\chi^{\mathfrak{e}_{7,2}}_{\rep{1539}}\chi^{\text{tI}}_\varepsilon~.
\end{align}
From this decomposition, one may deduce the modular matrix of the $\mathfrak{e}_{7,2}$ algebra: 
\begin{align}
\mathcal{S}_{\mathfrak{e}_{7,2}} = 
\frac{1}{2\sqrt{5}}
\begin{pmatrix}
\text{x}_1 & \text{x}_2 & \text{x}_3 & \text{x}_4 & \text{x}_1 & \text{x}_3 \\
\text{x}_2 & 0 & \text{x}_4 & 0 & -\text{x}_2 & -\text{x}_4 \\
\text{x}_3 & \text{x}_4 & -\text{x}_1 & -\text{x}_2 & \text{x}_3 & -\text{x}_1 \\
\text{x}_4 & 0 & -\text{x}_2 & 0 & -\text{x}_4 & \text{x}_2 \\
\text{x}_1 & -\text{x}_2 & \text{x}_3 & -\text{x}_4 & \text{x}_1 & \text{x}_3 \\
\text{x}_3 & -\text{x}_4 & -\text{x}_1 & \text{x}_2 & \text{x}_3 & -\text{x}_1 \\
\end{pmatrix}
\end{align}
where 
\begin{align}
\text{x}_1 &= -\zeta^{26}+\zeta^{34}~,&
\text{x}_2 &= \zeta^{7}-\zeta^{13}-\zeta^{23}+\zeta^{37}~,\nonumber\\
\text{x}_3 &= \zeta^{2}-\zeta^{18}~,&
\text{x}_4 &= -\zeta^{21}+\zeta^{29}-\zeta^{31}+\zeta^{39}~,
\end{align}
and $\zeta=\ee^{2\pi\ii/40}$.

\subsection{Mixed KMV extensions and fixed point resolutions}
\label{app:mixedKMV}
In the following, we illustrate the fixed point resolution of KMV simple current extensions with an example. We consider the direct product of the $\mathfrak{e}_7$ KM algebra at level $2$ and the tricritical Ising model. This KMV algebra of central charge $c=14$ has a $\Z_2\times\Z_2$ center symmetry, generated by the center of $\mathfrak{e}_7$ and the $\Z_2$ symmetry of the minimal model. The corresponding simple currents, which we denote by $J_{\mathfrak{e}_7}$ and $J_{\text{tI}}$, both have conformal dimension equal to $\frac{3}{2}$. The simple current $J_{\mathfrak{e}_7}$ belongs to the representation $\rep{1463}$ of $\mathfrak{e}_7$, and its fusion with the other $\mathfrak{e}_7$ primaries generates the permutation
\begin{align}
\{ \rep{1},\rep{56},\rep{133},\rep{912},\rep{1463},\rep{1539} \} \to \{ \rep{1463},\rep{56},\rep{1539},\rep{912},\rep{1},\rep{133} \}~.
\end{align}
The current $J_{\text{tI}}$ corresponds to the operator $\varepsilon''$ of the tricritical Ising model.\footnote{See appendix~\ref{app:Se7lvl2} for our minimal model conventions.} Its fusion with the other Virasoro primaries yields
\begin{align}
\{ I,\sigma,\varepsilon,\sigma',\varepsilon',\varepsilon''\} \to \{ \varepsilon'',\sigma,\varepsilon',\sigma',\varepsilon,I\}~.
\end{align}
The product $J_{\mathfrak{e}_7}J_{\text{tI}}$ is a spin-$3$ simple current of the KMV algebra. This mixed current can be used to extend the chiral algebra. After projecting onto neutral states, one finds four fixed points of the $\Z_2$ action, associated to the KMV representations
\begin{align}
&\rep{56}\otimes\sigma~,&
&\rep{56}\otimes\sigma'~,&
&\rep{912}\otimes\sigma~,&
&\rep{912}\otimes\sigma'~.
\end{align}
These fixed points needs to be resolved in order to obtain the modular properties of the extended theory.

After the fixed point resolution, the extended algebra has $16$ characters. Eight of them correspond to the neutral $\Z_2$ orbits of length $2$:
\begin{align}
\Chi_0 &= \chi^{\mathfrak{e}_7}_\rep{1}\chi^{\text{tI}}_{I}+\chi^{\mathfrak{e}_7}_\rep{1463}\chi^{\text{tI}}_{\varepsilon''}~,&
\Chi_1 &= \chi^{\mathfrak{e}_7}_\rep{1}\chi^{\text{tI}}_{\varepsilon''}+\chi^{\mathfrak{e}_7}_\rep{1463}\chi^{\text{tI}}_{I}~,\nonumber\\
\Chi_2 &= \chi^{\mathfrak{e}_7}_\rep{1}\chi^{\text{tI}}_{\varepsilon}+\chi^{\mathfrak{e}_7}_\rep{1463}\chi^{\text{tI}}_{\varepsilon'}~,&
\Chi_3 &= \chi^{\mathfrak{e}_7}_\rep{1}\chi^{\text{tI}}_{\varepsilon'}+\chi^{\mathfrak{e}_7}_\rep{1463}\chi^{\text{tI}}_{\varepsilon}~,\nonumber\\
\Chi_4 &= \chi^{\mathfrak{e}_7}_\rep{133}\chi^{\text{tI}}_{I}+\chi^{\mathfrak{e}_7}_\rep{1539}\chi^{\text{tI}}_{\varepsilon''}~,&
\Chi_5 &= \chi^{\mathfrak{e}_7}_\rep{133}\chi^{\text{tI}}_{\varepsilon''}+\chi^{\mathfrak{e}_7}_\rep{1539}\chi^{\text{tI}}_{I}~,\nonumber\\
\Chi_6 &= \chi^{\mathfrak{e}_7}_\rep{133}\chi^{\text{tI}}_{\varepsilon}+\chi^{\mathfrak{e}_7}_\rep{1539}\chi^{\text{tI}}_{\varepsilon'}~,&
\Chi_7 &= \chi^{\mathfrak{e}_7}_\rep{133}\chi^{\text{tI}}_{\varepsilon'}+\chi^{\mathfrak{e}_7}_\rep{1539}\chi^{\text{tI}}_{\varepsilon}~,
\end{align}
while the remaining ones originate from the four fixed points:
\begin{align}
\Chi_{8} &=\Chi_{9} = \chi^{\mathfrak{e}_7}_\rep{56}\chi^{\text{tI}}_{\sigma}~,&
\Chi_{10} &= \Chi_{11} = \chi^{\mathfrak{e}_7}_\rep{56}\chi^{\text{tI}}_{\sigma'}~,\nonumber\\
\Chi_{12} &= \Chi_{13} = \chi^{\mathfrak{e}_7}_\rep{912}\chi^{\text{tI}}_{\sigma}~,&
\Chi_{14} &= \Chi_{15} = \chi^{\mathfrak{e}_7}_\rep{912}\chi^{\text{tI}}_{\sigma'}~.
\end{align}
Let us denote by $\Phi_i$ the extended primaries associated to the characters $\Chi_i$. The conformal weights of the lowest Virasoro states in each extended family are given by
\begin{align}
h_0 &= 0~,& 
h_1 &= \tfrac{3}{2}~,& 
h_2 &= \tfrac{1}{10}~,& 
h_3 &= \tfrac{3}{5}~,& 
h_4 &= \tfrac{9}{10}~,& 
h_5 &= \tfrac{7}{5}~,& 
h_6 &= 1~,& 
h_7 &= \tfrac{3}{2}~,
\nonumber\\
h_8 &= \tfrac{3}{4}~,& 
h_9 &= \tfrac{3}{4}~,& 
h_{10} &= \tfrac{23}{20}~,& 
h_{11} &= \tfrac{23}{20}~,& 
h_{12} &= \tfrac{27}{20}~,& 
h_{13} &= \tfrac{27}{20}~,& 
h_{14} &= \tfrac{7}{4}~,& 
h_{15} &= \tfrac{7}{4}~.
\end{align}
They unambiguously define the modular $\mathcal{T}$ matrix. On the other hand, the $\mathcal{S}$ matrix is not entirely determined by the $\mathrm{SL}(2,\Z)$ properties of the KMV algebra: the entries associated to fixed points need to be worked out. After some computations, we find
\begin{align}
\mathcal{S} = \left(
\begin{array}{cccccccccccccccc}
\text{x}&\text{x}&\text{y}&\text{y}&\text{y}&\text{y}&\text{z}&\text{z}&\text{z}&\text{z}&\text{y}&\text{y}&\text{y}&\text{y}&\text{x}&\text{x}\\
\text{x}&\text{x}&\text{y}&\text{y}&\text{y}&\text{y}&\text{z}&\text{z}&-\text{z}&-\text{z}&-\text{y}&-\text{y}&-\text{y}&-\text{y}&-\text{x}&-\text{x}\\
\text{y}&\text{y}&-\text{x}&-\text{x}&\text{z}&\text{z}&-\text{y}&-\text{y}&\text{y}&\text{y}&-\text{z}&-\text{z}&\text{x}&\text{x}&-\text{y}&-\text{y}\\
\text{y}&\text{y}&-\text{x}&-\text{x}&\text{z}&\text{z}&-\text{y}&-\text{y}&-\text{y}&-\text{y}&\text{z}&\text{z}&-\text{x}&-\text{x}&\text{y}&\text{y}\\
\text{y}&\text{y}&\text{z}&\text{z}&-\text{x}&-\text{x}&-\text{y}&-\text{y}&\text{y}&\text{y}&\text{x}&\text{x}&-\text{z}&-\text{z}&-\text{y}&-\text{y}\\
\text{y}&\text{y}&\text{z}&\text{z}&-\text{x}&-\text{x}&-\text{y}&-\text{y}&-\text{y}&-\text{y}&-\text{x}&-\text{x}&\text{z}&\text{z}&\text{y}&\text{y}\\
\text{z}&\text{z}&-\text{y}&-\text{y}&-\text{y}&-\text{y}&\text{x}&\text{x}&\text{x}&\text{x}&-\text{y}&-\text{y}&-\text{y}&-\text{y}&\text{z}&\text{z}\\
\text{z}&\text{z}&-\text{y}&-\text{y}&-\text{y}&-\text{y}&\text{x}&\text{x}&-\text{x}&-\text{x}&\text{y}&\text{y}&\text{y}&\text{y}&-\text{z}&-\text{z}\\
\text{z}&-\text{z}&\text{y}&-\text{y}&\text{y}&-\text{y}&\text{x}&-\text{x}&-\text{x}&\text{x}&\text{y}&-\text{y}&\text{y}&-\text{y}&\text{z}&-\text{z}\\
\text{z}&-\text{z}&\text{y}&-\text{y}&\text{y}&-\text{y}&\text{x}&-\text{x}&\text{x}&-\text{x}&-\text{y}&\text{y}&-\text{y}&\text{y}&-\text{z}&\text{z}\\
\text{y}&-\text{y}&-\text{z}&\text{z}&\text{x}&-\text{x}&-\text{y}&\text{y}&\text{y}&-\text{y}&\text{x}&-\text{x}&-\text{z}&\text{z}&\text{y}&-\text{y}\\
\text{y}&-\text{y}&-\text{z}&\text{z}&\text{x}&-\text{x}&-\text{y}&\text{y}&-\text{y}&\text{y}&-\text{x}&\text{x}&\text{z}&-\text{z}&-\text{y}&\text{y}\\
\text{y}&-\text{y}&\text{x}&-\text{x}&-\text{z}&\text{z}&-\text{y}&\text{y}&\text{y}&-\text{y}&-\text{z}&\text{z}&\text{x}&-\text{x}&\text{y}&-\text{y}\\
\text{y}&-\text{y}&\text{x}&-\text{x}&-\text{z}&\text{z}&-\text{y}&\text{y}&-\text{y}&\text{y}&\text{z}&-\text{z}&-\text{x}&\text{x}&-\text{y}&\text{y}\\
\text{x}&-\text{x}&-\text{y}&\text{y}&-\text{y}&\text{y}&\text{z}&-\text{z}&\text{z}&-\text{z}&\text{y}&-\text{y}&\text{y}&-\text{y}&-\text{x}&\text{x}\\
\text{x}&-\text{x}&-\text{y}&\text{y}&-\text{y}&\text{y}&\text{z}&-\text{z}&-\text{z}&\text{z}&-\text{y}&\text{y}&-\text{y}&\text{y}&\text{x}&-\text{x}\\
\end{array}
\right)~,
\end{align}
where
\begin{align}
\text{x} &= \frac{5-\sqrt{5}}{20}~,&
\text{y} &= \frac{\sqrt{5}}{10}~,&
\text{z} &= \frac{5+\sqrt{5}}{20}~.
\end{align}
The patient reader can check that this modular matrix is symmetric, squares to the identity, and is compatible with the condition $\mathcal{S}^2=(\mathcal{S}\mathcal{T})^3$.

A few comments are in order. First, we remark that the extended primary $\Phi_6$ contains spin-$1$ currents transforming in the adjoint representation of $\mathfrak{e}_7$. This is not so surprising: the tricritical Ising model can be realized as the diagonal coset~\eqref{eq:e7coset}, so it should be possible to identify two commuting copies of the $\mathfrak{e}_7$ KM algebra at level $1$ in the product theory. These two factors survive the extension by the spin-$3$ simple current, and one can identify four combination of characters
\begin{align}
\chi^{\mathfrak{e}_7+\mathfrak{e}_7}_{\rep{1}\times\rep{1}} &= \Chi_0 + \Chi_6~,&
\chi^{\mathfrak{e}_7+\mathfrak{e}_7}_{\rep{56}\times\rep{56}} &= \Chi_1 + \Chi_7~,&
\chi^{\mathfrak{e}_7+\mathfrak{e}_7}_{\rep{1}\times\rep{56}} &= \Chi_8 + \Chi_{15}~,&
\chi^{\mathfrak{e}_7+\mathfrak{e}_7}_{\rep{56}\times\rep{1}} &= \Chi_9 + \Chi_{14}~,
\end{align}
which reconstruct the unitary representations of the $\mathfrak{e}_7+\mathfrak{e}_7$ current algebra. Let us stress that, for the (0,4) SCFT containing the present $c_\tL=14$ KMV subalgebra, the spin-$1$ currents contained in the extended primary $\Phi_6$ do not appear in the spectrum: only the $\mathfrak{e}_7$ KM symmetry associated to $\Phi_0$ is realized.

Second, we observe that the extended algebra admits three nontrivial simple currents, given by
\begin{align}
J &= \Phi_1~,&
J' &= \Phi_{14}~,&
J'' &= \Phi_{15}~.
\end{align}
While $J$ descends from the center of the original KMV algebra, the currents $J'$ and $J''$ are genuinely new and arise only after the extension. Their action on the extended primaries,
\begin{align}
J\times\Phi_i &= \Phi_{\sigma(i)}~,&
J'\times\Phi_i &= \Phi_{\sigma'(i)}~,&
J''\times\Phi_i &= \Phi_{\sigma''(i)}~,
\end{align}
can be determined using the Verlinde formula. One finds
\begin{align}
\sigma &= \{ 1,0,3,2,5,4,7,6,9,8,11,10,13,12,15,14 \}~,
\nonumber\\
\sigma' &= \{ 14,15,12,13,10,11,9,8,7,6,4,5,2,3,0,1 \}~,
\nonumber\\
\sigma'' &= \{ 15,14,13,12,11,10,8,9,6,7,5,4,3,2,1,0 \}~,
\end{align}
from which it follows that the center of the extended theory is isomorphic to $\Z_2\times\Z_2$.

\bibliographystyle{./utphys}
\bibliography{rationalBPSstrings}

\end{document}